%%%
%%% Paper on noncommutativity of fluxes in generalized cohomology theories
%%%
%%% March 18, 2005
%%%
%%% May 14, 2005
%%%
%%% June 8, 2005
%%%
%%%
%%% Major revisions, August 13, 2005
%%%
%%% August 20, 2005
%%% August 26, 2005

%%% Many revisions, Sept. 4, 2005
%%%
%%% Further revisions, Oct. 11, 2005
%%% Further revisions, Oct. 18, 2005
%%% Further revisions, Oct. 24, 2005
%%% Further revisions, Oct. 30, 2005
%%% Further revisions, Oct. 31, 2005
%%% Further revisions, Nov. 5, 2005
%%% Further revisions, Nov. 15, 2005

%%% Further small typos fixed, March 2, 2006

%%% Final version? March 16, 2006

%%% Some minor changes, May 15, 2006

\input epsf
\input harvmac.tex

%%AMS FONTS

%\input amssym.def
%\input amssym.tex

%%

%% Theorems, Definitions, etc.
\def\defn#1{\bigskip\noindent{\bf Definition #1} }
\def\thm#1{\bigskip\noindent{\bf Theorem #1} }

\def\expl#1{\bigskip\noindent{\bf Example #1} }
\def\rmk#1{\bigskip\noindent{\bf Remarks} }
%%

%%% macros for figures:

\def\figin{\epsfcheck\figin}\def\figins{\epsfcheck\figins}
\def\epsfcheck{\ifx\epsfbox\UnDeFiNeD
\message{(NO epsf.tex, FIGURES WILL BE IGNORED)}
\gdef\figin##1{\vskip2in}\gdef\figins##1{\hskip.5in}% blank space instead
\else\message{(FIGURES WILL BE INCLUDED)}%
\gdef\figin##1{##1}\gdef\figins##1{##1}\fi}
\def\DefWarn#1{}
\def\figinsert{\goodbreak\midinsert}
\def\ifig#1#2#3{\DefWarn#1\xdef#1{fig.~\the\figno}
\writedef{#1\leftbracket fig.\noexpand~\the\figno}%
\figinsert\figin{\centerline{#3}}\medskip\centerline{\vbox{\baselineskip12pt
\advance\hsize by -1truein\noindent\footnotefont{\bf
Fig.~\the\figno:} #2}}
\bigskip\endinsert\global\advance\figno by1}

%%%%

\def\IC{\relax{\rm I\kern-.18em C}}

\def\IL{\relax{\rm I\kern-.18em L}}
\def\IH{\relax{\rm I\kern-.18em H}}
\def\IR{\relax{\rm I\kern-.18em R}}
\def\IC{\relax\hbox{$\inbar\kern-.3em{\rm C}$}}
\def\IZ{\relax\ifmmode\mathchoice
{\hbox{\cmss Z\kern-.4em Z}}{\hbox{\cmss Z\kern-.4em Z}}
{\lower.9pt\hbox{\cmsss Z\kern-.4em Z}}
{\lower1.2pt\hbox{\cmsss Z\kern-.4em Z}}\else{\cmss Z\kern-.4em Z}\fi}
\def\CA{{\cal A}}
\def\CB {{\cal B}}
\def\CC {{\cal C}}

\def\CN {{\cal N}}
\def\CR {{\cal R}}

\def\CF {{\cal F}}
\def\CP {{\cal P }}
\def\CL {{\cal L}}

\def\CW {{\cal W}}
\def\CZ {{\cal Z}}
\def\CE {{\cal E}}
\def\CG {{\cal G}}
\def\CH {{\cal H}}
\def\CS {{\cal S}}
\def\CX {{\cal X}}

\font\manual=manfnt \def\dbend{\lower3.5pt\hbox{\manual\char127}}

\def\c{\check}
\def\IZ{\relax\ifmmode\mathchoice
{\hbox{\cmss Z\kern-.4em Z}}{\hbox{\cmss Z\kern-.4em Z}}
{\lower.9pt\hbox{\cmsss Z\kern-.4em Z}}
{\lower1.2pt\hbox{\cmsss Z\kern-.4em Z}}\else{\cmss Z\kern-.4em Z}\fi}
\def\half {{1\over 2}}

\def\p{\partial}

\def\CL {{\cal L}}

\def\CN {{\cal N}}

\def\CP {{\cal P }}
\def\CQ {{\cal Q }}
\def\CE{{\cal E }}

\def\CS {{\cal S }}

\def\CU{{\cal U}}

\def\CZ {{\cal Z }}
\def\ch{{\rm ch}}

\def\heis{{\rm Heis}}

% more macros, alphabetically

\def\bA{ {\bf A}}

\def\IZ{\relax\ifmmode\mathchoice
{\hbox{\cmss Z\kern-.4em Z}}{\hbox{\cmss Z\kern-.4em Z}}
{\lower.9pt\hbox{\cmsss Z\kern-.4em Z}}
{\lower1.2pt\hbox{\cmsss Z\kern-.4em Z}}\else{\cmss Z\kern-.4em
Z}\fi}
\def\IB{\relax{\rm I\kern-.18em B}}
\def\IC{{\relax\hbox{$\inbar\kern-.3em{\rm C}$}}}
\def\ID{\relax{\rm I\kern-.18em D}}
\def\IE{\relax{\rm I\kern-.18em E}}
\def\IF{\relax{\rm I\kern-.18em F}}
\def\IG{\relax\hbox{$\inbar\kern-.3em{\rm G}$}}
\def\IGa{\relax\hbox{${\rm I}\kern-.18em\Gamma$}}
\def\IH{\relax{\rm I\kern-.18em H}}
\def\II{\relax{\rm I\kern-.18em I}}
\def\IK{\relax{\rm I\kern-.18em K}}
\def\IP{\relax{\rm I\kern-.18em P}}

\def\IQ{\relax\hbox{$\inbar\kern-.3em{\rm Q}$}}
\def\IP{\relax{\rm I\kern-.18em P}}

\def\IB{\relax{\rm I\kern-.18em B}}
\def\ID{\relax{\rm I\kern-.18em D}}
\def\IE{\relax{\rm I\kern-.18em E}}
\def\IF{\relax{\rm I\kern-.18em F}}
\def\IG{\relax\hbox{$\inbar\kern-.3em{\rm G}$}}
\def\IGa{\relax\hbox{${\rm I}\kern-.18em\Gamma$}}
\def\IH{\relax{\rm I\kern-.18em H}}
\def\II{\relax{\rm I\kern-.18em I}}
\def\IJ{\relax{\rm I\kern-.18em J}}
\def\IK{\relax{\rm I\kern-.18em K}}
\def\IL{\relax{\rm I\kern-.18em L}}

\def\IN{\relax{\rm I\kern-.18em N}}
\def\IO{\relax{\rm I\kern-.18em O}}
\def\IP{\relax{\rm I\kern-.18em P}}
\def\IQ{\relax\hbox{$\inbar\kern-.3em{\rm Q}$}}
\def\IR{\relax{\rm I\kern-.18em R}}
\def\IT{{\bf T}}
\def\IW{\relax\hbox{$\inbar\kern-.3em{\rm W}$}}

\def\Im{{\rm Im}}

\def\inbar{\,\vrule height1.5ex width.4pt depth0pt}
\def\mod{\rm mod}

\def\p{\partial}

\font\cmss=cmss10 \font\cmsss=cmss10 at 7pt
\def\IR{\relax{\rm I\kern-.18em R}}

\def\Tr{\rm Tr}

\def\vol{{\rm vol}}

%% new macros

%
\def\inv{^{\raise.15ex\hbox{${\scriptscriptstyle -}$}\kern-.05em 1}}

\def\Dsl{\,\raise.15ex\hbox{/}\mkern-13.5mu D} %this one can be subscripted
\def\dsl{\raise.15ex\hbox{/}\kern-.57em\partial}

 \def\Tr{{\rm Tr}}

 %pound sterling

\def\R{{\bf R}}

\def\Z{{{\bf Z}}}
%\def\IZ{{\bf Z}}

%%%% More defs

%

%% Something to deal with sub-sub-sections

\def\unlockat{\catcode`\@=11}
\def\lockat{\catcode`\@=12}

\unlockat
%%% Something to deal with sub-sub-sections

\def\newsec#1{\global\advance\secno by1\message{(\the\secno. #1)}
\global\subsecno=0\global\subsubsecno=0\eqnres@t\noindent
{\bf\the\secno. #1}
\writetoca{{\secsym} {#1}}\par\nobreak\medskip\nobreak}
\global\newcount\subsecno \global\subsecno=0
\def\subsec#1{\global\advance\subsecno
by1\message{(\secsym\the\subsecno. #1)}
\ifnum\lastpenalty>9000\else\bigbreak\fi\global\subsubsecno=0
\noindent{\it\secsym\the\subsecno. #1}
\writetoca{\string\quad {\secsym\the\subsecno.} {#1}}
\par\nobreak\medskip\nobreak}
\global\newcount\subsubsecno \global\subsubsecno=0
\def\subsubsec#1{\global\advance\subsubsecno by1
\message{(\secsym\the\subsecno.\the\subsubsecno. #1)}
\ifnum\lastpenalty>9000\else\bigbreak\fi
\noindent\quad{\secsym\the\subsecno.\the\subsubsecno.}{#1}
\writetoca{\string\qquad{\secsym\the\subsecno.\the\subsubsecno.}{#1}}
\par\nobreak\medskip\nobreak}

\def\subsubseclab#1{\DefWarn#1\xdef
#1{\noexpand\hyperref{}{subsubsection}%
{\secsym\the\subsecno.\the\subsubsecno}%
{\secsym\the\subsecno.\the\subsubsecno}}%
\writedef{#1\leftbracket#1}\wrlabeL{#1=#1}}% Macros for boxes
\lockat

%% END MACROS
%%

% Macros for boxes

\def\boxit#1{\vbox{\hrule\hbox{\vrule\kern8pt
\vbox{\hbox{\kern8pt}\hbox{\vbox{#1}}\hbox{\kern8pt}}
\kern8pt\vrule}\hrule}}
\def\mathboxit#1{\vbox{\hrule\hbox{\vrule\kern8pt\vbox{\kern8pt
\hbox{$\displaystyle #1$}\kern8pt}\kern8pt\vrule}\hrule}}

%%%%%%%%%%%%%%%%%%%%%%%%%%%%%%
%%%%%%%%%%%%%%%%%%%%%%%%%%%%% REFERENCES
%%%%%%%%%%%%%%%%%%%%%%%%%%%%%%
%%%%%%%%%%%%%%%%%%%%%%%%%%%%%%%

%\AcharyaKV
\lref\AcharyaKV{
B.~S.~Acharya,
``A moduli fixing mechanism in M theory,''
arXiv:hep-th/0212294.
%%CITATION = HEP-TH 0212294;%%
}

\lref\evslin{A.Adams,J.Evslin,''The Loop Group of $E_8$ and
K-Theory from 11d,''  hep-th/0203218}

%
%\AlvarezES
\lref\AlvarezES{
  O.~Alvarez,
  ``Topological Quantization And Cohomology,''
  Commun.\ Math.\ Phys.\  {\bf 100}, 279 (1985).
  %%CITATION = CMPHA,100,279;%%
}
%

%\AlvarezGaumeVM
\lref\AlvarezGaumeVM{
  L.~Alvarez-Gaume, J.~B.~Bost, G.~W.~Moore, P.~Nelson and C.~Vafa,
  ``Bosonization On Higher Genus Riemann Surfaces,''
  Commun.\ Math.\ Phys.\  {\bf 112}, 503 (1987).
  %%CITATION = CMPHA,112,503;%%
}

\lref\APS{M.Atiyah,V.Patodi, I.Singer,
Math.Proc.Cambridge Phil.Soc.77(1975)43;405.}

\lref\AtiyahSegalII{M.F. Atiyah and G.B. Segal, ``Twisted K-theory
and cohomology,'' math.KT/0510674 }

%\BeasleyDB
\lref\BeasleyDB{
C.~Beasley and E.~Witten,
``A note on fluxes and superpotentials in M-theory compactifications on
%manifolds of G(2) holonomy,''
JHEP {\bf 0207}, 046 (2002)
[arXiv:hep-th/0203061].
%%CITATION = HEP-TH 0203061;%%
}

\lref\asv{M. F. Atiyah, I. M. Singer, ``The index of
elliptic operators: V'' Ann. Math. {\bf 93} (1971) 139.}

%\BekaertYP
\lref\BekaertYP{
  X.~Bekaert and M.~Henneaux,
  ``Comments on chiral p-forms,''
  Int.\ J.\ Theor.\ Phys.\  {\bf 38}, 1161 (1999)
  [arXiv:hep-th/9806062].
  %%CITATION = HEP-TH 9806062;%%
}

%\BelovZE
\lref\BelovZE{
  D.~Belov and G.~W.~Moore,
  ``Classification of abelian spin Chern-Simons theories,''
  arXiv:hep-th/0505235.
  %%CITATION = HEP-TH 0505235;%%
}

%\BelovHT
\lref\BelovHT{
  D.~Belov and G.~W.~Moore,
  ``Conformal blocks for AdS(5) singletons,''
  arXiv:hep-th/0412167.
  %%CITATION = HEP-TH 0412167;%%
}

%\BelovJD
\lref\BelovJD{
  D.~Belov and G.~W.~Moore,
  ``Holographic action for the self-dual field,''
  arXiv:hep-th/0605038.
  %%CITATION = HEP-TH 0605038;%%
}

 \lref\belov{D. Belov and G. Moore, to appear}

\lref\BRG{E.Bergshoeff, M.de Roo, M.B.Green, G.Papadopoulos,
P.K.Townsend,''Duality of II 7-branes and 8-branes'',
hep-th/9601150}

\lref\booth{P. Booth, P. Heath, C. Morgan, and R. Piccinini,
``Remarks on the homotopy type of groups of gauge transformations,''
C.R. Math. Rep. Acad. Sci. Canada {\bf 3}(1981)3}

\lref\botttu{R. Bott and L.W. Tu, {\it Differential Forms in Algebraic
Topology}, Springer Graduate Texts in Math, 82}

%\BoussoXA
\lref\BoussoXA{
R.~Bousso and J.~Polchinski,
``Quantization of four-form fluxes and dynamical neutralization of the
cosmological constant,''
JHEP {\bf 0006}, 006 (2000)
[arXiv:hep-th/0004134].
%%CITATION = HEP-TH 0004134;%%
}

\lref\bredon{G.E. Bredon, {\it Topology and Geometry}, Springer GTM 139}

%\BrunnerEG
\lref\BrunnerEG{
  I.~Brunner and J.~Distler,
  ``Torsion D-branes in nongeometrical phases,''
  Adv.\ Theor.\ Math.\ Phys.\  {\bf 5}, 265 (2002)
  [arXiv:hep-th/0102018].
  %%CITATION = HEP-TH 0102018;%%
}

%\BrunnerSK
\lref\BrunnerSK{
  I.~Brunner, J.~Distler and R.~Mahajan,
  ``Return of the torsion D-branes,''
  Adv.\ Theor.\ Math.\ Phys.\  {\bf 5}, 311 (2002)
  [arXiv:hep-th/0106262].
  %%CITATION = HEP-TH 0106262;%%
}

\lref\Brylinski{J.-L. Brylinski, {\it Loop spaces, Characteristic Classes
and Geometric Quantization}, Birkh\"auser 1993}
%

%\BurringtonUU
\lref\BurringtonUU{
  B.~A.~Burrington, J.~T.~Liu and L.~A.~Pando Zayas,
  ``Finite Heisenberg groups in quiver gauge theories,''
  arXiv:hep-th/0602094.
  %%CITATION = HEP-TH 0602094;%%
}
%\BurringtonAW
\lref\BurringtonAW{
  B.~A.~Burrington, J.~T.~Liu and L.~A.~Pando Zayas,
  ``Central extensions of finite Heisenberg groups in cascading quiver gauge
  theories,''
  arXiv:hep-th/0603114.
  %%CITATION = HEP-TH 0603114;%%
}
%\BurringtonPU
\lref\BurringtonPU{
  B.~A.~Burrington, J.~T.~Liu, M.~Mahato and L.~A.~Pando Zayas,
  ``Finite Heisenberg groups and Seiberg dualities in quiver gauge theories,''
  arXiv:hep-th/0604092.
  %%CITATION = HEP-TH 0604092;%%
}

\lref\cheegersimons{J. Cheeger and J. Simons, ``Differential characters and geometric invariants,''
in {\it Geometry and Topology}, Lect. Notes in Math. vol. {\bf 1167}, p. 50, Springer 1985}
%
%\CornishDB
\lref\CornishDB{
  N.~J.~Cornish, D.~N.~Spergel, G.~D.~Starkman and E.~Komatsu,
  ``Constraining the Topology of the Universe,''
  Phys.\ Rev.\ Lett.\  {\bf 92}, 201302 (2004)
  [arXiv:astro-ph/0310233].
  %%CITATION = ASTRO-PH 0310233;%%
}

\lref\Cvet{ M. Cvetic, H. Lu, C.N. Pope, K.S. Stelle,''
T-Duality in the Green-Schwarz Formalism,
and the Massless/Massive IIA Duality Map'',Nucl.Phys. B573 (2000) 149-176,
 hep-th/9907202}

\lref\daifreed{X. Dai and D.S. Freed, ``$\eta$-Invariants and
Determinant Lines,'' hep-th/9405012; D.S. Freed, ``Determinant
Line Bundles Revisited,'' dg-ga/9505002}
\lref\sevenauthor{Jan de Boer, Robbert Dijkgraaf, Kentaro Hori, Arjan Keurentjes,
John Morgan, David R. Morrison, Savdeep Sethi, ``Triples, Fluxes, and Strings,'' hep-th/0103170 ;
Adv.Theor.Math.Phys. 4 (2002) 995-1186}
%

%\DeligneQP
\lref\DeligneQP{
  P. Deligne {\it et al.},
  ``Quantum fields and strings: A course for mathematicians.  Vol. 1, 2,''
 See section 6.3 of "Classical Field Theory", on page 218-220.
%\href{http://www.slac.stanford.edu/spires/find/hep/www?irn=4231341}{SPIRES entry}
}
%

%\DiaconescuWY
\lref\DiaconescuWY{
D.~E.~Diaconescu, G.~W.~Moore and E.~Witten,
``E(8) gauge theory, and a derivation of K-theory from M-theory,''
arXiv:hep-th/0005090.
%%CITATION = HEP-TH 0005090;%%
}

%\DiaconescuWZ
\lref\DiaconescuWZ{
D.~E.~Diaconescu, G.~W.~Moore and E.~Witten,
``A derivation of K-theory from M-theory,''
arXiv:hep-th/0005091.
%%CITATION = HEP-TH 0005091;%%
}

%\DiaconescuBM
\lref\DiaconescuBM{
E.~Diaconescu,  D.~S.~Freed, and G.~Moore
``The M-theory 3-form and E(8) gauge theory,''
arXiv:hep-th/0312069.
%%CITATION = HEP-TH 0312069;%%
}

\lref\DHMunpub{E. ~Diaconescu, J. Harvey, and G. Moore,
``Differential K-theory and boundary string field theory,''
unpublished. }

%\DijkgraafPZ
\lref\DijkgraafPZ{
  R.~Dijkgraaf and E.~Witten,
  ``Topological Gauge Theories And Group Cohomology,''
  Commun.\ Math.\ Phys.\  {\bf 129}, 393 (1990).
  %%CITATION = CMPHA,129,393;%%
}

\lref\Esnault{H. Esnault and E. Viehweg, ``Deligne-Belinson
cohomology,'' in {\it Beilinson's conjectures on special values
of $L$-functions,}  pp. 43-91, Perspect. Math. {\bf 4},
Academic Press, 1988}
%

%\FerraraYX
\lref\FerraraYX{
  S.~Ferrara, J.~A.~Harvey, A.~Strominger and C.~Vafa,
  ``Second quantized mirror symmetry,''
  Phys.\ Lett.\ B {\bf 361}, 59 (1995)
  [arXiv:hep-th/9505162].
  %%CITATION = HEP-TH 9505162;%%
}

%\FrancoSM
\lref\FrancoSM{
  S.~Franco, A.~Hanany, D.~Martelli, J.~Sparks, D.~Vegh and B.~Wecht,
  %``Gauge theories from toric geometry and brane tilings,''
  arXiv:hep-th/0505211.
  %%CITATION = HEP-TH 0505211;%%
}

%\FreedVW
\lref\FreedVW{
  D.~S.~Freed,
  ``Classical Chern-Simons theory. Part 1,''
  Adv.\ Math.\  {\bf 113}, 237 (1995)
  [arXiv:hep-th/9206021].
  %%CITATION = HEP-TH 9206021;%%
}

%\FreedQB
\lref\FreedQB{
  D.~S.~Freed,
  ``Locality and integration in topological field theory,''
  arXiv:hep-th/9209048.
  %%CITATION = HEP-TH 9209048;%%
}

%\FreedTT
\lref\FreedTT{
  D.~S.~Freed and M.~J.~Hopkins,
  ``On Ramond-Ramond fields and K-theory,''
  JHEP {\bf 0005}, 044 (2000)
  [arXiv:hep-th/0002027].
  %%CITATION = HEP-TH 0002027;%%
}

\lref\FHTch{D. Freed, M. Hopkins, and C. Teleman, ``Twisted equivariant K-theory with
complex coefficients,'' arXiv:math.AT/0206257}

%\FreedYC
\lref\FreedYC{
  D.~S.~Freed and G.~W.~Moore,
  ``Setting the quantum integrand of M-theory,''
  arXiv:hep-th/0409135.
  %%CITATION = HEP-TH 0409135;%%
}

\lref\FMSi{D. Freed, G. Moore and G. Segal, ``The Uncertainty of
Fluxes,'' arXiv:hep-th/0605198.}

%\FreedTG
\lref\FreedTG{
D.~Freed, J.~A.~Harvey, R.~Minasian and G.~W.~Moore,
``Gravitational anomaly cancellation for M-theory fivebranes,''
Adv.\ Theor.\ Math.\ Phys.\  {\bf 2}, 601 (1998)
[arXiv:hep-th/9803205].
%%CITATION = HEP-TH 9803205;%%
}

\lref\freed{D. Freed,
Dirac Charge Quantization and Generalized Differential Cohomology,''
hep-th/0011220 }

%\FrenkelRN
\lref\FrenkelRN{
  I.~B.~Frenkel and V.~G.~Kac,
  ``Basic Representations Of Affine Lie Algebras And Dual Resonance Models,''
  Invent.\ Math.\  {\bf 62}, 23 (1980).
  %%CITATION = INVMB,62,23;%%
}

%\GawedzkiAK
\lref\GawedzkiAK{
  K.~Gawedzki,
  ``Topological Actions In Two-Dimensional Quantum Field Theories,''
In Cargese 1987, Proceedings, {\it Nonperturbative Quantum Field Theory},
pp. 101-141.
}

\lref\Gomi{
K. Gomi, ``Differential characters and the Steenrod squares,'' arXiv:math.AT/0411043
}
\lref\gomiII{K. Gomi, ``Projective unitary representations of smooth Deligne
cohomology groups,'' arXiv:math.RT/0510187}
%

%\GukovKN
\lref\GukovKN{
  S.~Gukov, M.~Rangamani and E.~Witten,
  ``Dibaryons, strings, and branes in AdS orbifold models,''
  JHEP {\bf 9812}, 025 (1998)
  [arXiv:hep-th/9811048].
  %%CITATION = HEP-TH 9811048;%%
}

\lref\BHarris{B. Harris, ``Differential characters and the Abel-Jacobi map,'' in
{\it Algebraic K-theory: connections with geometry and topology}, Nato. Adv. Sci.
Inst. Ser. C Math Phys. Sci. 279, Kluwer Acad. Publ. Dordrecht, 1989}
\lref\harveylawson{R. Harvey and B. Lawson, ``Lefshetz-Pontrjagin duality for differential
characters,'' An. Acad. Cienc. {\bf 73}(2001) 145;
R. Harvey, B. Lawson, and J. Zweck, ``The de Rham-Federer theory of
differential characters and character duality,'' Amer. J. Math. {\bf 125}(2003) 791}

\lref\hatcher{A. Hatcher, {\it Algebraic Topology},
http://www.math.cornell.edu/~hatcher/\# VBKT
}

\lref\hirzebruch{See F. Hirzebruch, Math Reviews, MR0198491, for a concise
summary}

%\HopkinsRD
\lref\HopkinsRD{
M.~J.~Hopkins and I.~M.~Singer,
``Quadratic functions in geometry, topology, and M-theory,''
arXiv:math.at/0211216.
%%CITATION = MATH-AT 0211216;%%
}

\lref\hori{K. Hori, ``D-branes, T-duality, and Index Theory,''
Adv.Theor.Math.Phys. 3 (1999) 281-342; hep-th/9902102  }

\lref\Hull{C.M. Hull,''Massive String Theories From M-Theory and F-Theory'',
JHEP 9811 (1998) 027,hep-th/
9811021 }
\lref\Town{C.Hull,P.Townsend,''Unity of Superstring Dualities,''
Nucl.Phys. B438 (1995) 109-137,hep-th/9410167}
\lref\GPR{ A. Giveon, M. Porrati, E. Rabinovici,
'' Target Space Duality in String Theory'', hep-th/9401139 }

\lref\klonoff{K. Klonoff, Thesis in preparation}

 \lref\KMW{A. Kitaev, G. Moore, and K. Walker,
``Noncommuting fluxes in a tabletop experiment,'' unpublished.}

\lref\Lavr{I.Lavrinenko, H.Lu, C.Pope,
T.Tran,``U-duality as general Coordinate Transformations, and Spacetime
Geometry'', hep-th/9807006 }

\lref\Piol{E.Kiritsis and B. Pioline, ``On $R^4$ threshhold corrections
in IIB string theory and (p,q) string instantons,''
Nucl. Phys. {\bf B508}(1997)509;
B. Pioline, H. Nicolai, J. Plefka, A. Waldron,
 $R^4$ couplings, the fundamental membrane and
exceptional theta correspondences, hep-th/0102123 }

\lref\lott{J. Lott, ``$\R/\Z$ index theory,'' Comm. Anal. Geom. {\bf 2} (1994) 279 }

%\MathaiYK
\lref\MathaiYK{
  V.~Mathai and D.~Stevenson,
  ``Chern character in twisted K-theory: Equivariant and holomorphic cases,''
  Commun.\ Math.\ Phys.\  {\bf 236}, 161 (2003)
  [arXiv:hep-th/0201010].
  %%CITATION = HEP-TH 0201010;%%
}

%\MathaiMU
\lref\MathaiMU{
  V.~Mathai and H.~Sati,
  ``Some relations between twisted K-theory and E(8) gauge theory,''
  JHEP {\bf 0403}, 016 (2004)
  [arXiv:hep-th/0312033].
  %%CITATION = HEP-TH 0312033;%%
}

\lref\mm{R. Minasian and G. Moore,``K Theory and Ramond-Ramond Charge,''
JHEP {\bf 9711}:002, 1997; hep-th/9710230.}
%

%\MooreCP
\lref\MooreCP{
  G.~W.~Moore and N.~Saulina,
  ``T-duality, and the K-theoretic partition function of typeIIA  superstring
  theory,''
  Nucl.\ Phys.\ B {\bf 670}, 27 (2003)
  [arXiv:hep-th/0206092].
  %%CITATION = HEP-TH 0206092;%%
}

%\MooreGB
\lref\MooreGB{
  G.~W.~Moore and E.~Witten,
  ``Self-duality, Ramond-Ramond fields, and K-theory,''
  JHEP {\bf 0005}, 032 (2000)
  [arXiv:hep-th/9912279].
  %%CITATION = HEP-TH 9912279;%%
}
%\MoorePN
\lref\MoorePN{
  G.~W.~Moore,
  ``Arithmetic and attractors,''
  arXiv:hep-th/9807087.
  %%CITATION = HEP-TH 9807087;%%
}

%\MooreJV
\lref\MooreJV{
  G.~W.~Moore,
  ``Anomalies, Gauss laws, and page charges in M-theory,''
  Comptes Rendus Physique {\bf 6}, 251 (2005)
  [arXiv:hep-th/0409158].
  %%CITATION = HEP-TH 0409158;%%
}

\lref\MW{ G.Moore, E. Witten, ``Integration over the u-plane in
Donaldson theory'',hep-th/9709193}

%\MooreGB
\lref\MooreGB{
G.~W.~Moore and E.~Witten,
``Self-duality, Ramond-Ramond fields, and K-theory,''
JHEP {\bf 0005}, 032 (2000)
[arXiv:hep-th/9912279].
%%CITATION = HEP-TH 9912279;%%
}

\lref\Mumford{D. Mumford, M. Nori, and P. Norman, {\it Tata Lectures on Theta III},
Birkh\"auser 1991}

\lref\Mathaii{P.Bouwknegt, V.Mathai,``D-branes, B-fields and
twisted K-theory'', ~~
JHEP 0003
(2000)007, hep-th/0002023}
\lref\fh{D. S. Freed, M. J. Hopkins,
``On Ramond-Ramond fields and K-theory,'' JHEP 0005 (2000) 044;
hep-th/0002027}
\lref\birreldavies{N.D. Birrell and P.C.W. Davies,
{\it Quantum Fields in Curved Space}, Cambridge Univ. Press, 1982}

%\MooreVF
\lref\MooreVF{
G.~Moore,
``K-theory from a physical perspective,''
arXiv:hep-th/0304018.
%%CITATION = HEP-TH 0304018;%%
}

\lref\pressleysegal{A. Pressley and G. Segal, {\it Loop Groups}, Oxford, 1986 }

\lref\Romans{L. Romans,''Massive IIA Supergravity in Ten Dimensions',
Phys.Let.169B(1986)374}

\lref\rudyak{Y.B. Rudyak, {\it On Thom Spectra, Orientability, and
Cobordism}, Springer-Verlag 1998}

%\SegalAP
\lref\SegalAP{
  G.~Segal,
  ``Unitarity Representations Of Some Infinite Dimensional Groups,''
  Commun.\ Math.\ Phys.\  {\bf 80}, 301 (1981).
  %%CITATION = CMPHA,80,301;%%
}

\lref\Sethi{ S. Sethi, C. Vafa, E. Witten,
``Constraints on Low-Dimensional String Compactifications'',
hep-th/9606122, Nucl.Phys. B480 (1996) 213-224 }

\lref\sutherland{W.A. Sutherland, ``Function spaces related to
gauge groups,'' Proc. Roy. Soc. Edinburgh, {\bf 121A}(1992)185}

\lref\stong{R. Stong, ``Calculation of
$\Omega_{11}^{spin}(K({\bf Z},4))$'' in
{\it Unified String Theories}, 1985 Santa Barbara
Proceedings, M. Green and D. Gross, eds. World Scientific 1986.}

\lref\szabo{K. Olsen and R.J. Szabo,
``Constructing $D$-Branes from $K$-Theory,''
hep-th/9907140.  }

\lref\Verl{E. Verlinde,''Global Aspects of Electric-Magnetic Duality'',
Nucl.Phys. B455 (1995) 211-228,hep-th/9506011}

\lref\vick{J.W. Vick, {\it Homology Theory}, Academic Press 1973}

\lref\Wits{E.Witten,''On S-Duality in Abelian Gauge Theory'',
hep-th/9505186}

\lref\zucchini{R. Zucchini, ``Relative topological integrals and
relative Cheeger-Simons differential characters,'' hep-th/0010110}

%\WittenWY
\lref\WittenWY{
E.~Witten,
``AdS/CFT correspondence and topological field theory,''
JHEP {\bf 9812}, 012 (1998)
[arXiv:hep-th/9812012].
%%CITATION = HEP-TH 9812012;%%
}

%\WittenED
\lref\WittenED{
E.~Witten,
``Quantum background independence in string theory,''
arXiv:hep-th/9306122.
%%CITATION = HEP-TH 9306122;%%
}

%\WittenWY
\lref\WittenWY{
  E.~Witten,
  ``AdS/CFT correspondence and topological field theory,''
  JHEP {\bf 9812}, 012 (1998)
  [arXiv:hep-th/9812012].
  %%CITATION = HEP-TH 9812012;%%
}

\lref\mooretalk{G. Moore,   IHES talk, June 2004. http://
www.physics.rutgers.edu/~gmoore/ .}

\lref\Warn{C.Isham,C.Pope,N.Warner,
''Nowhere-vanishing spinors and triality rotations in 8-manifolds,''
Class.Quantum Grav.5(1988)1297}
%
%\lref\fourflux{E. Witten, ``On Flux Quantization in $M$-Theory
%and the Effective Action,'' hep-th/9609122; Journal of
%Geometry and Physics, {\bf 22} (1997) 1.}

%
%\WittenMD
\lref\WittenMD{
E.~Witten,
``On flux quantization in M-theory and the effective action,''
J.\ Geom.\ Phys.\  {\bf 22}, 1 (1997)
[arXiv:hep-th/9609122].
%%CITATION = HEP-TH 9609122;%%
}
%\WittenBT
\lref\WittenBT{
E.~Witten,
``Topological Tools In Ten-Dimensional Physics,''
Int.\ J.\ Mod.\ Phys.\ A {\bf 1}, 39 (1986).
%%CITATION = IMPAE,A1,39;%%
}
%\WittenHC
\lref\WittenHC{
E.~Witten,
``Five-brane effective action in M-theory,''
J.\ Geom.\ Phys.\  {\bf 22}, 103 (1997)
[arXiv:hep-th/9610234].
%%CITATION = HEP-TH 9610234;%%
}
%\WittenVG
\lref\WittenVG{
E.~Witten,
``Duality relations among topological effects in string theory,''
JHEP {\bf 0005}, 031 (2000)
[arXiv:hep-th/9912086].
%%CITATION = HEP-TH 9912086;%%
}

%\WittenCD
\lref\WittenCD{
  E.~Witten,
  ``D-branes and K-theory,''
  JHEP {\bf 9812}, 019 (1998)
  [arXiv:hep-th/9810188].
  %%CITATION = HEP-TH 9810188;%%
}

\lref\wittenk{E. Witten, ``$D$-Branes And $K$-Theory,''
JHEP {\bf 9812}:019, 1998; hep-th/9810188.}

\lref\wittenduality{E. Witten, ``Duality Relations Among Topological Effects In String Theory,''
hep-th/9912086;JHEP 0005 (2000) 031}
\lref\wittenstrings{E. Witten, ``Overview of K-theory applied to
strings,'' hep-th/0007175.}

%\WittenHC
\lref\WittenHC{
  E.~Witten,
  ``Five-brane effective action in M-theory,''
  J.\ Geom.\ Phys.\  {\bf 22}, 103 (1997)
  [arXiv:hep-th/9610234].
  %%CITATION = HEP-TH 9610234;%%
}

%\AshtekarRG
\lref\AshtekarRG{
  A.~Ashtekar and A.~Corichi,
  ``Gauss linking number and electro-magnetic uncertainty principle,''
  Phys.\ Rev.\ D {\bf 56}, 2073 (1997)
  [arXiv:hep-th/9701136].
  %%CITATION = HEP-TH 9701136;%%
}

%\HenningsonWH
\lref\HenningsonWH{
  M.~Henningson,
  ``The quantum Hilbert space of a chiral two-form in d = 5+1 dimensions,''
  JHEP {\bf 0203}, 021 (2002)
  [arXiv:hep-th/0111150].
  %%CITATION = HEP-TH 0111150;%%
}

%%%%%%%%%%%%%%%%%%%%%%%%%%%%%%%%%%%%%%%%%%
%%%%%%%%%%%%%%%%%%%%%%%%%%%%%%%%%%%%%%%%%%%%
%%%%%%%%%%%%%%%%%%%%%%%%%%%%%%%%%%%%%%%%%%%%%

%%%%%%%%%%%%%%
%%%%%%%%%%%%%%
%%%%%%%%%%%%%%%%

%
\Title{\vbox{\baselineskip12pt \hbox{hep-th/0605200}
\hbox{NSF-KITP-05-118} } } {\vbox{\centerline{Heisenberg Groups and
Noncommutative Fluxes}
%\centerline{ in}
%\centerline{ Generalized Abelian Gauge  Theories}
}} \centerline{Daniel S. Freed$^{1}$, Gregory W. Moore$^{2}$, Graeme
Segal$^{3}$}

\bigskip

\centerline{$^{1}${\it Department of Mathematics, University of Texas at Austin, TX}}

\medskip

\centerline{$^{2}${\it Department of Physics, Rutgers University}}
\centerline{\it Piscataway, NJ 08854-8019, USA}

\medskip

\centerline{$^{3}${\it All Souls College, Oxford, UK} }

 \vskip.1in \vskip.1in \centerline{\bf Abstract}
 \medskip

\noindent We develop a group-theoretical approach to the formulation
of generalized abelian gauge theories, such as those appearing in
string theory and M-theory. We explore several applications of this
approach. First, we show that there is an uncertainty relation which
obstructs simultaneous measurement of electric and magnetic flux
when torsion fluxes are included. Next we show how to define the
Hilbert space of a self-dual field. The Hilbert space is
$\IZ_2$-graded and we show that, in general, self-dual theories
(including the RR fields of string theory) have fermionic sectors.
We indicate how rational conformal field theories associated to the
two-dimensional Gaussian model generalize to $(4k+2)$-dimensional
conformal field theories. When our ideas are applied to the RR
fields of string theory we learn that it is impossible to measure
the $K$-theory class of a RR field. Only the reduction modulo
torsion can be measured.

\Date{May 17,  2006 }

%\draftmode

\newsec{ Introduction }

Fluxes and D-branes have been playing a central role in string theory
and M-theory for almost 10 years now. Nevertheless, many important
mathematical issues remain open in the general theory of fluxes.
%
%A discussion of some of these, and their relevance to current topics
%of interest may be found in the discussion section *** below.
%
In this paper we focus on the question of the structure of the
Hilbert space of theories of fluxes. In particular, we consider the
question of how the Hilbert space is graded by electric and magnetic
flux. We will find that in general electric and magnetic fluxes
cannot be simultaneously diagonalized. A companion paper \FMSi\
presents the same material in a mathematically rigorous fashion.

As we explain below, our result holds for a rather broad class of
theories. These theories generalize  Maxwell's theory of
electromagnetism and are known as ``generalized abelian gauge
theories.'' Broadly stated, we will show that in these theories the
Hilbert space may be characterized as an irreducible representation
of a certain Heisenberg group. This point of view gives a
particularly elegant formulation of the Hilbert space of a self-dual
field. \foot{Another paper \belov\ explores some related issues in
type II supergravity from the Lagrangian point of view.} Moreover,
electric and magnetic flux sectors are defined by diagonalizing
subgroups of translations by flat fields. The translations by flat
fields and by their electromagnetic duals do not commute with each
other in the Heisenberg group, and therefore one cannot
simultaneously diagonalize the topological sector of both electric
and magnetic flux.

Perhaps   the most surprising implication of our general result is
for the RR fields of string theory.  Consider the Hamiltonian
formulation of a string theory of type II or type I on a spacetime
of the form $Y \times \IR$, where $Y$ is a compact  spin
$9$-manifold.   It is usually assumed that the flux sectors \foot{In
this paper we are only considering free theories. Thus the flux
sectors do not mix under time evolution. If we include
nonperturbative effects such as brane-antibrane
creation/annihilation then some of these sectors will mix. See
$\S{7.2}$ below.} in the theory are given by a grading of the
Hilbert space
\eqn\fluxsectors{
\CH \quad {\buildrel ? \over = } \quad \oplus_{x \in K(Y)} \CH_x
}
where $K(Y)$ is some $K$-theory group (thus $K^0(Y)$ for IIA,
$K^1(Y)$ for IIB, $KO^{-1}(Y)$ for type I, etc. These $K$-theory
groups are in general twisted by $B$-fields, and are equivariant if
$Y$ is a quotient space.) The group $K(Y)$ encodes both electric and
magnetic fluxes, because the RR field is self-dual. Thus, from the
general remarks above we cannot expect that an equation of the form
\fluxsectors\ can actually hold! The difficulty arises because
torsion electric and magnetic fluxes do not commute. \foot{There is
an unfortunate clash of terminology here. ``Torsion'' is also used
for the antisymmetric part of a connection on a tangent bundle, and
in string theory is often associated with the fieldstrength of the
$B$-field. In this paper we {\it never} use ``torsion'' in this
sense.} Recall that $K(Y)$ is an abelian group and therefore has a
canonical subgroup ${\rm Tors}(K(Y))$ of elements of finite order.
We will show that the  correct version of \fluxsectors\ is
\eqn\fluxsectorstrue{
\CH \quad =  \quad \oplus_{\bar x \in \bar K(Y) } \CH_{\bar x}
}
where $\bar K(Y):= K(Y)/{\rm Tors}(K(Y))$. This is not to say that the
torsion subgroup is irrelevant to the physics.
 If we regard the naive grading by
$K$-theory classes as a diagonalization of the action of the group
of flat fields $K(Y;\IR/\IZ)$ ($K^{-1}$ for IIA, $K^0$ for IIB,
$KO^{-1}$ for type I string theory) then the true situation is that
there is a central extension, which we might call a ``quantum
$K$-theory group''
\eqn\quantumk{
0 \rightarrow U(1) \rightarrow QK(Y) \rightarrow K(Y;\IR/\IZ) \rightarrow 0
}
and the Hilbert space is a representation of $QK(Y)$. Irreducible
representations of $QK(Y)$ where the $U(1)$ acts as scalar
multiplication are labelled \foot{albeit noncanonically} by $\bar
K(Y)$.

While torsion fluxes might seem somewhat esoteric to some readers,
the present result seems to us to be conceptually important because
it implies that the standard picture of a D-brane as a submanifold
of spacetime  with a vector bundle (and connection) should be
re-examined due to quantum effects. These quantum effects are not
suppressed in the large-distance limit. One might expect that  large
scale experiments are required to detect them, but in fact, they can
in principle be demonstrated with localized experiments, even in the
case of $3+1$-dimensional Maxwell theory \KMW.

Before we begin with technicalities we would like to make one further
general remark. As we have mentioned, the result of this paper actually applies to a rather
broad class of theories called ``generalized abelian gauge theories'' (GAGTs).
Let us pause to explain that term.

To put GAGTs in context let us recall some of the history of topology.
It is now recognized in string theory that singular cohomology theory
is an extremely important tool. In mathematics it was realized in
the 1940's that the cohomology functor $H^*$ can be characterized by
a system of axioms - the Eilenberg-Steenrod axioms for singular cohomology \vick.
These include various axioms of naturality and glueing (Meyer-Vietoris)
  together with
the  dimension axiom, which states that
\eqn\dimensionax{
H^\ell(pt; G) = \cases{  G & $\ell=0$ \cr
 0 & $\ell\not=0$\cr}
} where $G$ is an abelian group known as the coefficient group.
(When the coefficient group is $\IZ$ we will sometimes   write
$H^\ell(M)$ for $H^\ell(M;\IZ)$.) In the 1950's and 1960's it was
realized that one could drop the dimension axiom, retaining the
others and still have very useful topological invariants of spaces.
These became known as {\it generalized cohomology theories}.
Examples of generalized cohomology theories of interest in physics
include cobordism theory, K-theory, and elliptic  cohomology. These
theories enter physics in the context of topological sectors. On the
other hand, to do physics we also need {\it local} fields. There is
a useful new subject in mathematics known as {\it differential
cohomology} that elegantly combines the local field degrees of
freedom with nontrivial topology. These theories have their origin
in the work of Deligne (see, e.g. \Brylinski\Esnault) and of Cheeger
and Simons \cheegersimons. The foundational paper on the subject is
the work of Hopkins and Singer \HopkinsRD, which explains in what
sense these are indeed cohomology theories. They seem perfectly
suited to the needs of supergravity and string theory. See \freed\
and material below for an introduction to this subject and an
explanation of this statement. To a generalized cohomology theory
$E$ we may attach its differential version $\check E$. Examples
relevant to string theory include $\check H^3$, which is the proper
home for the isomorphism class of $B$-fields in type II  string
theory. Similarly, the $M$-theory $3$-form is related to $\check
H^4$ \DiaconescuBM. Moreover,  $\check K$ is the appropriate setting
for the RR fields of type II string theory, while $\check KO$ is the
appropriate setting for type $I$ theory. By definition, a {\it
generalized abelian gauge theory} (GAGT) is a field theory whose
space of gauge invariant field configurations is a differential
generalized cohomology group. \foot{It can be a torsor for such a
group, in the presence of a magnetic current.} The adjective
``abelian'' refers to the property that in these theories the space
of gauge inequivalent fields forms an abelian group. This property
will play an important role below.

There are some connections to previous works which should be pointed
out. A remark closely related to ours in the case of $3+1$
dimensional Maxwell theory was made in \GukovKN, as we explain
below. In  \MooreJV\mooretalk\  it was pointed out that the
Chern-Simons terms in M-theory leads to noncommutativity of the
7-form Page charges. Upon dimensional reduction this should be
related to the K-theoretic phenomenon discussed below \belov.
Similar phenomena have been well-known for some time in the theory
of 3-dimensional Maxwell theory with a Chern-Simons term (see
\BelovZE\  for a recent discussion). Also, similar phenomena appear
in the theory of   abelian 2-forms in 5-dimensions, ads/cft dual to
4dimensional Maxwell theory   \WittenWY\BelovHT. Finally,
applications of noncommuting fluxes to AdS/CFT duals of quiver gauge
theories have recently been investigated in
\BurringtonUU\BurringtonAW\BurringtonPU.

Let us now summarize the remainder of the paper. Section 2 is a
pedagogical review of the Deligne-Cheeger-Simons differential
cohomology theory. We describe the structure of the groups, and give
a heuristic explanation of Poincar\'e duality. Section 3 addresses
generalized Maxwell theories. We carefully define  electric and
magnetic flux sectors and explain the role of Heisenberg groups.
{\it  The main point of this paper is explained in equation $(3.9)$
and the subsequent paragraphs. } Section 4 applies some of the ideas
from sections 2 and 3 to describe the Hilbert space of a self-dual
field. Section 4.1 explains how the nonself-dual field is related to
the self-dual and anti-self-dual field  and explains how the
``rational torus'' of two-dimensional RCFT can be generalized to
CFT's in $(4k+2)$-dimensions. Section 5 applies the ideas to the RR
fields of type II string theory. Section 6 describes, from the
vantage point of Heisenberg groups,
 how the theory is modified when one includes
a Chern-Simons term in the action. Section 7 discusses issues
associated with the generalization to noncompact spaces $Y$. Section
8 mentions some possible future directions, and appendix A provides
some further mathematical background. Appendix B explains the
``tadpole constraint'' on flux sectors which applies in theories
with a Chern-Simons term.

\newsec{A Gentle Review of Differential Cohomology}

\subsec{Motivating differential cohomology}

In this section we review some basic mathematical facts about differential cohomology
which will be needed to make the physics point below. This discussion will also serve
as motivation for the discussion of the structure of differential K-theory in section 5.
Our intended audience is experts in supergravity and/or string theory with some
knowledge of topology, who wish to have a clear understanding of the   topological aspects
of  gauge theories of higher degree differential forms. The essential topological background can all be found
in \botttu\bredon\hatcher.

As mentioned, the paradigmatic example of  differential cohomology
theory is given by classical Maxwell theory, regarded as the theory
of a connection 1-form on a line bundle over spacetime. \foot{In the
companion paper \FMSi\  we are more careful and distinguish three
versions of Maxwell theory. The first is purely {\it classical}
Maxwell theory with no quantization of $[F]$ or of $[*F]$. The
version used in $\S{2.1}$ of this paper is the ``semiclassical
theory,'' in the terminology of \FMSi.
  }
If $M$ is spacetime, and we consider a fixed line bundle $L \to M$
then we can regard the Maxwell field as a connection on $L$. The
$2$-form fieldstrength $F$ is then the curvature of the connection.
The space of {\it gauge equivalence classes of  fields} is then
$\CA(L)/\CG$ where $\CA(L)$ is the space of connections, and $\CG$
is the gauge group, $\CG = Map(M,U(1))$. Of course, this is only the
description   in a definite topological sector, determined by the
first Chern class of $L$. The full space of gauge equivalence
classes of fields is actually
\eqn\gaufne{
\bigcup_{c_1\in H^2(M)} \CA(L_{c_1})/\CG
}
where we have made a choice of line bundle for each $c_1\in H^2(M)$ (at some cost in
naturality).

There is another viewpoint on the space \gaufne\ which is more natural
 and which generalizes nicely to   $\ell$-form fieldstrengths
with $\ell\not=2$.
The key remark is that the gauge invariant information carried by a connection $A$ is
precisely encoded by its holonomy function. That is, the function
\eqn\holonomy{
\Sigma \to \exp\left( 2\pi i \oint_{\Sigma} A \right)
}
taking a closed 1-cycle  $\Sigma$ to a phase may be regarded as a map
$\chi_A: Z_1(M) \to U(1)$, where $Z_1(M)$ is the group of all closed 1-cycles in
$M$.  Of course, $\chi_A$ is well-defined even if $A$ cannot be defined as a
smooth 1-form on all of $M$. Note that $\chi_A$ is in fact a group homomorphism,
but is not an arbitrary group homomorphism. It has the distinguishing property
that, if $\Sigma = \p \CB$ is a boundary in $M$ then
\eqn\fieldprop{
\chi_A(\Sigma) = \exp\left( 2\pi i \int_{\CB} F \right)
}
where $F\in \Omega^2(M)$ is a globally well-defined 2-form.
 {\it The space of all homomorphisms $Z_1(M) \to U(1)$ satisfying \fieldprop\
is equivalent to the space of gauge equivalence classes of connections on line bundles
on $M$.}  This   reformulation of   \gaufne\ defines
 the space of gauge inequivalent field configurations
in $2$-form gauge theory on $M$. We will denote this space $\check
H^2(M)$. In this paper it will be very crucial that there is an
abelian group structure on   $\check H^2(M)$. In terms of line
bundles with connection the group product is tensor product.

As a simple example, consider the Lens space $L_k = S^3/\IZ_k$. It has
a noncontractible cycle $\gamma$ with $k [\gamma] =0$. We can
define $\chi$ by $\chi(\Sigma)=1$ if $\Sigma$ bounds. It then follows that
$\chi(\gamma)$ is a $k^{th}$ root of unity. Choosing
$\chi(\gamma)=e^{2\pi i r/k}$ for some integer $r$  defines
line bundle with a flat connection. If $r\not=0~\mod k$ then this
line bundle  is nontrivial.

The generalization to gauge theories
with $\ell$-form fieldstrengths is now immediate:
\foot{The original definition in \cheegersimons\ uses a   different
grading, shifted by one, relative to   that used here.
Our grading behaves simply under the product of
differential characters described below. }

\bigskip
\defn{} The {\it  Cheeger-Simons group of differential characters}, denoted
$\check H^\ell(M)$, is the subgroup of all homomorphisms from closed
$(\ell-1)$-cycles to $U(1)$:
\eqn\dcsdf{
\check H^\ell(M) \subset {\rm Hom}(Z_{\ell-1}(M), U(1))
}
satisfying the following property: For each $\chi\in \check H^{\ell}(M)$ there is a
globally well-defined $F_\chi \in \Omega^\ell(M)$ such that, if $\Sigma = \p \CB$,
then
\eqn\defnfs{
\chi(\Sigma) = \exp\left( 2\pi i \int_{\CB} F \right) .
}
The group structure on $\c H^\ell(M)$ is simply pointwise
multiplication of characters.

By ``generalized Maxwell theory'' we mean a field theory such that the space
of gauge inequivalent fields is $\check H^\ell(M)$ for some $\ell$.
Ordinary Maxwell theory is the case $\ell=2$.

Let us take care of some notational matters. The space $\check
H^\ell(M)$ is an infinite-dimensional abelian group, and we will
describe its structure in detail. We sometimes work
multiplicatively, and sometimes additively. When we work
multiplicatively, we denote characters by $\chi$. When we work
additively we replace $U(1)$ by the isomorphic   group $\IR/\IZ$.
Also, we then denote elements  of $\check H^\ell(M)$ by $[\check
A]$. The notation is meant to suggest an $(\ell-1)$-form
``potential.'' However, it is important to stress that in general
there is no   globally well-defined $(\ell-1)$-form. (Indeed, the
superscript on $\c A$ is meant to remind us of that.) The square
brackets emphasize that we are working with gauge equivalence
classes. \foot{It is indeed possible to give meaning to $\c A$,
rather than its gauge equivalence class, as an object in a groupoid,
but we will not need this level of detail in most of the present
paper. } There is a class of differential characters, known as the
{\it topologically trivial} characters which {\it are} associated
with a globally well-defined potential $A \in \Omega^{\ell-1}(M)$.
These are defined in equation $(2.16)$ below.

Finally, we should note that there is a \v{C}ech approach to higher form
gauge theory. One chooses a good cover $\CU_\alpha$ of the manifold, and introduces
potentials $A_\alpha$ on $\CU_\alpha$ so that $d A_\alpha$ is globally well-defined.
One then has a system $A_\alpha - A_\beta = d A_{\alpha \beta}$ on $\CU_{\alpha\beta}$,
and so on. While perfectly valid, this is a very cumbersome approach to the
subject, and we will be able to understand everything we need without resorting to
patches.

\subsec{The structure of the differential cohomology groups}

We will now describe some of the general mathematical properties of
the groups $ \check H^\ell(M)$. These will be useful to us in describing
the structure of the Hilbert space of generalized Maxwell theories.

\ifig\twoboundaryi{ Integrating the fieldstrength over the
 two bounding manifolds $\CB$ and $\CB'$
must produce the same holonomy. Considering bounding manifolds that
differ slightly from each other implies   $dF=0$ . }
{\epsfxsize2.0in\epsfbox{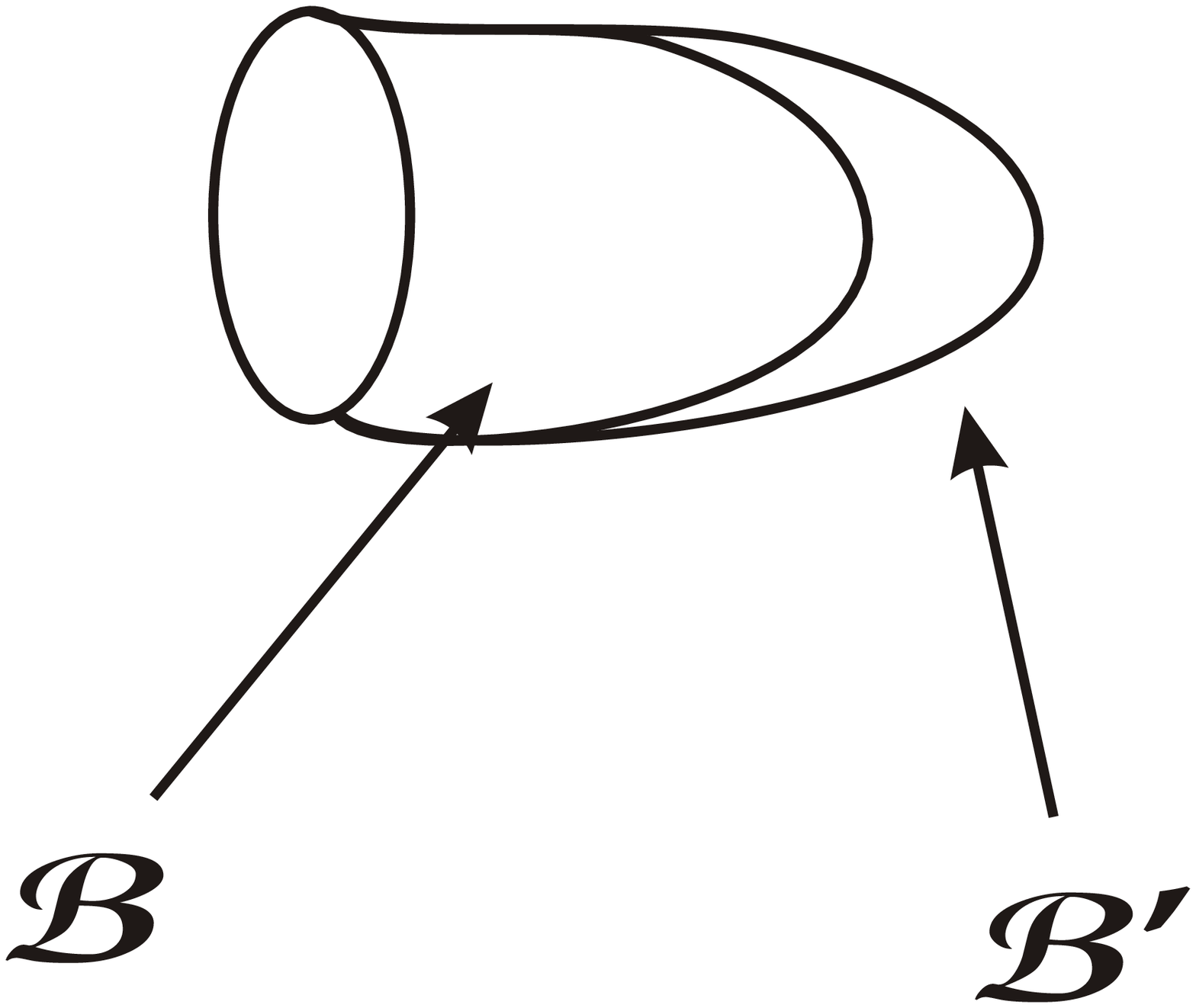}}

\ifig\twoboundaryii{ Integrating the fieldstrength over the
 two bounding manifolds $\CB$ and $\CB'$
must produce the same holonomy. Considering bounding manifolds such
that $\CB- \CB'$ is a closed cycle shows that $F$ must have integer
periods. } {\epsfxsize2.0in\epsfbox{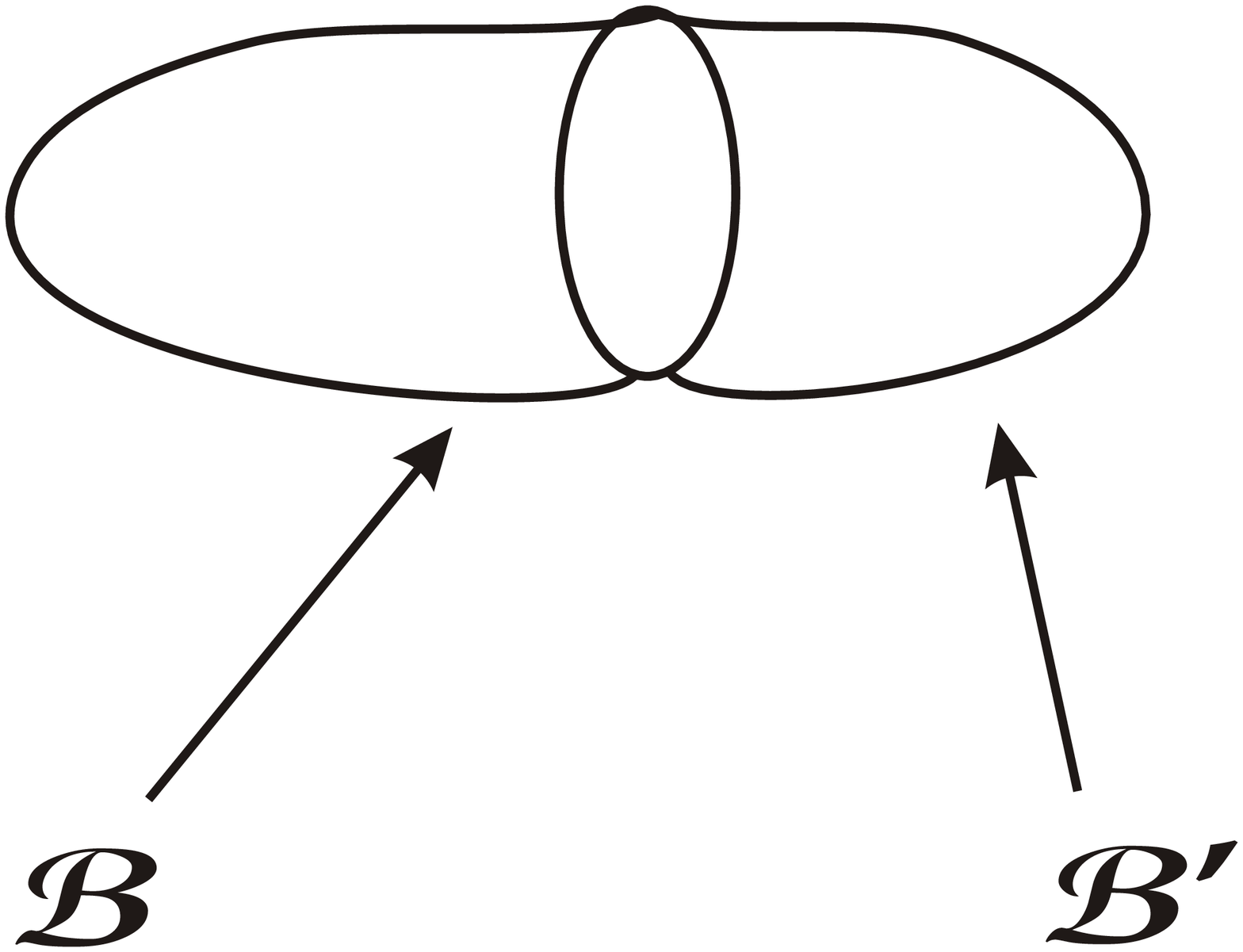}}

First of all, by definition, to a given character $ \chi \in \check
H^\ell(M)$, we may associate a fieldstrength $F_\chi$. We will often
drop the subscript on $F$ when the underlying character is
understood. We stress that $F$ is a globally well-defined form. Now,
slightly different boundaries $\CB$ and $\CB'$ as in \twoboundaryi,
lead to the same $\chi(\Sigma)$ and hence it follows that $F$ is in
the space of closed forms $F \in \Omega_d^\ell(M)$, i.e. $dF=0$.
Moreover, large changes in $\CB$, as in \twoboundaryii,  also lead
to the same holonomy. Since $\CB \cup (-\CB') $ will form a closed
$\ell$-cycle in $M$ we learn that $F$ must be closed with {\it
integral} periods. We denote this as $F \in \Omega^\ell_{ \IZ}(M)$.
\foot{The notation makes sense because if $\int_{\Sigma}F $ is
integral for every closed cycle then  $F$  is closed.} Characters
with $F=0$ are known as {\it flat}.

In discussions of gauge theories of differential forms physicists often stop
at this point, but one cannot stop here, because there is
gauge invariant information in a
character $\chi$ which is {\it not} contained in just its fieldstrength.
One way to realize this is to consider how one evaluates the
holonomy on a   cycle $\Sigma$ which
defines a torsion homology cycle. That is, suppose there is a natural
number $k$ such that $k \Sigma = \p \CB$ for an integral chain $\CB$
but $\Sigma$ is not itself a boundary.
We know that
\eqn\kpower{
\left( \chi(\Sigma)\right)^k = \exp \left( 2\pi i \int_\CB F \right)
}
but knowing $F$ alone is not sufficient to say how to take the $k^{th}$ root
of \kpower. The extra information about how to do this is encoded in the
  {\it characteristic class} of $\chi$, denoted  $a_\chi$. The characteristic
class   is an integral cohomology class $a_\chi \in H^\ell(M;\IZ)$. As with
the fieldstrength, we will often drop the subscript when the underlying
character is understood.
When $k \Sigma = \p \CB$, the
 characteristic class allows us to compute the holonomy of a character around $\Sigma$ via the
formula
\eqn\trueholo{
\chi(\Sigma) = \exp\biggl[ {2\pi i \over k} \bigl( \int_{\CB} F - \langle a, \CB\rangle \bigr) \biggr]
}
Here we have lifted $a$ to an integral cocycle in $C^k(M;\IZ)$, but
\trueholo\ does not depend on that  choice of lift, thanks to the
relation $k \Sigma = \p \CB$. Similarly, the formula is independent
of the choice of the chain $\CB$.  The reader should verify that
\trueholo\ in fact satisfies \kpower. Observe that if we denote $h =
{1\over 2\pi i } \log \chi$, then we may interpret \trueholo\ as
saying
\eqn\hsrels{
\delta h = F - a
}
which is  a key relation in \cheegersimons\ and the starting point for the
discussion in \HopkinsRD. Characters with $a=0$ are known as
{\it topologically trivial}.

In what follows it will help to bear in mind the following
mathematical remarks. The   cohomology group $H^\ell(M;\IZ)$ is a
finitely generated abelian group. It contains a torsion subgroup
${\rm Tors}(H^\ell(M;\IZ))$ consisting of the elements of finite
order. The quotient by this finite subgroup is a free abelian group
which will be denoted $\bar H^\ell(M;\IZ) $, thus
\eqn\torssbup{
0 \rightarrow {\rm Tors}(H^\ell(M;\IZ)) \rightarrow H^\ell(M;\IZ) \rightarrow  \bar H^\ell(M;\IZ) \rightarrow 0
}
The reduction modulo torsion $\bar H^\ell(M;\IZ)$  is isomorphic
to $\IZ^{b_{\ell}}$ where $b_{\ell}$ is the Betti number of $M$.
In general the sequence \torssbup\ does not split  naturally  so
that one cannot naturally speak of the ``torsion part'' of an integral cohomology class.
Still, loosely speaking, we can say that ``torsion information'' in $a$ is missing in the fieldstrength $F$.

We have now defined two maps, the fieldstrength map $F: \check H^\ell(M) \to \Omega^\ell_{\IZ}(M)$
and the characteristic class map $a: \check H^\ell(M) \to H^\ell(M;\IZ)$. It follows from \trueholo\ that
these maps are compatible. Namely, since $F$ is closed we can define a DeRham cohomology
class $[F]\in H^\ell_{DR}(M)$. Meanwhile, we can reduce modulo torsion by taking
$a \to \bar a \in   H^\ell(M;\IR)$. Using the isomorphism  $ H^\ell(M;\IR) \cong H^\ell_{DR}(M)$
we have $\bar a = [F]_{DR}$, in other words, the diagram
\input diagrams
\eqn\arrdiagr{
\diagram
  \check H^\ell(M) & \rTo &   \Omega^\ell_{\IZ}(M) \\
    \dTo &  & \dTo     \\
    H^\ell(M;\IZ) & \rTo &   H^\ell(M;\IR) \\
  \\
\enddiagram
} commutes. One might get the mistaken impression from \arrdiagr\
that $\check H^\ell(M)$ is the setwise fiber product $\CR := \{ (a,
F) \vert \bar a = [F]_{DR} \} \subset H^\ell(M;\IZ)\times
\Omega^\ell_{\IZ}(M)$.  This is not the case because there are
nontrivial characters which are both flat and topologically trivial.
In the Maxwell case these correspond to ``Wilson lines'' -- i.e.
flat connections on the trivial bundle --  which can be continuously
connected to the trivial connection. In general the space of
topologically trivial flat fields   is $H^{\ell-1}(M)\otimes
\IR/\IZ$. This space, which will be denoted $\CW^{\ell-1}(M)$,   is
a connected torus of dimension $b_{\ell-1}$. If we introduce a
metric on $M$  we can write this in terms of the harmonic forms
$\CH^{\ell-1}(M)$, namely
\eqn\wilsonlines{
\CW^{\ell-1}(M) := H^{\ell-1}(M)\otimes \IR/\IZ \cong \CH^{\ell-1}(M)/\CH^{\ell-1}_{\IZ}(M).
}

We are now in a position to describe a complete picture of the space $\check H^\ell(M)$.
The structure of this space is nicely summarized by two exact sequences.
The first exact sequence is   based on the fieldstrength map. A flat gaugefield
evidently defines a homomorphism from $H_{\ell-1}(M)$ into $\IR/\IZ$, and hence:
\eqn\flatseq{
0 \rightarrow \overbrace{H^{\ell-1}(M;\IR/\IZ)}^{\rm flat} \rightarrow \check H^\ell(M)
~~ {\buildrel {\rm fieldstrength}\over \longrightarrow} ~~  \Omega^{\ell}_{\IZ}(M) \rightarrow 0  .
}
If $\alpha \in H^{\ell-1}(M;\IR/\IZ)$ we will denote its image
in $\c H^\ell(M)$ by $\c \alpha$.
We must stress that flat fields are not necessarily topologically trivial.
These topologically nontrivial flat fields will play an important role in our discussion below,
so let us describe them in some detail.
In general the space of flat fields $H^{\ell-1}(M;\IR/\IZ)$ is a compact abelian group. The
connected component of the identity is the torus $\CW^{\ell-1}(M)$  we have just discussed.
In addition, there is
 a group of components, which is isomorphic to ${\rm Tors}(H^\ell(M;\IZ))$.
\foot{One proves this using the long exact sequence of cohomology groups associated with the
short exact sequence $0 \rightarrow \IZ \rightarrow \IR \rightarrow \IR/\IZ \rightarrow 0 $ of
coefficient groups. }
 Put differently
we have the exact sequence
\eqn\torsioncpt{ 0 \rightarrow \CW^{\ell-1}(M)  \rightarrow
H^{\ell-1}(M, \IR/\IZ) ~ {\buildrel \beta \over \rightarrow} ~  {\rm
Tors}(H^\ell(M)) \rightarrow 0 . }
where $\beta$ is known as the ``Bockstein map.''

Let us pause to note some simple examples of groups of flat fields with nontrivial components.
In Maxwell theory on a Lens space $L_k = S^3/\IZ_k$ we have $\ell=2$, while
$H^1(L_k;\IR/\IZ) \cong H^2(L_k;\IZ) \cong \IZ_k$.  In $M$-theory on a product of   Lens spaces
we will encounter
\eqn\example{
H^3(L_k \times L_k; \IR/\IZ).
}
where both the first and third terms in \torsioncpt\ are nonvanishing.
Heuristically, one can picture this group of flat fields as a
``sum'' of $k$ copies of a two-dimensional torus.

The second exact  sequence is based on the characteristic class map:
\eqn\ccseq{
0 \rightarrow \underbrace{\Omega^{\ell-1}(M)/\Omega^{\ell-1}_{\IZ}(M) }_{\rm Topologically\ trivial} \rightarrow \check H^\ell(M)
~~ {\buildrel {\rm char. class} \over \longrightarrow} ~~  H^{\ell}(M;\IZ) \rightarrow 0
}
The topologically trivial characters {\it are} in fact associated with a globally well-defined $(\ell-1)$-form
potential $A\in \Omega^{\ell-1}(M)$.
Indeed, given such a globally well-defined $A$ we can define the differential character:
\eqn\toptrvc{
\chi_A: \Sigma \to \exp\left( 2\pi i \int_{\Sigma} A \right)
}
When working additively we denote such characters by $[A]$, without the $\check{}$.
Clearly, the fieldstrength of such a character
is $F= dA $. Since $A$ is globally well-defined $[F]_{DR}=0$.   Comparing with
\trueholo\ we see that in fact the characteristic class is likewise equal to zero. Since $\check H^\ell(M)$ is
the space of {\it gauge inequivalent} fields we should be careful to identify the space of
topologically trivial characters with the quotient $\Omega^{\ell-1}(M)/\Omega^{\ell-1}_{\IZ}(M)$
since, after all, if we shift $A \to A+ \omega$, where $\omega \in \Omega^{\ell-1}_{\IZ}(M)$ is a
closed $(\ell-1)$-form with integral periods then the holonomy, and therefore the gauge equivalence
class, of the field has not changed.

In the physics literature, these shifts $A \to A+\omega$ constitute
the standard formulation of gauge transformations of form-valued
fields. In the literature, when $\omega$ is exact, $\omega =
d\Lambda$, it is referred to as a small gauge transformation, and
when $[\omega]$ is nontrivial, it is called a large gauge
transformation. In topologically nontrivial sectors one can choose a
``basepoint'' with associated fieldstrength $F_{\bullet}$   and
introduce a  global $A$ via
\eqn\basepoint{
F= F_{\bullet} + d A.
}
This is what is implicitly done in much of the literature. Of course, such a formulation relies on an
unnatural choice of basepoint, and isn't well-suited to including effects of torsion.
\foot{Moreover, it is precisely at this point that the viewpoint of $\c A$ as an object in a groupoid
becomes a significant improvement, for it has an automorphism group $H^{\ell-2}(Y,\IR/\IZ)$
with important physical consequences.
In physics, this group should be understood as the group of global gauge transformations,
and is closely connected with ``tadpole constraints'' and ``Gauss laws.''
 See \DiaconescuBM\MooreJV\BelovJD\  for further discussion.}

Note that the space of topologically trivial fields $\Omega^{\ell-1}(M)/\Omega^{\ell-1}_{\IZ}(M)$
is not a vector space. Rather we have the fibration:
\eqn\toptrvfib{
\matrix{ \Omega^{\ell-1}_d(M)/\Omega^{\ell-1}_{\IZ}(M) & \rightarrow & \Omega^{\ell-1}(M)/\Omega^{\ell-1}_{\IZ}(M)\cr
                                                       &              & \downarrow \cr
                                                        &              &        \Omega^{\ell-1}(M)/\Omega^{\ell-1}_{d}(M)\cr}
}
where $\Omega^{\ell-1}_{d}(M)$ is the space of {\it closed} forms. Note that
$\Omega^{\ell-1}_d(M)/\Omega^{\ell-1}_{\IZ}(M) $ is isomorphic to a torus. In fact it is
once again the torus of topologically trivial
flat fields
\eqn\trivtor{
\Omega^{\ell-1}_d(M)/\Omega^{\ell-1}_{\IZ}(M)
%
%\cong \CH^\ell(M)/\CH^{\ell}_{\IZ}(M)\cong H^{\ell-1}(M)\otimes \IR/\IZ
%
\cong
\CW^{\ell-1}(M).
}
Meanwhile, $\Omega^{\ell-1}(M)/\Omega^{\ell-1}_{d}(M)$ is a vector space. (Indeed, if   we introduce a metric, then
 we can identify it with the vector space ${\rm Im} d^\dagger$. )

\ifig\dcsspace{ As schematic picture of the group of differential
characters.  } {\epsfxsize2.0in\epsfbox{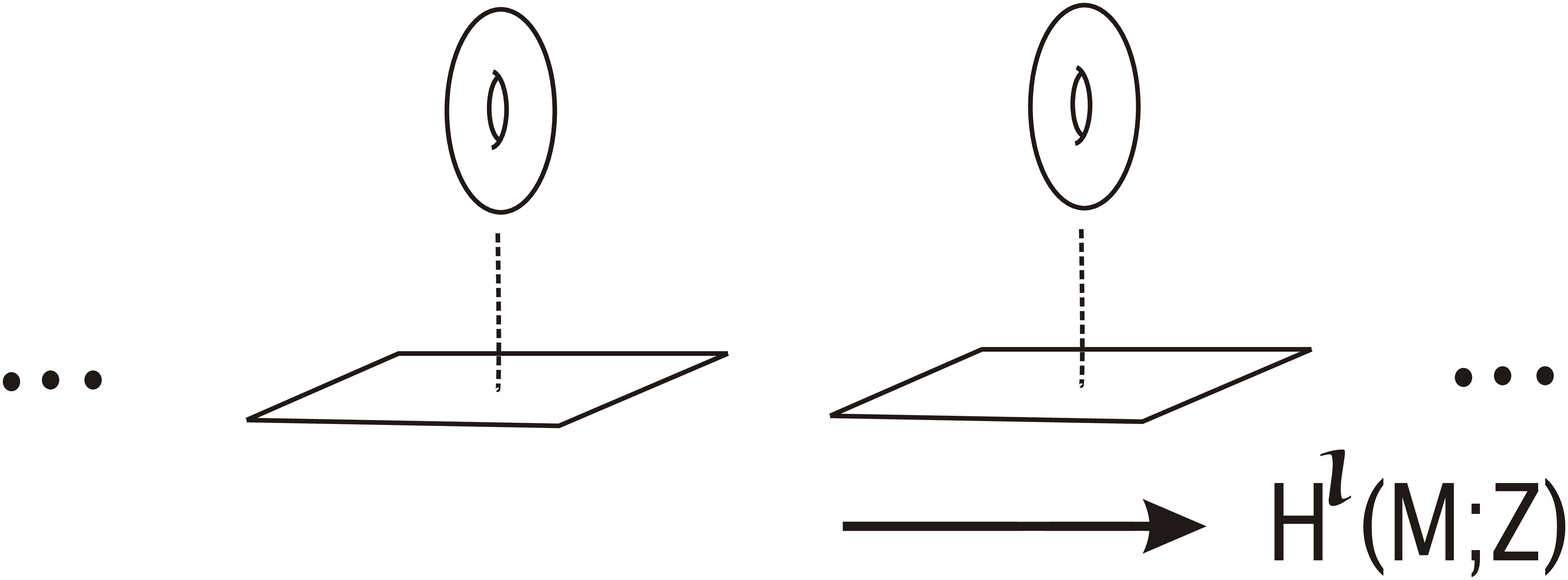}}

In summary, the infinite-dimensional abelian group $\check H^\ell(M)$ can be pictured as a union over components,
labelled by $H^\ell(M;\IZ)$, of a torus bundle over an infinite-dimensional affine space, as in \dcsspace.
By choosing a basepoint, as in \basepoint\ we may identify the affine space with the
vector space $ \Omega^{\ell-1}(M)/\Omega^{\ell-1}_{d}(M)$.

Let us give some examples of differential cohomology groups.

\item{1.} It is easy to show that the nonzero differential cohomology groups of a point are
$\check H^0(pt) = \IZ$ (given by the characteristic class)
and $\check H^1(pt) = \IR/\IZ$ (given by the topologically trivial flat fields).

\item{2.} For any manifold $M$, $\check H^1(M)$ is exactly the same thing as the space of differentiable
maps $f: M \to U(1)$.  The characteristic class is $f^*([d \theta])$ where $[d \theta]$ is the fundamental
class of $S^1$ and $F= {1\over 2\pi i } d (\log f) $.

\item{3.} A very significant special case of the previous example is $\c H^1(S^1)$, which may be
thought of as the fieldspace of a periodic scalar field on a circle,
familiar from string theory. The topological components are the
winding number. The topologically trivial fields are
$\Omega^0(S^1)/\Omega^0_{\IZ}(S^1) = \IT \times \Omega^0/\IR$.
\foot{We denote the circle $S^1$ by $\IT$ when we wish to emphasize
the $U(1)$ group structure. Thus, $S^1 = U(1) = \IT \cong \IR/\IZ$
are four notations used for  the circle group. } The flat fields are
just the factor $\IT$ of constant maps. The vector space
$V=\Omega^0/\IR$ are the loops in $\IR$ with center of mass equal to
zero.  Altogether we have
\eqn\scalari{
\c H^1(S^1) \cong \IT \times \IZ \times V
}
This corresponds to the explicit decomposition,
\eqn\scalarii{
\varphi(\sigma) = \exp 2\pi i \biggl[  \phi_0 +   w \sigma + \sum_{n\not=0} {\phi_n \over n} e^{2\pi i n \sigma}  \biggr]
}
where $\sigma \sim \sigma +1$ parametrizes $S^1$ and $\bar \phi_n = \phi_{-n}$.

\item{4.} As we have seen, $\check H^2(M)$ is the group of isomorphism classes of line bundle with connection
over $M$.

\item{5.} The $B$-field of type II string theory defines a class in $\check H^3(M)$. Differences
of $C$-fields in $M$-theory are valued in $\check H^4(M)$ (for more information see \DiaconescuBM.)

\item{6.} The WZ functional in the 2D WZW model is based on an element in $\check H^3(G)$ where $G$ is
the group manifold of the WZW model. This was indeed the occasion for one of the first
occurances of differential cohomology in the physics literature  \AlvarezES\GawedzkiAK.
For some other early papers introducing differential cohomology into the physics literature
see \DijkgraafPZ\FreedQB.

\subsec{Product}

There is a product on Cheeger-Simons characters:

\eqn\product{
\check H^{\ell_1}(M) \times \check H^{\ell_2}(M) \rightarrow \check H^{\ell_1 + \ell_2}(M).
}
The fieldstrength of a product of characters $\chi_1 \star \chi_2$ is simply $F_1 \wedge F_2$,
and the characteristic class is $a_1\smile a_2$. However, the holonomy is somewhat subtle
to define. (For discussions from different viewpoints
 see \BHarris\harveylawson\HopkinsRD\DeligneQP).
The product \product\  is not to be confused with the pointwise product making $\c H^{\ell}(M)$ an
abelian group.
In fact, the product $\chi_1\star \chi_2$ in \product\ induces a ring structure on $\c H^*(M)$.
%In
%particular  $(\chi_1\cdot \chi'_1)*\chi_2 = (\chi*\chi_2)\cdot (\chi'*\chi_2)$.
%
The distributive law is more transparently written in additive notation:
\eqn\ringstruct{
([\c A_1] + [\c A_1'])\star [\c A_2] = [\c A_1]\star  [\c A_2] + [\c A_1']\star [\c A_2]
}
In fact $\c H^*(M)$ is a graded commutative ring (with the grading we have adopted)
\cheegersimons\Gomi.

There are two important cases where the
product is readily described. These occur when one of the factors is flat or topologically
trivial. If,   say,
 $\chi_1 = \chi_{A_1}$ is
topologically trivial, then so is the product character $\chi_{1}\star \chi_2$. Indeed the
product character only depends on the fieldstrength $F_2$ of $\chi_2$. Written
additively $[A_1]\star [\c A_2] = [ A_1 \wedge F_2]$.  In particular if both are
topologically trivial
\eqn\simpleprod{
[A_1]\star [A_2] = [A_1 d A_2].
}
Note in particular that the product of two fields which are topologically trivial and flat must
vanish. We will use this fact later.

On the other hand, if $\chi_1$ is flat
and hence of the form $\chi_1 =\c \alpha_1$ for  $\alpha_1\in H^{\ell_1-1}(M;\IR/\IZ)$  then the product is also flat,
and hence in the image of $ H^{\ell_1+\ell_2-1}(M;\IR/\IZ)$. Indeed, the product only depends
on $\chi_2$ through its characteristic class $a_2$ and is given by the image of
$\alpha_1\smile a_2 \in H^{\ell_1+\ell_2-1}(M;\IR/\IZ)$.

As an exercise, the reader might wish to contemplate the simplest
nontrivial product, namely $\check H^1(M) \times \check H^1(M) \to
\check H^2(M)$. That is, from two circle-valued functions on $M$ one
must produce a line bundle with connection (up to isomorphism).

\subsec{Integration}

If $M$ is compact and oriented, and of dimension $\dim M =n$ then evaluation on the fundamental
cycle $[M]$ defines  an integration
map
\eqn\integral{
\int_M^{\check H}: \check H^{n+1}(M) \to \c H^1(pt) \cong \IR/\IZ
}
We stress that this is {\it not} the integral of the fieldstrength
(which has the wrong degree) but the ``holonomy around $M$.'' Any
class in $\c H^{n+1}(M)$ must be topologically trivial, hence
represented by a globally defined form $A\in \Omega^n(M)$ and the
integral \integral\ is just $\int A \, \mod\,  \IZ$. More generally,
suppose $\CX$ is a {\it family } of spacetimes with a projection
$\pi: \CX \to S$ with typical fiber $M$. In physical applications
$S$ might parametrize metrics or other data on the spacetime $M$. In
this case there is an integration
\eqn\intfamily{
\int^{\check H}_M: \check H^k(M) \to \check H^{k-n}(S)
}
and \integral\ is the special case of $S=pt$.

For examples we have
\eqn\intri{
\int^{\check H}_{pt}: \check H^1(pt) \to \IR/\IZ
}
is just evaluation of the $\IR/\IZ$ valued function on the point.
As a second example:
\eqn\intrii{
\int^{\check H}_{S^1}: \check H^2(pt) \to \IR/\IZ
}
is ${1\over 2\pi i } \log hol(A) $ where $hol(A)$ is the holonomy of
the connection around $S^1$. As a third example, if $\dim M = 2p+1$,
and $[\check A] \in \check H^{p+1}(M)$, then, when $p$ is odd,
\eqn\csterm{
\int^{\check H}_M [\check A ] \star  [\check A ]
}
is a Chern-Simons term. Note that from \simpleprod\ it follows that
if $[A]$ is topologically trivial then \csterm\ becomes the familiar
expression $\int A dA$.  In the physics literature, when ``$A$'' is
not globally well-defined, this integral is often defined by
extending $F$ to an extending field $\tilde F$ on a $2p+2$
dimensional bounding manifold $\CB$ and taking $\int_{\CB} \tilde F
\wedge \tilde F$.  \foot{Such a definition is in fact ambiguous,
since one can always add to the extending field $\tilde F$ a
compactly supported closed form on $\CB$ vanishing at the boundary.
Rather, one  must extend the entire differential character, and not
merely $F$. Moreover, there can be obstructions to making such an
extension.  } When $p$ is even \csterm\ is $\IZ_2$-valued, as
discussed below.

We would like to place  this integration procedure in a more general  context.
In any generalized cohomology theory there is an integration procedure.
For a generalized cohomology theory $E$ we can
define an ``integration''
\eqn\intgam{
\int^{E}_M : E^{q}(M) \to E^{q-n}(M)
}
where $n=\dim M$. This integral will  be defined if $M$ is
``$E$-oriented.''   Typically, an $E$-orientation involves some
extra structures on $M$. In order to define the integral \intgam\ we
embed $M$ into a large sphere $S^N$. The normal bundle of $M$ must
be $E$-oriented, which simply
 means there is a Thom isomorphism
\eqn\thomone{ \Phi_1: E^{q}(M) \to E^{q+N-n}_{cpt}(\CN). }
Now, $\Phi_1(x)$ can be extended from the tubular neighborhood $\CN$ of $M$ in $S^N$ to all of $N$
by ``extension by zero.'' Call this extension $\widetilde{\Phi_1(x)}$.
  Next, choose any point $P$ not in the tubular neighborhood. The normal
bundle of $P$ is just the ball $B^N$ around $P$. There is also
a Thom isomorphism $\Phi_2: E^s(pt) \to E^{s+N}(B^N)$. Finally we restrict $\widetilde{\Phi_1(x)}$ to
$B^N$ and then apply the inverse Thom isomorphism. This defines the integral:
\eqn\integdef{
\int^E_M x :=
\Phi_2^{-1}\widetilde{\Phi_1(x)}\vert_{B^N} \in E^{q-n}(pt)
}
More generally, if we have a family $\CX \to S$ and a class $x \in
E^q(\CX)$ then $\int^{E}_{\CX/S} x  \in E^{q-n}(S)$ is integration
along the fibers. As examples, $\int^H_M$ is defined if $M$ is
oriented, and corresponds to the usual notion of integration.
Meanwhile $\int^K_M$ is defined if $M$ has a
   ${\rm Spin}^c$ structure. If $\dim M$ is even then
$\int^K_M x = {\rm Index}(\Dsl_x)$ by the Atiyah-Singer index theorem.
The case when $\dim M$ is odd will be used and explained below.
By combining the above  integration for  generalized cohomology theories
with integration of ordinary differential forms one can define
an integration procedure $\int^{\check E}$ on
differential generalized cohomology theories \HopkinsRD.
  We have already introduced
$\int^{\c H}$ above and will have occasion to use $\int^{\c K}$ below
(an explicit formula is given in equation $(A.1)$ below.)

\subsec{Poincar\'e-Pontrjagin duality}

Finally, Poincar\'e duality for differential cohomology plays a very important
role in what follows. Poincar\'e duality states that if
 $M$ is compact and oriented, of dimension $\dim M=n$,
 then we have a perfect pairing
\eqn\pddiff{
\check H^{\ell}(M) \times \check H^{n+1-\ell}(M) \to \IR/\IZ
}
The pairing is given by multiplication of characters followed by the integration
described above:
\eqn\pairng{
\langle \chi_1, \chi_2\rangle := \int^{\c H}_M \chi_1 \star  \chi_2
}
The adjective ``perfect'' means that every homomorphism $\check
H^\ell(M) \to \IR/\IZ$ is given by pairing with an element of
$\check H^{n+1-\ell}(M)$. \foot{We are glossing over some
technicalities here. We are using the smooth Pontrjagin dual; we
refer to   \harveylawson\ for a rigorous discussion.
%
%The important point for us is that the Pontrjagin dual,
%namely the space of smooth homomorphisms $\c H^{\ell}(M)\to U(1)$ is isomorphic
%to $\c H^{n+1-\ell}(M)$.
}

In order to justify Poincar\'e duality we consider   two exact sequences.
The first is the fieldstrength sequence in degree $\ell$
\eqn\flatseqp{
0 \rightarrow  H^{\ell-1}(M;\IR/\IZ)  \rightarrow \check H^\ell(M) \rightarrow
  \Omega^{\ell}_{\IZ}(M) \rightarrow 0
}
together with the characteristic class sequence in degree $n+1 - \ell$:

\eqn\ccseqp{
0 \rightarrow  \Omega^{n-\ell}(M)/\Omega^{n-\ell}_{\IZ}(M)   \rightarrow \check H^{n+1-\ell}(M) \rightarrow
  H^{n+1-\ell}(M;\IZ) \rightarrow 0  .
}
Now note that there is a standard perfect pairing $H^{\ell-1}(M;\IR/\IZ)\times H^{n+1-\ell}(M;\IZ)\to \IR/\IZ$
given by multiplication and integration. On the other hand there is also a perfect pairing
\eqn\othepair{
\Omega^{\ell}_{\IZ}(M)  \times \biggl(
\Omega^{n-\ell}(M)/\Omega^{n-\ell}_{\IZ}(M) \biggr) \to \IR/\IZ
}
given by $(F, A_D) \mapsto   \int_M F \wedge A_D~ \mod \IZ $.

Now we use the following general fact. Given two short exact sequences
\eqn\twosespp{
\diagram
 0  & \rTo    &  A     & \rTo   &  B        & \rTo     & C        & \rTo    & 0   \\
    &         &  \dTo  &        &  \dTo     &       &   \dTo       &         &     \\
0   & \rTo    &   A'   & \rTo   &  B'        & \rTo     &  C'      & \rTo    & 0    \\
  \\
\enddiagram
} if the first and third vertical arrows are isomorphisms then the
middle arrow is an isomorphism. Apply this to \flatseqp\ and the
dual of \ccseqp (recall that ${\rm Hom}(\cdot, \IR/\IZ)$ is an exact
functor).

\rmk{}

\item{1.} Let us return to the example of the periodic scalar, i.e. $\c H^1(S^1)$ discussed in
\scalari. An explicit formula for the pairing $\langle \varphi^1, \varphi^2 \rangle$ can be given
as follows. Let $\varphi = \exp(2\pi i \phi)$ where $\phi: \IR \to\IR$ satisfies
$\phi(s+1) = \phi(s) +w$. Here $w\in \IZ$ is the winding number and $\sigma = s \mod 1$.
The pairing is then \cheegersimons\
\eqn\corrscalarpair{ \langle \varphi^1, \varphi^2 \rangle  =
\int_0^1 \phi^1 {d \phi^2\over ds} ds  - w^1 \phi^2(0)  \quad \mod
\IZ. }
To prove this consider the outer product $\varphi^1 \times \varphi^2
\in \c H^2(S^1 \times S^1)$, where we pull back and then multiply.
This describes a line bundle with connection on $S^1 \times S^1$. We
can identify $ \langle \varphi^1, \varphi^2 \rangle $ with (the
inverse of ) the holonomy $h$ along the diagonal. That holonomy can
be computed by considering the triangle $\Delta_{\sigma_0}$ obtained
by projecting the triangle in $\IR^2$ bounded by $\CC_1 = \{ (s_0 +
x, s_0 +x )\vert 0 \leq x \leq 1\}$, $\CC_2 = \{ (s_0 + x, s_0)
\vert 0 \leq x \leq 1\}$, and $\CC_3 = \{ (s_0 + 1, s_0 +x )\vert 0
\leq x \leq 1\}$. Then
\eqn\around{ h(\CC_1) = \exp\biggl(2\pi i \int_{\Delta_{s_0}} F
\biggr) h(\CC_2)^{-1} h(\CC_3)^{-1} }
Now $F= d \phi_1 \wedge d \phi_2$ so
\eqn\intcurv{ \int_{\Delta_{s_0}} F = \phi^1(s_0) w_2 - \int_0^1
\phi^1(s_0 + x) {d\over dx} \phi^2(s_0+x) dx }
On the other hand, $h(\CC_2) $ can be computed by noting that if
$\pi: S^1 \times \{ \sigma_0\} \to \{ (\sigma_0, \sigma_0) \}$ then
$h(\CC_2) = \pi_*(\varphi^1 \times \pi^*\varphi^2(\sigma_0)) =
(\varphi^2(\sigma_0))^{w_1}$ and similarly $h(\CC_3) =
(\varphi^1(\sigma_0))^{-w_2}$. Combining with \around\ we obtain an
expression which is independent of $s_0$. Setting $s_0=0$ gives
\corrscalarpair.

\item{2.}
Note that Poincar\'e duality gives a simple way of characterizing
the differential character induced by a source (i.e. a ``brane'')
wrapping a cycle $\Sigma \in Z_{n-\ell}(M)$. Such a cycle defines a
homomorphism $\check H^{n+1-\ell}(M) \to U(1)$ by evaluation and
hence defines $\check \delta(\Sigma) \in \check H^\ell(M)$. That is,
$\langle \c \delta(\Sigma), \chi\rangle = \chi(\Sigma) $. (Of
course, this differential character is {\it not} smooth.)

\newsec{Hamiltonian Formulation of Generalized Maxwell Theory}

Let us now consider spacetimes of the form $M= Y \times \IR$. In generalized Maxwell theory
the gauge equivalence classes of fields are $[\c A] \in \c H^\ell(M)$. We take the action to
be simply
\eqn\action{
S = \int_M \pi R^2 F*F
}
where $R^2$ can be thought of as ${1\over 2 e^2}$, with $e$ the
coupling constant. Alternatively $R$ can be viewed   as a ``target
space'' radius. In this paper we put $c=1$, but we do {\it not} put
$\hbar =1$. Maxwell theory is the case $\ell=2$.  The quantization
of this theory is completely straightforward. The phase space is
$T^*\c H^\ell(Y)$. The Hilbert space is $\CH = L^2(\c H^\ell(Y))$,
and a typical state will be denoted $\psi(\c A)$.\foot{In this paper
we are not careful about questions of analysis such as the
definition of $L^2$ on the infinite-dimensional space $\c
H^\ell(Y)$. Since we are dealing with free field theory these points
can be dealt with rigorously \FMSi. }

Now, there is a natural grading of the Hilbert space by (the topological class of the) magnetic flux:
\eqn\magflx{
\CH = \oplus_{m} \CH_m   \qquad\qquad m\in H^\ell(Y,\IZ).
}
This is immediate, since $m$ labels the connected components of the configuration
space $\c H^\ell(Y)$. On the other hand, generalized Maxwell theory
 enjoys electro-magnetic duality. A completely equivalent theory is based on
a dual potential $[\c A_D]\in \c H^{n-\ell}(M)$ with action
\action\ with
\eqn\dualcoupling{
R R_D = \hbar.
}
It follows immediately that there must   also be
a grading by (the topological class of the) electric flux:
\eqn\elecflx{
\CH = \oplus_{e} \CH_e  \qquad \qquad e\in H^{n-\ell}(Y,\IZ),
}
where, for the moment, we think of $e$ as classifying the connected
component of the dual field space. Having said this, the question
naturally   arises whether we can simultaneously measure both electric and magnetic
flux. That is, can we find a simultaneous grading:
\eqn\magelecflx{
\CH \quad {\buildrel ? \over =} \quad  \oplus_{e,m } \CH_{e,m}.
}

The natural response of the reader to the question \magelecflx\
might be that the answer is ``yes!''
  The reason is that  one generally thinks of measuring
magnetic flux via the period integrals $\int_{\Sigma_1} F$ where $\Sigma_1$ is a closed
$\ell$-cycle. Similarly, one measures electric flux via
$\int_{\Sigma_2} *F$ where $\Sigma_2$ is a closed
$n-\ell$-cycle. Now in the Legendre transform to the Hamiltonian version of the
 theory the canonical momentum conjugate to $[\c A]$ is   $\Pi = 2\pi R^2 (*F)_Y$
(where $*F$ is restricted to the spatial slice $Y$). Using the standard quantum
mechanical commutation relations we have
\eqn\ccrs{
\eqalign{
[\int_{\Sigma_1} F, \int_{\Sigma_2} *F] & = 2\pi R^2 [\int_{\Sigma_1} F, \int_{\Sigma_2}   \Pi ] \cr
& = 2\pi R^2  [\int_Y \omega_1 F , \int_Y \omega_2   \Pi ] \cr
& = 2\pi i \hbar   R^2  \int_Y \omega_1 d \omega_2 = 0 \cr}
}
where $\omega_i $ are closed forms Poincar\'e dual   to $\Sigma_i$.
Thus, it would seem, electric and magnetic fluxes clearly commute.
The problem with this argument is that the above period integrals
only measure the flux modulo torsion. After all, a real number like
$\int_{\Sigma} F$ cannot be torsion: If $N\int_{\Sigma} F=0 $ for
some integer $N$ then $\int_{\Sigma} F=0$. But the   fluxes $e,m$ in
\elecflx,\magflx\ are elements of abelian groups. As we have
stressed, these groups in general have nontrivial torsion subgroups.

In order to understand what happens at the torsion level we need to understand the grading by
electric flux more deeply. We motivated it above by appealing to the electromagnetic dual
formulation of the theory, but we need to understand it directly in the formulation in terms
of $[\c A]$. In order to do this let us note that diagonalizing $(*F)_Y$ means diagonalizing $\Pi$,
but the momentum $\Pi$ is the generator of translations. This motivates the following definition:

\defn{} A {\it state $\psi$ of definite electric flux $\c \CE\in \c H^{n-\ell}(Y)$ }
 is a translation eigenstate on $\c H^\ell(Y)$, i.e.
\eqn\defelst{
\forall \c \phi \in \c H^\ell(Y) \qquad \qquad \psi(\c A + \c \phi) = \exp\biggl( 2\pi i \int^{\c H}_Y \c \CE \star \c \phi   \biggr) \psi(\c A)
}

In general, working with states of definite electric flux is not useful. We are often only really interested
in working with states of definite {\it topological} sector of electric flux. Observe that  $\c \CE_1$ and $\c \CE_2$
are continuously connected  (i.e. have the same characteristic class) if and only if
\eqn\smaesec{
\int^{\c H}_Y \c \phi_f \star  \c \CE_1 = \int^{\c H}_Y \c \phi_f \star  \c \CE_2  \qquad \qquad \forall \phi_f \in H^{\ell-1}(Y,\IR/\IZ)
}
(To see this consider the discussion above on the Poincar\'e dual pairing.) Therefore we can make the
more useful definition:
\defn{} A {\it state of definite topological class of electric flux} is an   eigenstate under translation by
flat characters  $H^{\ell-1}(Y,\IR/\IZ)\subset \c H^\ell(Y)$, i.e.
\eqn\defelstp{
\forall  \phi_f \in H^{\ell-1}(Y,\IR/\IZ) \qquad \qquad \psi(\c A + \c \phi_f) = \exp\biggl( 2\pi i \int_Y   e \phi_f  \biggr) \psi(\c A)
}
for some $e\in H^{n-\ell}(Y;\IZ)$.

Note that if $\psi$ is in a
state of electric flux $\c \CE$ then $e=a_{\c \CE}$, and the phase
 in \defelst\ reduces to the phase in \defelstp\ for translations by flat fields.

It follows from the definition of \defelstp\ that
  \elecflx\ is the decomposition of $\CH$ in terms of characters of the group
of translations by $H^{\ell-1}(Y,\IR/\IZ)$. Dually, the decomposition by magnetic flux can be understood
as the diagonalization of the group of flat dual connections $H^{n-\ell-1}(Y, \IR/\IZ)$.

We are now in a position to see that in fact the
simultaneous decomposition \magelecflx\ is in general impossible. Suppose
the two translation groups could be simultaneously diagonalized and that
$\psi$ is in a state of definite magnetic flux $m\in H^\ell(Y,\IZ)$. That is, suppose the support of the
wavefunction $\psi$ is entirely contained in the topological sector labelled by $m$. {\it Such
a state cannot be in an eigenstate of translations by flat characters  if there are any
flat, but topologically nontrivial characters}. This is simply because translation by
such a topologically nontrivial flat character must translate the support to a different
topological magnetic flux sector. In this way we conclude that one cannot in general measure
both electric and magnetic flux. This is the main point of this paper, but we would like to
understand it more deeply. Doing so requires some further mathematical background.

\subsec{ Heisenberg groups and their representations }

We have have just argued that in general electric and magnetic fluxes do not commute.
 In order to understand this phenomenon better and more systematically
we must digress and review the theory of Heisenberg groups. We will
be somewhat heuristic and imprecise at some points below. For a more
precise discussion see appendix A of \FMSi\ and references therein.

To begin, let $S$ be any topological abelian group with a measure.
\foot{We will consider $S=\c H^\ell(Y)$ in our application. This
does not have a measure, but by considering wavefunctions with
suitable Gaussian falloff it can be thought of as a group with a
measure. We assume the measure is sufficiently well-behaved under
translation. }
 Consider $\CH = L^2(S)$.
On the one hand, $\CH$ is a representation of $S$. After all, for all $s_0 \in S$ we can define the
translation operator:
\eqn\transop{
(T_{s_0}\psi)(s) := \psi(s+ s_0) .
}
On the other hand, $\CH$ is also a representation of the
Pontrjagin dual group of characters, denoted $\hat S$.
If $\chi \in \hat S$ is a character on $S$  then we define the multiplication
operator $M_\chi$ on $\CH$ via
\eqn\multop{ (M_\chi \psi)(s) := \chi(s) \psi(s). }
Note that $\CH$ is {\it not} a representation of $S \times \hat S$. This simply follows from the
easily-verified relation
\eqn\noncomm{
T_{s_0} M_{\chi} = \chi(s_0) M_\chi T_{s_0}.
}
The proper way to interpret this relation is that $\CH$ is a representation of a central
extension of $S \times \hat S$. We therefore pause to  review the theory of central extensions

A central extension of a group $G$ by an abelian group $A$ is a group $\tilde G$ such that
\eqn\ceone{
0 \rightarrow A \rightarrow \tilde G \rightarrow G \rightarrow 0
}
with $A$ in the center of $\tilde G$.
We can think of $\tilde G$ concretely as a set of pairs $(g,a)$ with a twisted multiplication:
\eqn\twistmul{
(g_1, a_1) (g_2, a_2) = (g_1 g_2, c(g_1,g_2) a_1 a_2) .
}
In order to satisfy the associativity law the function
$c: G \times G \to A$ must satisfy the {\it cocycle relation}
$ c(g_1, g_2) c(g_1 g_2, g_3) = c(g_1, g_2 g_3)c(g_2,g_3) $. Cocycles that differ by coboundaries, i.e.,
$c'(g_1,g_2) = {f(g_1 g_2)\over f(g_1) f(g_2)}  c(g_1,g_2) $ for some function $f: G \to A$,
 define isomorphic central extensions.

In this paper we are concerned with the special case that $G$ is abelian and $A= U(1)$.
The formulae are a little clearer if we write $G$ additively, and $\tilde G$ multiplicatively.  In this case
the group commutator is easily shown to be:
\eqn\groupcom{
[(x_1, z_1), (x_2, z_2)] = (0, {c(x_1,x_2)\over c(x_2,x_1)} )
}
and therefore the group is abelian iff $c$ is symmetric.
Let $s(x_1,x_2) = {c(x_1,x_2)\over c(x_2,x_1)}$, be the commutator function.
Note that

\item{1.} $s$ is {\it skew }: $s(x,y) = s(y,x)^{-1}$.

\item{2.} $s$ is {\it alternating}: $s(x,x)=1$.

\item{3.} $s$ is {\it bimultiplicative}: $s(x_1+x_2,y) = s(x_1,y)
s(x_2,y)$ and $s(x,y_1 + y_2) = s(x,y_1) s(x,y_2)$.

The third property follows by applying the cocycle identity. If $s$
is nondegenerate, that is, if for all $x$ there is a $y$ with
$s(x,y)\not=1$,  we say  that $\tilde G$ is a (nondegenerate) {\it
Heisenberg group}. In this case $\tilde G$ is maximally
noncommutative in the sense that its center $Z(\tilde G)$ is the
group $U(1)$. If $s$ is degenerate we sometimes speak of a
``degenerate Heisenberg group.'' Often a ``Heisenberg group'' is
understood to be nondegenerate.
%
%Equivalently, we can define a
%Heisenberg group by saying that $s$  defines an isomorphism of $G$ with $\hat G$.
%

We now need some basic facts about  central extensions and
Heisenberg groups. (See \Mumford\FMSi\ for a careful discussion and
references for these well-known results.)

\bigskip
\thm{1} Let $G$ be a topological abelian  group. \foot{There are
some restrictions on $G$. See \FMSi, appendix A.}  The isomorphism
classes of central extensions of $G$ by $U(1)$ are in one-one
correspondence with continuous bimultiplicative maps
\eqn\commfun{ s: G \times G \to U(1) } which are alternating (and
hence skew).
\bigskip

In otherwords, given the commutator function $s$ one can always find
a corresponding cocycle $c$. \foot{In fact, a stronger statement is
that the category of extensions with given commutator is equivalent
to the category of cocycles with coboundaries as morphisms.}  This
theorem is useful because $s$ is invariant under change of $c$ by a
coboundary, and moreover the bimultiplicative property is simpler to
check than the cocycle identity. (In fact, one can show that it is
always possible to find a cocycle $c$ which is bimultiplicative.
This property automatically ensures the cocycle relation.) It is
important to realize that $s$ only characterizes $\tilde G$ up to
{\it noncanonical} isomorphism: to give a definite group one must
choose a definite cocycle.

The next result we need is

\bigskip
\thm{2} If $\tilde G$ is a central extension of a locally compact
topological abelian group  $G$ by $U(1)$,  then the unitary irreps
of $\tilde G$ where $U(1)$ acts by scalar multiplication are in
$1-1$ correspondence with the unitary irreps of the center of
$\tilde G$, where $U(1)$ acts by scalar multiplication.

In particular, applying this result to Heisenberg groups we obtain
the Stone-von Neumann theorem, which guarantees the essential
uniqueness of the irreducible representation of a Heisenberg group
$\rho$ with $\rho(0,z) = z$. We will actually need a generalization
of this statement to infinite-dimensional groups (which are not
locally compact). In this case, one needs to add the data of a
polarization. Physically, this is a positive energy condition. See
\FMSi.

Returning now to equation \noncomm\ and our discussion of $\CH =
L^2(S)$ for an abelian group $S$, we note that there is a natural
Heisenberg group extending  $S \times \hat S$. It is defined so that
\eqn\heishom{
( (s, \chi), z)\to z T_s M_\chi
}
defines a homomorphism into the group of  invertible operators on $L^2(S)$. This in turn forces us
to  choose the
cocycle
\eqn\cocyclechoice{
c\bigl( (s_1, \chi_1), (s_2, \chi_2) \bigr) = {1\over \chi_1(s_2)}
}
whose antisymmetrization is
\eqn\cocyclechoicep{
s\bigl( (s_1, \chi_1), (s_2, \chi_2) \bigr) = {\chi_2(s_1)\over \chi_1(s_2)} .
}

As an  example,   consider the case $S= \IR$, then we can define an
isomorphism $\IR \cong \hat \IR$ so that $\chi_{r}(s) = \exp[ i
\hbar r s ]$.    If we interpret $r\in \IR$ as a coordinate
(corresponding to the operator $e^{i r \hat x}$)  and $s\in \hat
\IR$ as a momentum (corresponding to the operator $e^{i s \hat p}$)
then the
 commutator function gives the canonical symplectic pairing on the
phase space $\IR \times \IR$. We now recognize the above theorem as
the Stone-von Neuman
theorem as usually stated in quantum mechanics texts.

In section 6 we will need the following construction of
the unique irrep of the
Heisenberg group.
Let
\eqn\ceone{
0 \rightarrow U(1) \rightarrow \tilde G \rightarrow G \rightarrow 0
}
be a Heisenberg extension of an abelian group $G$. Choose a maximal isotropic
subgroup $L\subset G$. This is a maximal subgroup of $G$ on which the restriction
$s\vert_{ L \times L }=1$. The inverse image $\tilde L \subset \tilde G$
is a maximal commutative subgroup of $\tilde G$. The Heisenberg representation is then
an induced representation of $\tilde G$
based on a character of $\tilde L$ such that $\rho(x,z) =z\rho(x)$. Note that such a
character must satisfy
\eqn\split{
\rho(x) \rho(x') = c(x,x') \rho(x+x') \qquad\qquad \forall x,x' \in L
}
(Note that $c$ need not be trivial on $L$, but it does define an
abelian extension $\tilde L$ which, by theorem 1, must be a product
$L\times U(1)$, albeit noncanonically. Different choices of $\rho$
are different choices of splitting. )  The induced representation
is, geometrically, just the space of $L^2$-sections of the
associated line bundle $\tilde G \times_{\tilde L} \IC$ defined by
$\rho$. The representation is independent of the choice of $\rho$,
and any two choices are related by an automorphism of $L$ given by
the restriction of an inner automorphism of $\tilde G$.

More concretely, our representation is the space $\CF$ of functions
$f: \tilde G \to \IC$ such that
\eqn\equivrnt{
f\left( (x,z)(x',z') \right) = \rho(x',z' ) f\left( x,z \right) \qquad \forall (x',z') \in \tilde L
}
Note that this implies $f(x,z) = z f(x,1)$, so defining $F(x):=f(x,1)$ we
can simplify the description of $\CF$ by identifying it with the
space of functions   $F: G \to \IC$ such that:
\eqn\equivrnt{ F\left( x+x' \right) = \rho(x' ) c(x,x' )^{-1} F(x)
\qquad \forall  x'  \in  L . }
The $L^2$ condition states that
\eqn\elltwo{
\int_{G/L} \vert F(x) \vert^2 dx < + \infty.
}
The group action is simply left-action of $\tilde G$ on the functions $f(x,z)$.
When written in terms of $F(x)$ the representation of $(x,z)\in \tilde G$
is:
\eqn\gpact{
(\rho(x,z)\cdot F)(y) = z  c(x,y)F(x+y) .
}

As an example, if $G = P\oplus Q$ is a vector space with $P=Q^*$ and
canonical symplectic form $\Omega$, then we can choose the cocycle
$e^{i \pi \Omega}$. Then we can take $L=P\oplus 0 $. The
equivariance condition \equivrnt\ reduces the function $F$ to a
function $\psi$ depending only on $Q$ and one easily checks that
\gpact\ reduces to the familiar translation and multiplication
operators:
\eqn\trmlt{
\eqalign{
(\rho(0\oplus q,1)\cdot \psi)(q')  & = \psi(q' + q) \cr
(\rho(p\oplus 0 ,1)\cdot \psi)(q) & = e^{2\pi i \Omega(p,q)} \psi(q) \cr}
}

\subsec{Application to   Generalized Maxwell Theory}

Let us apply the remarks of the previous section to $S = \check H^\ell(Y)$. By Poincar\'e duality
we know that the Pontrjagin dual group is $\hat S = \check H^{n-\ell}(Y)$. Thus, we may
nicely characterize the Hilbert space of the theory as the unique irrep of the
Heisenberg group
\eqn\heisgp{
\heis :=
{\rm Heis}\bigl(\check H^\ell(Y)\times \check H^{n-\ell}(Y)\bigr)
}
defined by the cocycle \cocyclechoice. Thus, by  \pairng\
\eqn\sfunct{
 s( ([\c A_1], [\c A_1^D]), ([\c A_2],[\c A_2^D]) )   = \exp\biggl[ 2\pi i \bigl(
\langle [\c A_2],[\c A_1^D] \rangle - \langle [\c A_1],[\c A_2^D] \rangle \bigr)\biggr]   .
}
  As we have mentioned, there
is a unique representation (after a choice of polarization)
 on which the central elements $(0,z)$ are
realized as multiplication by $z$.

Now let us recall our interpretation of electric  and magnetic
fluxes as the characters of the group of translation by  flat fields
of the   potential ($H^{\ell-1}(Y, \IR/\IZ)$) and of the dual
potential ($H^{n-\ell-1}(Y, \IR/\IZ)$), respectively. When lifted to
subgroups of  \heisgp\ these two groups do not commute, and hence
electric and magnetic flux sectors cannot be simultaneously
measured. We can quantify this as follows: If $\phi_f \in
H^{\ell-1}(Y,\IR/\IZ)$ let $U_{el}(\phi_f)$ be the unitary operator
on $\CH$ whose value on $\CH_e$ is $\exp\left(2\pi i \int_Y e
\phi_f\right)$, and similarly for $\phi_f^D\in
H^{n-\ell-1}(Y,\IR/\IZ)$ let $U_{mg}(\phi_f^D)$ be $\exp\left(2\pi i
\int_Y m \phi_f^D\right)$ on $\CH_m$.  Then it follows from \sfunct\
that
\eqn\uncertainty{ [U_{el}(\phi_f), U_{mg}(\phi_f^D)] = \exp 2\pi i
\int_Y \phi_f \beta(\phi_f^D) = \exp 2\pi i T\left(\beta(\phi_f),
\beta(\phi_f^D)\right) .}
On the left-hand side we have a group commutator. On the right hand
side $\beta$ is the Bockstein map of \torsioncpt, and $T$   - which
is defined by the middle equality - is the pairing
\eqn\torsionpairing{ {\rm Tors}\left( H^\ell(Y)\right) \times {\rm
Tors}\left(H^{n-\ell}(Y) \right) \rightarrow U(1) }
known as the {\it torsion pairing} or the {\it link pairing}. In
topology textbooks it is shown that this is a perfect pairing
\bredon. \foot{The central fact is that  $H^\ell(Y,\IZ) \otimes
H^{n-\ell-1}(Y, \IR/\IZ) \to \IR/\IZ$ is a perfect pairing. Now
consider \torssbup\ and \torsioncpt (with $\ell \to n-\ell)$.
Observe that $\bar H^{n-\ell}(Y,\IR)/\bar H^{n-\ell}(Y,\IZ)\cong
\CW^{n-\ell-1}$ and since $\bar H^\ell(Y)\times \bar H^{n-\ell-1}(Y)
\to \IZ $ is perfect we learn that $\CW^{n-\ell-1} \times \bar
H^\ell(Y) \to \IR/\IZ$ is perfect. Therefore, by \twosespp\ the
torsion pairing is perfect.} Note especially that there is no factor
of $\hbar$ on the right-hand-side of this uncertainty relation!

Despite the uncertainty relation \uncertainty\ there is a useful
simultaneous decomposition of the Hilbert space into electric and
magnetic sectors in the following sense. It follows from the
discussion around \simpleprod\ that the cocycle defining the
Heisenberg pairing in \heisgp\ is equal to one on topologically
trivial flat fields. Therefore we {\it can} simultaneously
diagonalize the translations by $\CW^{\ell-1}(Y)\times
\CW^{n-\ell-1}(Y)$. The character group is $\bar H^{n-\ell}(Y)
\times \bar H^{\ell}(Y)$, and hence we  get the decomposition
\eqn\decomflxsc{
\CH = \oplus_{\bar e, \bar m} \CH_{\bar e, \bar m}
}
in harmony with \ccrs.

The theory of Heisenberg representations also allows us to
understand more deeply  the summand $\CH_{\bar e, \bar m}$. Let
$\heis_0$ denote the connected component of the identity in $\heis$.
We have
\eqn\conncompt{
0\rightarrow \heis_0 \rightarrow \heis \rightarrow H^\ell(Y,\IZ) \times H^{n-\ell}(Y,\IZ) \rightarrow 0
}
Now recall that each connected component of $\c H^\ell(Y)$ is a principal homogeneous space for
$\Omega^{\ell-1}(Y)/\Omega^{\ell-1}_{\IZ}(Y)$, which in turn is a torus bundle over a vector space
as in \toptrvfib.
%
%\eqn\torbundle{
%0 \rightarrow Z^{\ell-1}(Y)/Z^{\ell-1}_{\IZ}(Y) \rightarrow \Omega^{\ell-1}(Y)/\Omega^{\ell-1}_{\IZ}(Y)
%\rightarrow \Omega^{\ell-1}(Y)/Z^{\ell-1}(Y)\rightarrow 0
%}
%
%and of course $Z^{p-1}(Y)/Z^{p-1}_{\IZ}(Y) = H^{p-1}(Y,\IZ)\otimes \IR/\IZ$, is the connected
%component of $H^{p-1}(Y,  \IR/\IZ)$.
%
Thus we have
\eqn\hesnought{
0 \rightarrow \CW^{\ell-1}(Y)  \times \CW^{n-\ell-1}(Y) \rightarrow \heis_0
\rightarrow  \heis_0'  \rightarrow 0
}
where $\heis_0'$ is the Heisenberg extension of the vector space:
\eqn\heisvs{ \Omega^{\ell-1}(Y)/\Omega^{\ell-1}_d(Y) \times
\Omega^{n-\ell-1}(Y)/\Omega^{n-\ell-1}_d(Y) . } The Heisenberg group
$\heis_0'$   is the usual infinite dimensional Lie group whose Lie
algebra is  formed by the commutators $[A(x), E(x')]=i \hbar
\delta(x-x')$ familiar from standard discussions of the quantization
of the free electromagnetic field. By the Stone-von Neumann theorem
there is a unique irreducible representation $\CH_{\bar 0, \bar 0} $
of $\heis_0'$. Meanwhile, as we have said,  the tori
$H^{\ell-1}(Y,\IZ)\otimes \IR/\IZ $ and $ H^{n-\ell-1}(Y,\IZ)\otimes
\IR/\IZ $ have zero pairing under the Pontrjagin pairing and hence
by Theorem 2 of $\S{3.1}$ the irreducible representations of
$\heis_0$ are labelled by $(\bar e,\bar m) \in  \bar
H^{n-\ell}(Y;\IZ)\times \bar H^\ell(Y;\IZ)$. These irreducible
representations are the summands in \decomflxsc. To be quite
explicit, $\CH_{\bar e, \bar m}$ consists of ``$L^2$ functions'' on
$\Omega^{\ell-1}/\Omega^{\ell-1}_{\IZ}$ with representation
\foot{This formula fixes some sign ambiguities we have hitherto left
unspecified.}
\eqn\explrep{
\eqalign{
(\rho([B]\cdot \psi)([A]) & = \psi([A+B]) \qquad\qquad\qquad\qquad\quad  [B]\in \Omega^{\ell-1}/\Omega^{\ell-1}_{\IZ} \cr
(\rho([B^D]\cdot \psi)([A]) & = e^{2\pi i \int_Y B^D (\bar m + dA)}\psi([A]) \qquad\qquad [B^D]\in \Omega^{n-\ell-1}/\Omega^{n-\ell-1}_{\IZ} \cr}
}
which satisfy, in addition
\eqn\additns{
\psi([A] + [a_f] ) = e^{2\pi i\int_Y \bar e a_f } \psi([A]) \qquad\qquad [a_f] \in \Omega^{\ell-1}_d/\Omega^{\ell-1}_{\IZ}
}

Equation \decomflxsc\ should not be construed to mean that there is
no information in the Hilbert space about the torsion fluxes.
Indeed, since the pairing \torsionpairing\ is perfect  we may
introduce the  {\it  finite}  Heisenberg group \eqn\finheis{ {\rm
Heis}({\rm Tors}(H^\ell(Y))\times {\rm Tors}(H^{n-\ell}(Y) ) ). } We
can lift elements of ${\rm Tors} (H^\ell(Y))$ and ${\rm
Tors}(H^{n-\ell}(Y) ) $ to flat fields, and then lift these to  the
group $\heis$ in \heisgp. When acting on $\CH_{\bar 0, \bar 0} $ the
choice of lifting does not matter, and we conclude that the summand
$\CH_{\bar 0, \bar 0} $    is naturally a representation of this
finite Heisenberg group.
%
%%The other summands are also representations of this finite Heisenberg group, but the representation
%depends on the choices above.
%

\expl{}

We can give a simple example already in the ``physical'' case of $n=3, \ell=2$,
i.e. $3+1$-dimensional Maxwell theory.  Suppose $Y= S^3/\IZ_n$ is any
3-dimensional Lens space. In this case $H^2(Y, \IZ)=\IZ_n$.
The Heisenberg group ${\rm Heis}({\rm Tors}(H^2(Y))\times {\rm Tors}(H^{2}(Y) ) ) $ is
the standard extension of $\IZ_n \times \IZ_n$ by $U(1)$. In this case we can measure magnetic
flux or electric flux, but never both simultaneously.
%
%\foot{In this case we may directly check that the torsion pairing is a perfect pairing
%for   $n$ prime as follows.
%Let $\alpha$ be the  generator of $H^1(Y,\IZ_n) \cong \IZ_n$. Then applying the
%Bockstein operator, $\beta(\alpha)$ is the generator
%of $H^2(Y,\IZ_n)$ and is the reduction modulo $n$ of a generator of $H^2(Y,\IZ)$. By Poincar\'e duality
%$\langle \alpha \beta(\alpha), [Y]\rangle$ is a generator of $H^3(Y;\IZ_n) \cong \IZ_n$.}
%

\rmk{}

\item{1.} {\it A Gedankenexperiment}.
It has occasionally been suggested that the topology of our
(spatial) universe is not $\IR^3$ but rather a 3-manifold of
nontrivial topology. Of the various candidates that have been
considered the simplest are the Lens spaces. See, e.g. \CornishDB.
This raises the amusing question of whether our universe is in a
state of definite electric or magnetic flux. In such a universe the
phenomenon we are discussing could in principle be demonstrated
experimentally as follows. One first adiabatically  transports an
electron around a nontrivial loop, and compares the phase of the
wavefunction before and after transport. This measures the magnetic
flux via  the Aharonov-Bohm phase. The universe is then in a
definite state of magnetic flux. One then does the same thing with a
magnetic monopole. If one then repeats the experiment with the
electron then all possible magnetic fluxes will be measured with
equal probability.  It is, of course, more interesting to find  more
realistic manifestations of the basic phenomenon. One proposal for
how this could be done is in \KMW.

\item{2.} {\it Relation to AdS/CFT}.  A closely related observation to what we have explained,
discovered in the special case of Maxwell theory on a Lens space, was made
by Gukov, Rangamani, and Witten in   \GukovKN.   The
authors of \GukovKN\ pointed out that it is nontrivial to define an 't Hooft loop
for $U(1)$ gauge theory in the presence of nontrivial topology.
%
%They give a
%definition  and argue
%  the  Wilson and 't Hooft loops do not simultaneously commute. We will see in the
%next remark that this is not quite correct. Nevertheless, the authors
In string theory, the effect we have discussed implies that   the
number of absorbed $F1$ and $D1$ strings in a $D3$ brane cannot be
simultaneously measured, as required by the AdS/CFT correspondence
\GukovKN. Recently, generalizations of this remark have been
investigated in \BurringtonUU\BurringtonAW\BurringtonPU.

\item{3.} {\it 't Hooft loops and Wilson loops}. The present formalism is well-suited to
discussing 't Hooft and Wilson loops taking into account subtleties of torsion.
Let $\Sigma\in Z_{\ell-1}(Y)$ be a closed cycle. Then we can define an operator $W(\Sigma)$
on the representation of \heisgp\ by realizing it as $\CH = L^2(\c H^\ell(Y))$ and
simply using the evaluation map:
\eqn\wilsonloop{
(W(\Sigma)\psi)([\c A]) := \exp (2\pi i \int_{\Sigma} [\c A] ) \psi([\c A])
}
The 't Hooft loop $W_d(\Sigma_d)$, associated to a cycle
$\Sigma_d\in Z_{n-\ell-1}(Y)$ is defined exactly the same way if we
realize the Hilbert space as $L^2(\c H^{n-\ell}(Y))$. The two
representations are related, of course, by a formal Fourier
transform. Using this one finds the action of the 't Hooft loop
operator in the first realization is
\eqn\thooftloop{ (W_d(\Sigma_d)\psi)([\c A]) :=   \psi([\c A]+ \c
\delta(\Sigma_d) ) }
where $\c \delta(\Sigma_d) $ was defined in Remark 2 of $\S{2.5}$.
(This generalizes the original description of 't Hooft using
singular gauge transformations.) Assuming that $\Sigma$ and
$\Sigma_d$ do not intersect the commutation relations are
\eqn\thwls{
W_d(\Sigma_d) W(\Sigma) = e^{2\pi i \langle \c \delta(\Sigma), \c \delta(\Sigma_d) \rangle} W(\Sigma) W_d(\Sigma_d)
}
If $\Sigma$ and $\Sigma_d$ are disjoint then we can choose nonintersecting
tubular neighborhoods around them, and find representative cocycles for
$\c \delta(\Sigma)$ and $\c \delta(\Sigma_d)$ with support in these neighborhoods.
It follows that the pairing is in fact {\it zero} and the 't Hooft and Wilson
loop operators in fact always commute in $U(1)$ gauge theory:
\eqn\thwls{
W_d(\Sigma_d) W(\Sigma) =   W(\Sigma) W_d(\Sigma_d) .
}
(There is a small discrepancy with \GukovKN. The discussion in that
paper did not take account of the fact that acting by an 't Hooft
operator does not commute with restricting the domain of the
wavefunction to the flat fields.) The unexponentiated Wilson and 't
Hooft operators were considered in \AshtekarRG\ and shown not to
commute. But their commutator is given by a Gauss linking number so
that the exponentiated operators do commute.

\item{4.} {\it Vertex operators}. In the special case of $\c H^1(S^1)$ the Hilbert space
$L^2(\c H^1(S^1))$ is familiar from the quantization of a periodic
scalar field. In this case the ``Wilson operator'' corresponding to
$p\in S^1$ is a vertex operator changing the momentum by one unit,
while the ``'t Hooft operator'' at $p\in S^1$ is a vertex operator
changing winding number by one unit. The theory of 't Hooft and
Wilson loops can be viewed as a generalization of the vertex
operator calculus.

\item{5.}
Quite generally, if  $G' \subset G$ is a normal subgroup then $G'' =
G/G'$ acts on $Irrep(G')$, the set of isomorphism classes of
irreducible representations of $G'$. (Lift elements of $G''$ to $G$.
Conjugation by these elements induces an automorphism of $G'$ which
acts on the representations of $G'$.) Apply this to $G'= \heis_0$
and $Irrep(G') = \bar H^{n-\ell}(Y;\IZ)\times \bar H^{\ell}(Y;\IZ)$.
The action of $(e,m)\in G''$ on $(\bar e_1, \bar m_1)\in Irrep(G')$
is $(\bar e_1, \bar m_1) \to (\bar e_1 + (-1)^{n-\ell} 2 \bar e,
\bar m_1 - 2 \bar m)$. Note that the torsion information in $(e,m)$
drops out.

\item{6.} {\it Refinements of the grading}.
One can give a finer grading than \decomflxsc. One need only choose a maximal
commutative subgroup of the Heisenberg group. Thus, for example, choosing
$H^{\ell-1}(Y)\otimes \IR/\IZ \times H^{n-\ell-1}(Y,\IR/\IZ)$ one can grade the
Hilbert space by pairs $(\bar e, m)$.  As we explain in the following remark,
 this grading is natural if we formulate
the Hilbert space in terms of wavefunctions $\psi([\c A])$. Nevertheless,
there is no natural
``largest abelian subgroup,'' so the natural grading is \decomflxsc.

\item{6.} {\it The quantum Gauss law}. The quantization of electric flux
can be understood from another point of view. We work in a definite
component of magnetic flux $m\in H^\ell(Y)$, and represent the gauge field by $A \in \Omega^{\ell-1}(Y)$,
for example, by means of a basepoint as in \basepoint.
Let us denote the generator of translations $A \to A+ \omega $ by
$\CG(\omega) = 2\pi i \int \omega \hat \CP $, where $\hat \CP$ is an operator
valued in $\Omega^{n-\ell}(Y)$.
 The quantum mechanical Gauss law, which states that the
physical  wavefunction must be gauge invariant,  is
\eqn\quantgl{
\exp\biggl(2\pi i \int_Y \omega \hat \CP \biggr) \psi = \psi \qquad\qquad \forall \omega \in \Omega^{\ell-1}_{\IZ}(Y)
}
The spectrum of the operator $\hat \CP$ is therefore in
$\Omega^{n-\ell}_{\IZ}(Y)$.  In particular  the Hilbert space is
graded by $( \bar e,m) \in \bar H^{n-\ell}(Y,\IZ) \times
H^{\ell}(Y,\IZ)$. Heuristically, we may think of $\hat \CP $ as the
fieldstrength of the electric flux $\c \CE$ defined in \defelst.

\item{7.} {\it The semiclassical limit.} In the classical
theory the   equation of motion is
$$
d\bigl( 2\pi R^2 * F \bigr) =0
$$
where $*$ is Hodge star on $M$.  It follows that the  spatial
components $(*F)_Y $ are closed, and the equation of motion shows
that ${d \over dt} (*F)_Y   $ is exact. Thus $[(*F)_Y  ] \in
H^{n-\ell}_{DR}(Y)$ is a conserved quantity, but it is not quantized
in the classical theory. Indeed, $[*F]$ obviously cannot be
quantized since it varies continuously with the metric. However, in
the   Legendre transform to the Hamiltonian formalism the momentum
is $\Pi =   2\pi R^2 (*F)_Y\in \Omega^{n-\ell}(Y)$.
 Semiclassically,  $  - i \hbar {\delta \over \delta A}\to \Pi $ and therefore in
the semiclassical limit there is a correspondence between the
generator of translations with $*F$, or, more precisely, as $\hbar
\to 0$,
\eqn\semiclassical{
\hat \CP \to {R^2 \over  \hbar} (*F)\vert_Y .
}
This means that, semiclassically,  the quanta of $[(*F)_Y]$ are
proportional to  $\hbar$ (and depend on the metric), but have a
discrete spectrum. It is only in this sense that $[*F]$ is quantized
in the semiclassical theory. There is an uncertainty relation
between the topologically trivial flat fields and the cohomology
class $[(*F)_Y]$, which can be made arbitrarily small for $\hbar \to
0$. However, there is no tunable $\hbar$ for the pairing between the
topologically nontrivial flat fields. Put differently, the basic
pairing \sfunct\ which we use to define the Heisenberg group does
not admit continuous deformations, unlike the symplectic form on a
vector space.  Another aspect of the semiclassical limit is that it
can be equivalent to the large distance limit. Note that
 a constant conformal scaling of the background metric $g \to \Omega^2 g$
is equivalent to a scaling $\hbar \to \hbar \Omega^{2\ell-n}$ so for
$\ell < \half n$ scaling the metric up is equivalent to the
semiclassical limit. Therefore, the noncommutativity of electric and
magnetic flux sectors is a quantum effect which cannot be suppressed
by taking the limit of large distances: It is a ``macroscopic''
quantum effect. A final aspect of the semiclassical limit that
deserves attention is that the limit  is best taken with coherent
states. These will have Gaussian wavefunctions in $A$ and hence will
be linear combinations of plane wave states of definite electric
flux.

\newsec{The self-dual field}

In the previous section we defined  the Hilbert space of generalized
Maxwell theory in terms of the unique irreducible representation of
\heisgp. This viewpoint has the elegant property that it is a
manifestly electric-magnetic dual formulation. This viewpoint also
allows us to give a very crisp formulation of the Hilbert space of a
self-dual field. For other descriptions of the Hamiltonian
quantization of the self-dual field see \BekaertYP\HenningsonWH\
and references therein. Our discussion differs significantly from
previous treatments. In particular, in part because the phase space
is disconnected, the standard principles of quantization are
insufficient to give a complete definition of the quantum theory. We
thus are forced to invent some new rules. Further justification that
our definition fits   into a coherent physical picture compatible
with known results on the partition function of self-dual fields is
left for the future.

Suppose $\dim Y =2p-1$.  We can then consider imposing the
(anti)-self-dual constraint $F=\pm
*F$ on the $p$-form fieldstrength, where $*$ is the Hodge operator
in the $2p$-dimensional spacetime. \foot{In Minkowskian signature we
must have $p=1\mod 2$ so that $*^2=1$. In physical applications with
Minkowskian signature one often imposes self-duality constraints
where $p=0\mod 2$ and the fieldstrengths are valued in a symplectic
vector space. After all, one could always perform a Kaluza-Klein
reduction from a theory in $2p_1$ dimensions to a theory in  $2p_2$
dimensions where $p_1> p_2$. If $p_1$ is odd and $p_2$ is even then
we will encounter self-dual theories of this type. This is in fact
what happens in certain supergravities. For example, if the type IIB
string is reduced on a Calabi-Yau manifold $\CZ$ then the
4-dimensional theory has a self-dual abelian gauge theory with $F$
valued in $H^3(\CZ;\IR)$. For simplicity, we will simply take $p$
odd in what follows.  A general framework for a self-dual field is
described in \FMSi.  }

Two heuristic arguments suggest that the quantization of the (anti)-self-dual field should be
based on the definition of a Heisenberg extension of a {\it single copy} of $\c H^p(Y)$.
First, the  two factors
\eqn\nsd{
\c H^{\ell}(Y) \times \c H^{n-\ell}(Y)
}
from which we form our Heisenberg group for the nonself-dual field
represent translations in the field and the dual field,
respectively.  When $n=2p, \ell=p$ we might expect the translations
by self-dual fields to be the diagonal subgroup, consisting of
elements lifting   $(\chi,\chi)$. Similarly, we might expect the
translations by the anti-self-dual fields to be translation by the
subgroup defined by $(\chi, \overline{\chi})$, where the bar denotes
complex conjugation. Heuristically, the  wavefunction of a self-dual
field should be ``invariant'' (perhaps up to a phase) under
translation by an anti-self-dual field. Roughly speaking, the
quotient of \nsd\ for $n=2p, \ell=p$ by the anti-diagonal leaves a
{\it single} copy of $\c H^p(Y)$. In $\S{4.1}$ we will see this
intuition is only roughly correct, but it gives the right idea.

A second line of thought which leads to the same idea is to note that,
due to the self-duality equations, the phase space of a self-dual field,
that is, the space of gauge-inequivalent solutions of the equations of
motion may be identified with $\c H^p(Y)$.
\foot{This is to be contrasted with the nonself-dual field, where the phase space
is $T^*\c H^\ell(Y)$, and not $\c H^\ell(Y) \times \c H^{n-\ell}(Y)$. The two spaces are very
similar but are definitely different. Thus, the phase space of the non-self-dual field
does {\it not} decompose as a product of phase-spaces for the self-dual and
anti-self-dual fields.  }
 This manifold is a Poisson
manifold and we wish to quantize it.

With this motivation, let us try  to define a Heisenberg extension of $\c H^p(Y)$
in the case where $p=2k+1$ is odd (so $\dim M = 4k+2$). The natural guess is that
we should take the commutator function to be the pairing of Poincar\'e duality:
\eqn\selfdualess{
s_{\rm trial}([\c A_1], [\c A_2])
%~~ {\buildrel ? \over =} ~~
= \exp 2\pi i \langle [\c A_1], [\c A_2] \rangle .
}
In terms of topologically trivial fields, we may write:
\eqn\selfdualess{
s_{\rm trial}([A_1], [A_2])
%~~ {\buildrel ? \over =} ~~
=\exp \left( 2\pi i \int_Y A_1 d A_2 \right)  .
}
Because the   product \product\ is graded commutative it follows that
$s_{\rm trial}$ is skew, and Poincar\'e duality implies that it is nondegenerate.
However, as we saw in $\S{3.1}$ a commutator function must also be
  alternating, i.e.
$s([\c A], [\c A])=1$ for all $[\c A]$. Because the product is
graded $[\c A] \star [\c A] = - [\c A] \star [\c A]$ and this only suffices
to show that $[\c A]\star [\c A]$ is two-torsion. In fact, it
 turns out that $s_{\rm trial}$
is {\it not} alternating! Rather, one can show \Gomi\ that
\eqn\alterncond{ s_{\rm trial}([\c A], [\c A]) = (-1)^{\int_Y
\nu_{2k} \smile a } }
where $\nu_{2k}$ is a characteristic class of $Y$, valued in
$H^{2k}(Y,\IZ_2)$, known as the Wu class. In general the Wu classes
can be written as polynomial expressions in the Stiefel-Whitney
classes $w_i(TY)$. For  an oriented manifold the first few Wu
classes are
\eqn\wuclasses{ \eqalign{ \nu_0 & = 1 \cr \nu_2 & = w_2 \cr \nu_4 &
= w_4 + w_2^2.\cr} }

In order to produce a commutator function we introduce a {\it $\IZ_2$-grading},
that is, a homomorphism $\epsilon: \c H^{p}(Y) \to \IZ_2$. We define
\eqn\zeetwo{ \epsilon([\c A]) = \cases{ 0 & $\int a\smile
\nu_{2k}=0\,  \mod \, 2 $ \cr
 1 &  $\int a\smile  \nu_{2k}=1\,  \mod \, 2 $ \cr}
}
and then define the commutator function
\eqn\truecommt{
s ([\c A], [\c A']) := \exp 2\pi i\bigl[  \langle [\c A], [\c A'] \rangle - \half \epsilon([\c A]) \epsilon([\c A']) \bigr] .
}
One may check this is skew, alternating, and nondegenerate.
In this way we define a   Heisenberg group   extension of $\c H^p(Y)$,
denoted $\heis_{SD}$.
Similarly, the anti-self-dual group $\heis_{ASD}$ is obtained by replacing $s \to s^{-1}$.
Note that the Heisenberg group we have obtained is in fact $\IZ_2$-graded, in the
sense that there is a homomorphism $\epsilon: \heis_{SD} \to \IZ_2$.

We can now invoke the essential uniqueness of the irreducible
$\IZ_2$-graded representation of the $\IZ_2$-graded Heisenberg
group. \foot{We say ``essential'' because one must choose a
polarization. The theorems of $\S{3.1}$ are easily extended to the
$\IZ_2$-graded case. } By this we mean that the representation space
$\CH$ is $\IZ_2$ graded and that the representation is compatible
with the $\IZ_2$ grading on ${\rm End}(\CH)$. The groups
$\heis_{SD}, \heis_{ASD}$ have unique irreducible representations
$\CH_{SD}$ and $\CH_{ASD}$, respectively. We will take these as the
{\it definitions} of the Hilbert spaces of the self-dual and
anti-self-dual fields.

\expl{} The first example, $p=1$ is already somewhat nontrivial.
This is the  case of   the self-dual scalar field. $L^2(\c H^1)$ is
the standard Hilbert space of the non-selfdual scalar field, and
represents $\heis(\c H^1 \times \c H^1)$. On the other hand,
$\heis(\c H^1)$ is nothing other than a central extension of the
loop group $L\IT$,  which is studied as an ungraded central
extension in \SegalAP\pressleysegal. In this case the Wu class is
$\nu_0=1$ and hence, referring
 to the decomposition \scalarii,
we have the degree
\eqn\scalardeg{
\epsilon(\varphi ) = \cases{ 0 & $w=0~ \mod 2 $ \cr
 1 &  $w=1 ~ \mod 2 $ \cr}
}
It follows from \corrscalarpair\ that
\eqn\psnegns{
s(\varphi^1, \varphi^2) =
\exp 2\pi i \biggl( \half \int_0^1 \bigl(  \phi^1 {d \phi^2\over ds} -  \phi^2 {d \phi^1\over ds}  \bigr) +
\half (\phi^1(0) w^2 - \phi^2(0) w^1) \biggr)
}
Here we have used
\eqn\scalarpairi{
  \int_0^1 \phi^1 {d \phi^2\over ds} ds  - w^1 \phi^2(0)- \half w^1 w^2  =
 \half \int_0^1 \bigl(  \phi^1 {d \phi^2\over ds} -  \phi^2 {d \phi^1\over ds}  \bigr) +
\half (\phi^1(0) w^2 - \phi^2(0) w^1).
}
 Substituting
the explicit decomposition \scalarii\ we have
\eqn\scalarpairii{ s( \varphi^1, \varphi^2 )  = \phi_0^1 w^2 -
\phi_0^2 w^1 + 2\pi i \sum_{n\not=0} {1\over n} \phi_n^1 \bar
\phi_n^2 .}
Thus we recognize the level one central extension of the loop group.
Note that  the winding number and
the zeromode are paired by the standard Pontrjagin dual pairing of $\IT$ with $\IZ$.

The   representation of
the oscillator modes is   straightforward.
As for the  zero-modes the  Heisenberg group extension of $\IT \times \IZ$
has the representation $L^2(\IT) \cong L^2(\IZ)$.  Note that, according to \scalarpairi\ the operators
$e^{2\pi i w \phi(\theta)}$ {\it anticommute} for odd winding number.
This is, of course, the correct answer, since the $1+1$ dimensional
self-dual scalar field is equivalent to a free fermion. The natural
$\IZ$-grading on $L^2(\IZ)$ is the grading by fermion number.

\rmk{}

\item{1.}
As this example shows, in forming the ``level one'' Heisenberg group
and its representations one must introduce extra structures. While
the $\IZ_2$-graded irrep is unique up to isomorphism, there are
choices in defining an explicit representation, and these choices
affect, for example, the Hamiltonian. In the case $p=1$ one must
introduce a spin structure. The higher dimensional case requires the
introduction of a generalization of a spin structure, in keeping
with the discussion in \WittenHC\HopkinsRD\BelovJD. We hope to
return to this rather subtle issue elsewhere.

\item{2.} In a recent preprint \gomiII\ central extensions of $\c H^{2k+1}(Y)$ for
$\dim Y = 4k+2$ and their representations were independently considered.
These representations are at ``level two'' and involve
slightly degenerate Heisenberg groups, and should be distinguished from
the ``level one'' representations needed for the self-dual field.

\item{3.} Note in particular that the above discussion shows that
vertex operators in self-dual theories in higher dimensions will
sometimes be {\it fermionic}. This is natural since the
compactification   of a self-dual theory from $4k+2$ dimensions to
$2$ dimensions along a $4k$-manifold will be a chiral conformal
field theory. The parity of vertex operators will be related to the
intersection form on the $4k$-manifold. As an example consider the
self-dual field in six dimensions on $Y=S^1 \times K$, where $K$ is
a compact $4$-manifold with $b^1=b^3=0$ and no torsion in its
cohomology. Then the flat fields $H^{2}(Y,\IR/\IZ)$ lift to a
commutative subgroup of $\heis(\c H^3(Y))$ and hence we can
decompose $\CH = \oplus_{x\in H^2(K,\IZ)} \CH_x$ where $\CH_x$ is
the sector with eigenvalue $x d\theta  $. The decomposition into
even and odd sectors is
\eqn\evensoddsector{ \CH^+ = \oplus_{x^2=0\, \mod\,2} \CH_x
\qquad\qquad \CH^- = \oplus_{x^2=1\, \mod\,2} \CH_x }
This example is related to the discussion in   \WittenVG.

\subsec{Relating the self-dual field to the non-self-dual field in $(4k+2)$-dimensions }

When working with fieldstrengths, or with   oscillator modes,
 it is rather straightforward to decompose the nonself-dual field into a
theory of self-dual and anti-self-dual fields. However, when one takes into
account nontrivial topology the situation becomes quite subtle.
The formulation of the selfdual field  is already
rather delicate (but well-known) for the case   of the periodic scalar field
($k=0$).
We will now show how chiral splitting fits into our formalism, and how that
may be used to approach the theory of the self-dual field.

Quite generally, if $S$ is a locally compact  abelian group, with a
nondegenerate Heisenberg extension $\tilde S$ then there is an
$\tilde S \times \tilde S$ equivariant isomorphism
\eqn\mumfthm{
L^2(S) \cong \CH^* \otimes \CH
}
where $\CH$ is the Heisenberg representation of $\tilde S$. (See \Mumford\ Proposition 1.6
for a clear explanation of this.) This is {\it not} the chiral splitting of
the nonself-dual Hilbert space but it is related to it.
In order to explain   chiral splitting we must generalize \mumfthm\ somewhat.
Let us identify $S$   with its Pontrjagin
dual $\hat S$ by a perfect pairing $\langle \cdot, \cdot \rangle$. Then,
on the irrep of $\heis(S\times S)$, realized as $L^2(S)$,
we may define translation and multiplication operators
\eqn\transmult{
\eqalign{
(T_a\psi)(b):= \psi(b+a) \cr
(M_a\psi)(b):= e^{2\pi i \langle b,a\rangle } \psi(b).\cr}
}
For any pair of integers $k,l\in \IZ$ the  operators
\eqn\rhokl{
\rho_{k,l}(a):= T_{ka} M_{la }
}
 define a ``level'' $2kl$ representation of $\heis(S)$. That is
$\rho_{k,l}$ represent a central extension of $S$ with commutator function
\eqn\klcomm{
s_{k,l}(a,b) = e^{2\pi i 2kl\langle a, b \rangle } .
}
Let us denote by $\CG_{k,l}$ the group of operators on $L^2(S)$ generated by $\rho_{k,l}$.
Then the groups $\CG_{k,l}$ and $\CG_{k',l'}$ are in each others
commutant  if $kl' + k' l=0$. In particular $\CG_{k,l}$ and
$\CG_{k,-l}$ commute. The Hilbert space $\CH$ may be decomposed into
irreducible representations of $\CG_{k,l} \times \CG_{k,-l}$.

Let us now consider the case $S=\c H^{2k+1}(Y)$. \foot{Once again,
this group is not locally compact, but we believe the following
considerations can be made rigorous.} The groups $\CG_{k,l}$ are
groups of vertex operators forming a level $2kl$ extension of $\c
H^{2k+1}(Y)$. In general these vertex operators have {\it nothing}
to do with the self-dual and anti-selfdual degrees of freedom.
However, in special circumstances they are related to self-dual and
anti-selfdual degrees of freedom.

In order to explain the chiral splitting it is necessary to consider
the {\it dynamics} of the fields. In general the nonself-dual field
with action \action\ has a  phase space $T^*\c H^\ell(Y)$ with its
canonical symplectic structure. We may identify a cotangent vector
with an element of $\Omega^{n-\ell}_d(Y)$ and we will denote it by
$\Pi$. Assuming a product metric on $M=Y\times \IR$, the Hamiltonian
is (Hodge $*$ here and below refers to the metric on $Y$):
\eqn\hamilton{
H = \int_Y {1\over 4 \pi R^2} \Pi * \Pi + \pi R^2 F * F .
}
The equations of motion are
\eqn\eogms{
\eqalign{
{d\over dt}  [\c A] & = {1\over 2\pi R^2} * \Pi \cr
* {d\over dt}  \Pi & = - 2\pi R^2 d^\dagger F . \cr}
}
(In the first equation ${d\over dt}  [\c A]$ denotes the tangent vector to the
curve.)
Upon quantization these become operator equations in the Heisenberg picture.

When $\ell = 2k+1, n= 4k+2$ the wave operator splits
as
\eqn\waveop{
\p_t^2 + d^\dagger d = (\p_t - *d)( \p_t + * d)
}
Note that $*d: \Omega^{2k}(Y) \to \Omega^{2k}(Y)$. Moreover, this
operator  has a purely imaginary spectrum since $d^\dagger = - *d*$.
We can decompose tangent vectors  $\delta A$ to solutions of the
equations of motion into self-dual (``left-moving'') and
anti-self-dual (``right-moving'') parts $\delta A = \delta A_L +
\delta A_R$ where
\eqn\leftright{
\eqalign{
(\p_t - *d)  \delta A_L & = 0 \cr
(\p_t + *d )  \delta A_R & =  0 \cr}
}
From the Hamiltonian equations of motion it follows that
\eqn\pidar{
\Pi = 2\pi R^2 * {d\over dt}  A = 2\pi R^2 ( dA_L - d A_R).
}

Let us now consider the subgroups of operators $\CG_{q,p}$
acting on $L^2(\c H^{2k+1}(Y))$. In particular, let us
focus on  the vertex operators $\rho_{q,p}([A_0])$
where $[A_0]$ is topologically trivial.
It is straightforward  to show that
\foot{In fact, we conjecture that by broadening the
geometrical interpretation of $\c A$ we can define operators
$\c A_L, \c A_R$ such that the general vertex operator is
$$
\rho_{q,p}([\c A_0]) = \exp 2\pi i \biggl[
  \bigl( \hbar^{-1}R^2 q-p\bigr) \langle \c A_0, \c  A_L \rangle
- \bigl( \hbar^{-1}R^2 q+p\bigr) \langle \c A_0, \c  A_R \rangle \biggr]
$$ }
\eqn\rhopqa{
\rho_{q,p}([A_0]) = \exp 2\pi i \Biggl[
\int_Y A_0 \biggl\{ \bigl( \hbar^{-1}R^2 q-p\bigr) d A_L
- \bigl( \hbar^{-1}R^2 q+p\bigr) d A_R \biggr\} \Biggr]
}
Thus, when
\eqn\rattior{
R^2 = {p\over q} \hbar
}
 the vertex operators in the
level $2pq$ extension $\CG_{q,p}$ are right-moving, or
anti-chiral. Similarly, $\CG_{q,-p}$ are left-moving, or chiral.
Moreover, $\CG_{q,p}\times \CG_{q,-p}$ is of finite
index in $\heis(\c H^{2k+1}(Y) \times \c H^{2k+1}(Y))$.
Therefore, the Hilbert space of the nonself-dual field will
be decomposed into a finite sum of irreducible representations
of this subgroup.

The center of   $\CG_{q,p}$  is $U(1) \times T_{2pq}$ where
$T_{2pq}$ is the set of $2pq$-torsion points in $H^{2k}(Y,\IR/\IZ)$.
We wish to represent the $U(1)$ canonically. Thus, by theorem 2 of
$\S{3.1}$ the group of irreps is the finite group $\hat T_{2pq}$.
Accordingly, the chiral decomposition of the Hilbert space takes the
form
\eqn\sumbreps{
\CH = \oplus_{\alpha,\beta}  N_{\alpha\beta}\CH_{\alpha}^* \otimes \CH_{\beta}
}
where $\alpha, \beta$ run over $\hat T_{2pq}$, and $N_{\alpha \beta}$ accounts
for degeneracies.
\foot{From the example of the the scalar field we know that $N_{\alpha\beta}$
depends on the factorization of the integer $K=2pq$. }

The Hamiltonian evolution likewise splits.
Let $\omega^i$, $i=1,\dots, b_{2k+1}(Y)$  be a basis of $\CH^{2k+1}(Y)$
with integral periods.
Then we split the fields into zeromodes and oscillator modes
as
\eqn\oscill{
\eqalign{
\Pi & = \omega^i p_i + \Pi_{osc} \cr
F & = \omega^i w_i + F_{osc} \cr}
}
where $w_i$ are integral
and $p_i = 2\pi \hbar n_i$, with $n_i$ integral. Then the Hamiltonian is
$H= H_L + H_R $ where
\eqn\splithim{ \eqalign{ H_L & = {\pi \over 2} R^2 \left( w_i +
{\hbar \over R^2} n_i \right) \left( w_j + {\hbar \over R^2} n_j
\right)g^{ij} + H_{L,osc} \cr H_R & = {\pi \over 2} R^2 \left( w_i -
{\hbar \over R^2} n_i \right) \left( w_j - {\hbar \over R^2} n_j
\right)g^{ij} + H_{R,osc} \cr} }
where $g^{ij} = \int_Y \omega^i * \omega^j$.

For completeness we give the expressions for the Hamiltonians in
oscillator modes. Let  $\{ \lambda_n\} $ be  the spectrum of
$d^\dagger d$ on $\Im d^\dagger \subset \Omega^{2k}(Y)$. Each
eigenvalue is (generically) two-fold degenerate. We can introduce
complex eigenmodes diagonalizing $*d$,
\eqn\eigenmodes{
*d e_n^+ = i \sqrt{\lambda_n} e_n^+ \qquad\qquad *d e_n^- = -i
\sqrt{\lambda_n} e_n^- }
with $\int_Y e_n^+ * e_m^- = \lambda_n^{-1/2} \delta_{n,m}$ and the
other overlap integrals vanishing. Then the usual oscillator
expansions are:
\eqn\alfet{ \eqalign{ A_{L,osc} &= {1\over \sqrt{4\pi R^2}}
\sum_{n=1}^\infty e^{i\sqrt{\lambda_n}t} e_n^+ a_n +
e^{-i\sqrt{\lambda_n}t} e_n^- a_n^\dagger \cr A_{R,osc} &={1\over
\sqrt{4\pi R^2}} \sum_{n=1}^\infty e^{i\sqrt{\lambda_n}t} e_n^-
\tilde a_n + e^{-i\sqrt{\lambda_n}t} e_n^+ \tilde a_n^\dagger \cr}}
With $[a_n^\dagger, a_n] = \hbar \delta_{n,m}$, etc. and
\eqn\Hosc{
 H_{osc} =   \sum_{n=1}^\infty \sqrt{\lambda_n} ( a_n^\dagger
 a_n +\tilde a_n^\dagger
\tilde a_n). }

The above discussion generalizes the well-known chiral splitting of the
Gaussian model at the rational conformal field theory points. However,
to define the theory of a self-dual scalar we must go further.
When $p=2,q=1$, i.e. $R^2= 2\hbar$, (or by \dualcoupling, when
$R^2 = \half \hbar $)
we can define the self-dual theory by taking a
double cover of the ``target space'' $U(1)$. This
 defines a {\it level one} extension of  $\c H^{2k+1}(Y)$,
which is the desired level for the theory of a self-dual field.

Let us recall the well-known case of   $k=0$.  The radius $R^2=
2\hbar$ (equivalently $R^2 = {\hbar \over 2}$ ) is known as the free
fermion radius, \foot{Note, in particular, that the theory of a
self-dual field is {\it not} obtained by factorizing the theory at
the self-dual radius!} and the theory of the self-dual field is
equivalent to the theory of a chiral (Weyl) fermion. From this point
of view, the dependence of the theory on spin structure is obvious.
The four irreducible representations of the chiral algebra at level
four are associated with the chiral vertex operators $1, e^{\pm
{i\over 2} \phi}, e^{i \phi}$ where we normalize $\phi$ so that
$e^{i\phi}$ is the chiral fermion. The Neveu-Schwarz sector of the
fermion is
\eqn\nssector{ \CH_{NS} = \CH_{1} \oplus \CH_{e^{i \phi}} }
where the summands on the right hand side of \nssector\ are $\IZ_2$
even and odd, respectively. Similarly, the Ramond sector is
\eqn\rsector{ \CH_{R} = \CH_{e^{{i\over 2} \phi }} \oplus
\CH_{e^{-{i\over 2} \phi }} }
where again the two summands are even and odd, respectively.

 We end
this section with a conjecture. We conjecture that the example of
the self-dual field in $1+1$ dimensions generalizes to $k>0$ as
follows: The selfdual theory is obtained by ``factorizing'' the
nonself-dual theory at $R^2= 2\hbar$, or, equivalently, at $R^2 =
{\hbar \over 2}$. (Note this is precisely the normalization occuring
in the Lagrangian in \WittenHC.) When $R^2 = 2 \hbar $ the group
$\CG_{1,2}$ at level four has irreducible representations labelled
by the Pontrjagin dual $\hat T_4$ of $T_4 \subset
H^{2k}(Y,\IR/\IZ)$. The vertex operators $T_{\c A_1} M_{\c A_2}$
admit a formal decomposition into a product of chiral and
anti-chiral parts. Roughly speaking we would like to use the chiral
parts of these vertex operators to define the vertex operators of
the self-dual theory. The fusion rules of the nonself-dual theory
are simply given by the abelian group law on $\hat T_4$. Therefore,
the operators associated to the elements of $\hat T_4$ of order two
form a closed subalgebra. By general principles of conformal field
field theory there is a state-operator correspondence. Therefore, we
conjecture that the ``chiral algebra'' of the self-dual theory may
be identified with $\oplus \CH_{\alpha}$ where the sum runs over the
elements in $(\hat T_4)$ of order $2$. Denote this group by $(\hat
T_4)_2$. The analog of NS and R representations - i.e. the ``spin
representations'' -
 are then labelled by $\hat T_4/(\hat T_4)_2$. This group may be
identified with 2-torsion elements of $H^{2k}(Y;\IZ)\otimes
\IR/\IZ$. From \splithim\ the  Hamiltonian will be (up to a Casimir
shift in the energy)
\eqn\chiral{ H_{SD} = \hbar \pi (n_i + \theta_i) (n_j + \theta_j)
g^{ij} + \sum \sqrt{\lambda_n} a_n^\dagger a_n }
where $\theta \in \CH^{2k+1}$ has half-integer periods and projects to the
corresponding 2-torsion element in  $H^{2k}(Y;\IZ)\otimes \IR/\IZ$.

\newsec{Ramond-Ramond fields and $K$-theory}

In this section we consider the implications of the above results
for   type II string theory. \foot{A related discussion of the
Hilbert space and its conserved charges, starting with an action
principle for type II RR fields,  and carefully deriving the
Hamiltonian formulation of the theory from Legendre transformation
is being carried out in joint work with D. Belov and will appear
elsewhere \belov.  } The RR fields of type II string theory are
classified topologically by (twisted)
 $K$-theory, and
hence it is generally assumed that on $\IR \times Y$ where $Y$ is a
compact $9$-manifold the Hilbert space is likewise graded by
$K$-theory:
\eqn\naivegrd{ \CH \quad {\buildrel  ? \over = } \quad \oplus_{x \in
K^{\epsilon, B}(Y) } \CH_x }
where  $\epsilon=0$ for $IIA $ and $\epsilon=1$ for $IIB$ theory, and $B$ is the twisting. The K-theory element
$x$ encodes both the electric {\it and} the magnetic fluxes, and hence the
considerations of this paper strongly suggest that this naive statement must be
corrected.

The gauge invariant classes of RR fields are identified with
elements of  twisted differential $K$-theory, $\check
K^{\epsilon,\check B }(Y)$. We will denote RR fields in this group
as $[\c C]$.    Here the twisting is by a ``$B$-field,'' regarded
\foot{A subtle point is that the
 group $\check K^{\epsilon,\check B }(Y)$ depends on the choice of ``cocycle'' $\c B$,
but the isomorphism class of this group only depends on $[\c B]$.
That is, automorphisms of $\c B$ act nontrivially on $\check
K^{\epsilon,\check B }(Y)$. Similarly, in the topological case, if
$B$ denotes a cocycle representing $h$, then we write $K^{\epsilon,
B}(M)$ for the twisted $K$-theory.  }
 as an element  $[\c B] \in
\check H^3(Y)$. We denote its characteristic class by $h\in
H^3(Y,\IZ)$ and its fieldstrength by $H\in \Omega^3_{\IZ}(Y)$.

The precise definition of $\check K^{\epsilon,\check B }(M)$, for any manifold $M$,
 can be found in \freed\HopkinsRD\
(for the untwisted case. See also appendix A.).
It satisfies properties closely analogous to those
of differential cohomology.
First of all we have a characteristic class map
\eqn\charclssk{
\check K^{\epsilon,\check B }(M) \rightarrow K^{\epsilon, B}(M)
}
mapping to ordinary twisted $K$-theory. Here $B$ is a cocycle for
$h$ derived from a differential cocycle $\c B$.  In order to define
the fieldstrength map we need some notation.  Let $R= \IR[u,u^{-1}]$
be a graded ring where $u$ is the inverse Bott element of degree
$2$. Let $d_H :=d-u^{-1} H$ be the twisted DeRham differential of
degree $1$. The space of $d_H$-closed differential forms of total
degree $j$ is denoted
\eqn\totaldf{ \Omega(M;R)^j_{d_H}. }
In physics, an element of this space is the total fieldstrength. For
example, if $j=0$, as in IIA theory,
\eqn\totlfsf{
  F =   F_0 + u^{-1}   F_2 + u^{-2}   F_4 + u^{-3}   F_6 + u^{-4}   F_8 + u^{-5}   F_{10}
}
and $d_H   F=0$ is the standard supergravity Bianchi identity.
\foot{In the supergravity literature one often finds the notation $\tilde F$ reserved for the
fieldstrength we have used. We will not try to introduce a second fieldstrength, closed under $d$.
This is an unnatural (albeit common) maneuver. }
 In order to state the analog of
\arrdiagr\ we note that in twisted K-theory there is an analog of the Chern character map:
\eqn\cherncr{
ch_{B}: K^{\epsilon, B}(M) \rightarrow H^{\epsilon}_{d_H}(M)
}
(here $\epsilon$ is a $\IZ_2$-grading. $H^{\epsilon}$ denotes the even/odd
cohomology for $\epsilon=0/1$). This Chern character was defined in \MathaiYK\FHTch\AtiyahSegalII,
and is briefly recalled in appendix A.
The Dirac quantization law on the fluxes is \MooreGB\FreedTT\freed\MathaiMU
\eqn\diracquant{
[  F]_{d_H} = ch_{B}(x) \sqrt{\hat A}.
}
for some $x\in K^{\epsilon, B}(M)$. This is the analog of \arrdiagr.
We denote by
\eqn\totaldfqtz{ \Omega(M;R)^j_{d_H,\IZ} }
the subspace of \totaldf\ whose periods are quantized according to \diracquant.

As before, the differential K-theory is an extension of the setwise fiber product
\eqn\fibnr{
 \CR^j = \{ (  F ,x):  ch_{B}(x) \sqrt{\hat A} = [  F]_{d_H}  \}\subset \Omega(M;R)^j_{d_H,\IZ}\times K^{j,B}(M)
}
  by the torus of topologically
trivial flat RR fields:
\eqn\topsnt{
0 \rightarrow K^{j-1, B}(Y) \otimes \IR/\IZ \rightarrow \check K^{j,\check B}(Y)
\rightarrow \CR^j \rightarrow 0 .
}
As before there are two exact sequences, analogous to \flatseq\ and \ccseq:
\eqn\flatseqk{
0 \rightarrow \overbrace{K^{\ell-1, B }(M;\IR/\IZ)}^{\rm flat} \rightarrow \check K^{\ell,\c B} (M) \rightarrow
  \Omega(M;R)^{\ell}_{d_H,\IZ} \rightarrow 0
}
\eqn\ccseqk{ 0 \rightarrow
\underbrace{\Omega(M;R)^{\ell-1}/\Omega(M;R)^{\ell-1}_{d_H,\IZ}
}_{\rm Topologically\ trivial} \rightarrow \check K^{\ell,\c B} (M)
\rightarrow
  K^{\ell,B}(M) \rightarrow 0
}
Note that it follows easily from the sequences that
\eqn\simplespec{
\check K^0(pt) = \IZ \qquad \qquad \c K^{-1}(pt) = \c K^{+1}(pt) = \IR/\IZ.
}
There is a periodicity $\c K^{\ell+2}(M) \cong \c K^{\ell}(M)$.

In the application to the RR field,
 topologically trivial fields are denoted by $[C]$ where $C \in \Omega^{1/0}(M;R)$, in the
$IIA/IIB$ cases, and $C$ is globally well-defined. The gauge
invariance is $C \to C+ \omega$ where $d_H \omega =0$, and moreover
$\omega$ satisfies a quantization condition: $ \omega \in
\Omega(M;R)^{1/0}_{d_H,\IZ} $. The fieldstrength of a topologically
trivial field is $ F([C]) = d_H C $. The {\it flat} RR fields are
$K^{\epsilon}(M;\IR/\IZ)$ where $\epsilon=-1$ for $IIA$ and
$\epsilon =0$ for $IIB$. This is again a compact abelian group and
is a sum of tori:
\eqn\kflat{
0 \rightarrow \CW^{\epsilon}(M) \rightarrow K^{\epsilon}(M;\IR/\IZ) \rightarrow {\rm Tors} K^{\epsilon+1}(M) \rightarrow 0
}
where
$ \CW^{\epsilon}(M) = H^{\epsilon}(M;\IR)/\Lambda$ and
where $\Lambda$ is a full lattice given by the image of the Chern character.

As in the differential cohomology case there  is a product
\eqn\kproduct{ \c K^{\ell_1,\c B_1}(M) \times \c K^{\ell_2,\c
B_2}(M) \to \c K^{\ell_1+\ell_2,\c B_1 + \c B_2}(M) } such that
\eqn\ktrivprd{
[C_1] \star [ \c C_2] = [ C_1 \wedge F(\c C_2)] .
}

As we mentioned in $\S{ 2.4 }$  by general principles for $M$
compact and $\c K$-oriented there is an integral \foot{We won't give
a precise definition of a $\c K$-orientation, which can be found in
\HopkinsRD. If $M$ is a Riemannian spin$^c$ manifold then it admits
a $\c K$-orientation, the obstruction being $W_3$.}
\eqn\kitns{
\int^{\c K}_{M} : \c K^{\ell}(M) \to \c K^{\ell-n}(M).
}
For topologically trivial fields:
\eqn\ktins{ \int^{\c K}_M [C] = u^{[n/2]}\int_M  C ~ \mod \IZ }
where $\dim M =n$.

Thus far, the discussion has been completely parallel to that for the generalized Maxwell field. Now, however,
we must take into account that the RR field is self-dual.   Here we will simply argue by analogy with the
formulation of the self-dual field discussed in $\S{4}$.

The key will be to introduce a perfect pairing on $\check K^{\epsilon,\check B}(Y)$, where $Y$ is the spatial slice.
In this case the integral is valued in:
\eqn\difflkintgrl{ \int^{\check K}_Y : \check K^{0}(Y) \to \check
K^{-9}(pt) \cong  \check K^{-1}(pt) \cong   \IR/\IZ }
and we can therefore define a pairing:
\eqn\cocyldf{
\langle [\c C_1] ,[\c C_2] \rangle := \int^{\check K}_Y [\c C_1] \star  \overline{[\c C_2] }.
}
Here $\bar y$ denotes the ``complex conjugate'' of $y$. This is an
automorphism of $\check K^{\epsilon,\check B}(Y)$ which takes the
characteristic class to its complex conjugate,   acts on the
fieldstrength by $u \to -u$, and reverses the twisting $\c B \to -
\c B$.  We now have a key

\thm{}  Let $Y$ be odd-dimensional, compact, and $\c K$-oriented.
 Then
\eqn\kpairng{
\c K^{\ell, \c B}(Y) \times \c K^{\ell, \c B}(Y)\to U(1)
}
defined by \cocyldf\ is a perfect pairing.
\foot{The remarks concerning distributions in footnote 14 again apply here.}

As in the case of differential cohomology this can be justified by considering the
two exact sequences \flatseqk\ and \ccseqk. For example, in
 the case of $\ell=0$ (type IIA) we have a perfect pairing
\eqn\fieldpair{ \Omega(Y;R)^0_{d_H,\IZ} \times \biggl( \Omega(Y;R)^{
1}/\Omega(Y;R)^{ 1}_{d_H,\IZ} \biggr) \to U(1) }
given by
\eqn\simpel{
(  F, c') \rightarrow \exp[2\pi i \int_Y    F \wedge \bar c' ]
}
On the other hand, there is a standard perfect pairing (see remark 1 below):
\eqn\kpari{
K^0(Y) \times K^{-1}(Y,\IR/\IZ) \rightarrow U(1)
}
In the IIB theory we exchange gradings $0,-1$ above. $\spadesuit$

When $\dim Y = 3 \mod 4$ the pairing \cocyldf\ is symmetric, and when $\dim Y = 1 \mod 4$
(in particular, when $\dim Y =9$)
the pairing is antisymmetric. This is easily seen at the level of
topologically trivial fields where the pairing is simply
\eqn\toptrpr{
\exp\bigl( 2\pi i \int c d_H \overline{c'} \bigr) .
}

As in $\S{4}$ we introduce a trial pairing:
\eqn\trsint{
s_{\rm trial}([\c C_1], [\c C_2]) = \exp 2\pi i \langle [\c C_1] ,[\c C_2] \rangle
}
This is skew, but in general it is not alternating. (An example
where it is not alternating can be given using remark 3 below.)
As before we define a degree $\epsilon([\c C])\in \{0,1\}$ by
\eqn\degnet{
\langle [\c C] ,[\c C] \rangle = \half \epsilon([\c C])
}
The degree  is defined modulo two and only depends on the
characteristic class. That is, it is an element of ${\rm
Hom}(K^\ell(Y), \IZ_2)$. \foot{The degree can be identified with the
mod-two index of the $9$-dimensional Dirac operator coupled to $x
\otimes \bar x$.
%We do not know of a standard topological
%construction analogous to the Wu class which gives the degree in
%this case.
}  We can therefore define a $\IZ_2$-graded Heisenberg
group (up to noncanonical isomorphism) by
\eqn\ehsitnek{
s([\c C_1], [\c C_2]) = \exp 2\pi i\Biggl( \langle [\c C_1] ,[\c C_2] \rangle - \half \epsilon([\c C_1]) \epsilon([\c C_2]) \Biggr)
}

This Heisenberg extension   has, (up to noncanonical isomorphism)   a unique irreducible
$\IZ_2$-graded representation. Thus we take as definition:

\defn{} The Hilbert space of the RR field $\CH_{RR}$ is the unique
$\IZ_2$-graded irrep of $\heis (\c K^{\epsilon, \c B}(Y))$ defined
by \ehsitnek\
 where $\epsilon = 0,1$ for IIA,IIB.

Now we can return to the issues raised in the introduction. The
hypothetical grading of $\CH_{RR}$ by topological sectors $x \in
K^{\epsilon, B}(Y)$ would be the grading induced by diagonalizing
the translation by the flat RR fields $K^{\epsilon-1,
B}(Y,\IR/\IZ)$. However, elements in this group do not commute in
the Heisenberg extension of $\c K^{\epsilon,\c B}(Y)$.  As in the
case of differential cohomology, the cocycle pairs trivially on
topologically trivial flat fields (this follows   from
\fieldpair\simpel). In this way we arrive  at \fluxsectorstrue.
Moreover, as in the case of differential cohomology,  the pairing on
$K^{\epsilon-1, B}(Y,\IR/\IZ)$ descends to the  torsion pairing on
${\rm Tors}(K^{\epsilon, B}(Y))$, and this is known to be a perfect
pairing. For more details see  \rudyak\ and Appendix B of \FMSi.

\rmk{}

\item{1.} {\it The torsion pairing in $K$-theory}. The   pairing
\eqn\prfctpair{
{\rm Tors} K^{\epsilon,B}(Y) \times  {\rm Tors} K^{\epsilon,B}(Y)  \to U(1)
}
  is defined as usual by multiplication and integration. The product
lives in ${\rm Tors} K^0(Y)$ so the main task is to understand the
 $K$-theory integral $\int^{K}_Y: {\rm Tors}(K^0(Y)) \to \IR/\IZ$. This can be
constructed as follows:  If $x = [E - N ]$ is torsion (where $N$ is
the rank of $E$ and also stands for the trivial  bundle of rank $N$)
then for some $k$ there is an isomorphism $\psi: E^{\oplus k} \to kN
$. Choose any connection $\nabla$ on $E$ and let $\nabla_0$ denote
the trivial connection with zero holonomy. Then we define
\eqn\paring{
\int^{K} x :=
\eta(\Dsl_{\nabla}) - \eta(\Dsl_{\nabla_0}) -
{1\over k} \int_{Y} CS(\psi^*(\nabla^{\oplus k} ) , \nabla_0^{\oplus k} ) \hat A(Y)~~  \mod \IZ
}
where $\eta$ is the eta-invariant of Atiyah-Patodi-Singer.  The
expression on the right-hand side is independent of the choice of
$E, k,\psi,\nabla$ and the metric used to define $\eta$.

\item{2.} As a simple example let us consider $Y= L_k \times S^6$ where $L_k$ is a three-dimensional
Lens space. Then ${\rm Tors}(K^0(Y)) = \IZ_k \oplus \IZ_k$.
Natural generators are given by   $\CL-1$, where $\CL \to L_k$ is the line bundle whose first
Chern class generates $H^2(L_k)$, and $(\CL-1)\otimes E$, where $E\to S^6$ has index $1$ and rank $0$.
Physically, these may be thought of as classes of the $RR$ 2-form and 8-form fieldstrength,
respectively. We have seen that in the quantum theory one cannot simultaneously specify these two
components!

\item{3.} As a more elaborate example, consider $Y = S^{2n+1}/\IZ_q$ where $\IZ_q$ acts on the unit sphere in
$\IC^{n+1}$ by multiplication by $e^{2\pi i/q}$. Then  we have \hirzebruch\
\eqn\lensprime{
K^0(S^{2n+1}/\IZ_q) = \IZ + (\IZ_{q^{s+1}})^r + (\IZ_{q^s})^{q-r-1}
}
where $n = s(q-1) + r$, $0 \leq r < q-1$. Let $\sigma = H-1$ be the pullback
from $\IC P^n$. Then $\sigma, \sigma^2, \dots, \sigma^{q-1}$ generate the
$(q-1)$ factors. We may apply this to $n=4$ to obtain various examples of nontrivial Heisenberg groups.

\item{4.} In type I string theory the space of ``RR fields'' is a
torsor for $\check{KO}^{-1}(M)$ (the field itself should be viewed
as a trivialization of a background magnetic current: see \freed).
The fieldstrength is obtained by replacing $R \to \IR[u^2,u^{-2}]$,
and the structure of $\check{KO}^{-1}(M)$ then follows the same
pattern as \flatseqk\ccseqk. It would be very worthwhile to
investigate the properties of the relevant Heisenberg groups for
general orientifolds of type IIB strings.

\newsec{Generalized Maxwell theory with a Chern-Simons term}

One important way to modify generalized Maxwell theory is by adding
a Chern-Simons like term. In this section we consider the case of
odd-dimensional Maxwell-Chern-Simons theory from the viewpoint of
the present paper. We will see that this modification has dramatic
consequences for the grading of the Hilbert space by flux sectors.

The action for odd-dimensional Maxwell-Chern-Simons theory is:
\eqn\maxactthrd{
S = \pi R^2 \int_{Y \times \IR}   F*F + 2\pi k \int^{\c H}_{Y\times \IR} [\c A]\star  [\c A]
}
Here $\dim Y = 2p$ and $[\c A] \in \c H^{p+1}(M)$,    $k$ is an
integer, and we assume $p$ is odd. \foot{If we choose extra
structures on $Y$ then we can consider fractional $k$. In
particular, in three dimensions, if we choose a spin structure then
$k$ can be half-integral.  See \BelovZE\ for a recent discussion of
the Hilbert space of the theory in this case.}

In the Hamiltonian framework, where we consider a path integral with
boundary $Y$ (at fixed time), the Chern-Simons term should be
interpreted as a section of  a line bundle with connection $(\CL_k,
\nabla)$ over $\bA$ with $\bA=\check H^{2p}(Y)$. \foot{One way to
understand this is as follows. We can form a universal
``connection'' $[\c A]$ on $M \times \c H^{p+1}(M)$. If $M$ is
compact and we multiply and integrate over $M$ we obtain an element
of $\c H^1(\c H^{p+1}(M))$. This $U(1)$-valued function of the gauge
field is of course the exponentiated Chern-Simons term.   If we
apply the same construction over $Y$ we obtain an element of $\c
H^2(\bA)$. This is the line bundle with connection mentioned above.
}
  The phase space is
accordingly $T^*\bA$ but with symplectic form:
\eqn\syplm{
\Omega = \Omega_{c} + \CF(\nabla)
}
where $\Omega_c $ is the canonical symplectic form on $T^*\bA $.

There is an important subtlety introduced by the ``tadpole
condition.'' There are only gauge invariant wavefunctions on the
component (or components) with  $ka=0$ where $a$ is the
characteristic class of $[\c A]$. \foot{When $k$ is half-integral
the condition sets $a$ to be a certain torsion class associated with
a quadratic refinement. See \WittenVG \HopkinsRD \BelovJD\ for a
discussion.  } This replaces the usual grading by magnetic flux. We
give an explanation of this condition in Appendix B.

The Hilbert space of the theory is, heuristically, the space of $L^2$ sections of $\CL_k$,
restricted to the $ka=0$ component(s),
and the Hamiltonian is the Laplacian. Once again, the Hilbert space can be elegantly formulated
as a representation of a Heisenberg group. Our goal in this section
is to  describe that Heisenberg group.

It is more convenient to work in the framework of Hamiltonian
reduction from a space of gauge potentials.
For simplicity we will assume that $ka=0$ implies $a=0$. (This holds in particular for the
important cases of $p=1$ or $k=1$.) The tadpole constraint then restricts our
attention to the topologically trivial fields.
We can therefore formulate the theory in terms of Hamiltonian reduction, beginning
with the phase space $ \Omega^p(Y) \times \Omega^p(Y)$ of pairs $(\tilde \Pi , \alpha)$
with a gauge group $\Omega^p_{\IZ}(Y)$ acting by $(\tilde \Pi , \alpha) \to (\tilde \Pi, \alpha + \lambda)$
for $\lambda\in \Omega^p_{\IZ}(Y)$.
The Chern-Simons
term  defines a line bundle with connection $(\CL_k, \nabla)$  on the space $\Omega^p(Y)$ of
all gauge potentials. In physical notation the curvature of $\nabla$ is
\eqn\curvat{
\CF(\nabla) = 2\pi k \int \delta A \wedge \delta A .
}
Thus, the symplectic form can be written explicitly:
\eqn\symplef{
\Omega\bigl( (\tilde \Pi_1, \alpha_1), (\tilde \Pi_2, \alpha_2)\bigr)
= \int_Y \tilde \Pi_1 \alpha_2 - \tilde \Pi_2 \alpha_1 + 4\pi k \int_Y \alpha_1 \wedge \alpha_2
}
The moment map for the action of the gauge group is
\eqn\momentmap{
\mu = d\tilde \Pi + 4\pi k F
}
Note that $\mu=0$ is the restriction of the equations of motion to the spatial components.

We will now  present a rather general construction, and then
specialize it to the case relevant for Maxwell-Chern-Simons theory.
First, quite generally, suppose  $(M,\Omega) $ is a symplectic
manifold  equipped with a ``prequantum  line bundle.'' This is a
line bundle $L \to M$ with connection $\nabla$ such that the
curvature $F(\nabla)$ is the symplectic form $\Omega$. In this
situation we can consider the group ${\rm Ham}(M,L,\nabla)$ of all
automorphisms of the structure $(M,L,\nabla)$, i.e. all pairs
consisting of a diffeomorphism $f$ of $M$ and a bundle map $\hat f:
L \to L$  covering it which preserves the connection $\nabla$ on
$L$. Clearly such an $f$ must be a symplectomorphism, $f\in {\rm
Symp}(M)$. In general, given $f\in {\rm Symp}(M)$ we can define the
flat bundle $L_{ f} =  f^*L \otimes L^{-1}$. The map  $ f\to L_{ f}$
gives us a homomorphism ${\rm Symp}(M) \to H^1(M;\IT)$ which fits
into the  exact sequence
\eqn\hamaust{ 1 \rightarrow H^0(M,\IT) \rightarrow {\rm
Ham}(M,L,\nabla) \rightarrow {\rm Symp}(M) \rightarrow   H^1(M,\IT)
  . }
%
%where ${\rm Symp}(M)$ is the group of symplectomorphisms of $M$.

Now suppose that $M$ is an affine space for a connected abelian
group $G$ and that $\Omega$ is translation-invariant. Then we can
define the subgroup ${\rm Ham}(M,L,\nabla)_{aff}$  of ${\rm
Ham}(M,L,\nabla)$ consisting of elements for which $f\in G$. Now,
${\rm Ham}(M,L,\nabla)_{aff}$  is a Heisenberg group, and will be
the Heisenberg group we seek. (When $G$ is not connected there are
further subtleties, but we will not need this case. )

As a simple example, if we take $M$ to be $\IR^{2n}$ with the usual
symplectic structure and connection given by $d + i pdq$, where $(p,q)\in \IR^{2n}$
are Darboux coordinates, then
 ${\rm Ham}(M,L,\nabla) $ is the usual Heisenberg group. To see this, note that
 the translations $T_{q_0}: (p,q) \to (p,q+q_0)$ and $T_{p_0}: (p,q) \to (p+p_0, q)$
lift to automorphisms of the bundle with connection:
\eqn\simplauts{
\eqalign{
\hat T_{q_0}:(p,q,z) & \to (p,q+q_0,z) \cr
\hat T_{p_0}:(p,q,z) & \to (p+p_0,q,e^{ip_0\cdot q } z) \cr}
}
Clearly, these generate the Heisenberg group.

We now would like to understand ${\rm Ham}(M,L,\nabla)_{aff}$ in the
special case that  $M$ is a quotient of a linear space. Let $V$ be a
vector space equipped with a symplectic form $\Omega$. Using
$\Omega$ we define a central extension
\eqn\cents{
0 \rightarrow \IT \rightarrow \tilde V \rightarrow V \rightarrow 0 .
}
We can view $\tilde V$ as the set of pairs $(v,z)$ with the product
given by a   cocycle
\eqn\conclyc{
(v,z)(v',z') = (v+v', zz' c(v,v')  )
}
such that
$$
{c(v,v')\over c(v',v) } = e^{2\pi i \Omega(v,v')}.
$$
It is natural to take $c(v,v') = e^{i \pi \Omega(v,v')}$,
corresponding to a distinguished trivialization of the principal $U(1)$
bundle $\tilde V\to V$. The bundle $\tilde V$ has, moreover,  a natural connection.
The connection is defined by writing the parallel transport
from $(w,z)\in   \tilde V $ along a straightline path
$p_{w,v} := \{ w+ t v\vert 0 \leq t \leq 1\}$. The parallel transported
point is simply given by left-multiplication by $(v,1)$. Thus, using
\conclyc\ we have the parallel transport law:
\eqn\partrsp{
U(p_{w,v}): (w,z) \Rightarrow   (w+v, c(v,w) z)
}
It is straightforward to check that the curvature of this connection is precisely $\Omega$.

Now suppose that we have a subgroup $\Lambda\subset V$ with $\Omega$
integral when restricted to $\Lambda$. The subgroup $\Lambda$ can be
a lattice (not necessarily of full rank) or a product of a lattice
and a vector subspace. Roughly speaking, we want to repeat the above
construction on the quotient $M=V/\Lambda$, but we must be a little
careful.

Consider first the case that $\Lambda$ is a lattice.
We can construct a principal $U(1)$ bundle with connection over $V/\Lambda$. To do this
we split the sequence \cents\ over $\Lambda$. That is, we choose a function
$\epsilon: \Lambda \to U(1)$ with
\eqn\splitt{ \epsilon_v \epsilon_{v'} = e^{i \pi \Omega(v,v')}
\epsilon_{v+v'}  \qquad \forall v,v' \in \Lambda }
Since $\Omega$ is integral on $\Lambda\times \Lambda$ we may regard
$\epsilon$ as a quadratic refinement of the symmetric bilinear form $e^{i \pi \Omega(v,v')}$.
Now  $v \to (v, \epsilon_v)$ defines a homomorphism $\Lambda \to \tilde V$ embedding $\Lambda$ in
the group $\tilde V$. Let us call $\Lambda_\epsilon$ the image of this homomorphism.
It now makes sense to take the quotient $\tilde V/\Lambda_\epsilon$. This
space is a principal $U(1)$ bundle over
$V/\Lambda$. Moreover, if we define $V/\Lambda$ as a space
of  right-cosets then it is clear that the connection
\partrsp\ descends to a connection on $\tilde V/\Lambda_\epsilon$ since it is defined by
{\it left}-multiplication.  The curvature of this connection is
again $\Omega$. As a simple exercise, let us compute the holonomy
  at $[w]\in V/\Lambda$ around a closed curve lifting to
$p_{w,\lambda}$, where $\lambda\in \Lambda$. We compute
\eqn\isnttn{ \eqalign{(\lambda,1)(w,z) & = (w+\lambda, e^{i \pi
\Omega(\lambda, w)} z) \cr & = (w, e^{2\pi i \Omega(\lambda,w)}
\epsilon_{\lambda}^{-1} z) (\lambda, \epsilon_\lambda) \cr} }
Since $(\lambda, \epsilon_\lambda) \in \Lambda_\epsilon$ we conclude
that
\eqn\holon{ Hol(p_{w,\lambda}) = e^{2\pi i \Omega(\lambda,w)
}\epsilon_\lambda^{-1} . }

When the subgroup $\Lambda$ is not a lattice, and has a nontrivial connected component
of the identity, denoted
$\Lambda_0 \cong {\rm Lie}(\Lambda)$,
 the above construction must be modified because we identified $T(V/\Lambda)$ with $V$ in the
 above discussion. In general, the   tangent space to $V/\Lambda$ is
$V / \Lambda_0$. Left-multiplication by $([v],1)$ on $[(w,z)]$ is
only well-defined  when $v$ and $w$ are both in $V_0$, the zeroset
of the moment map. Explicitly, $V_0$ is given by:
\eqn\vnought{
V_0 := \{ v\in V \vert \Omega(v,\lambda)=0 \qquad \forall \lambda \in \Lambda_0 \}.
}
Thus, we must actually choose $M= V_0/\Lambda$. Note that $\Lambda
\subset V_0$ since $\Omega$ is integral on $\Lambda$ and $\Lambda_0$
is contractible to zero.  Now, to repeat the above construction we
must also choose $\epsilon$ so that $\epsilon_\lambda=1$ for
$\lambda\in \Lambda_0$. Then we can form $\tilde
V_0/\Lambda_\epsilon$ as a principal $U(1)$ bundle with connection
over $V_0/\Lambda$. We will denote by $L$ the associated line
bundle. The formula \holon\ remains unchanged.

We are interested in automorphisms of $L$ preserving the connection
and covering translations in $V_0/\Lambda$.  Let $\xi \in
V_0/\Lambda$, and $\lambda\in \Lambda$. Consider the holonomy,
around a curve lifting to $p_{w,\lambda}$, of the bundle $\xi^*
L\otimes L^{-1}$. From \holon\ it follows that this holonomy is
\eqn\holodiff{ e^{2\pi i \Omega(\lambda, w+ \xi)}
\epsilon_\lambda^{-1} e^{-2\pi i \Omega(\lambda,w)} \epsilon_\lambda
= e^{2\pi i \Omega(\lambda, \xi)} }
Note that the choice of lift of $\xi$ into $V$ does not matter. Also note that
the $w$-dependence in \holon\ has cancelled, as expected, since
$\xi^* L\otimes L^{-1}$ is flat. Thus, the holonomies are preserved if $\xi$
lifts to an element in
\eqn\defnlambprp{
\Lambda^\perp:= \{ v\in V_0 \vert  \Omega(\lambda,v) \in \IZ \qquad \forall \lambda\in \Lambda \} .
}
Indeed, one can   define what is manifestly an automorphism of the
bundle $\tilde V_0/\Lambda_\epsilon$, together with its connection,
by using {\it right-multiplication} by the subgroup of $\tilde V_0$
which commutes with $\Lambda_\epsilon$. The reason is simply that
the connection was defined by left-multiplication. The inverse image
$\tilde \Lambda^\perp \subset \tilde V_0$ of $\Lambda^\perp$ under
\cents\  (with $V $ replaced by $  V_0$) is easily checked to be the
commutant of $\Lambda_\epsilon$ in $\tilde V_0$. Of course,
right-multiplication by $\Lambda_\epsilon$ acts trivially on $\tilde
V_0/\Lambda_\epsilon$, so we finally find that the group of affine
linear Hamiltonian automorphisms is
\eqn\thegroup{ {\rm Ham}(V_0/\Lambda, L, \nabla)_{aff} = \tilde
\Lambda^\perp/\Lambda_\epsilon. }
The symplectic form $\Omega$ descends to a nondegenerate symplectic
form on $V_0/\Lambda_0$, and $\Lambda/\Lambda_0$ is a lattice in
this vector space (not necessarily of full rank). Using this fact it
is easy to show - using a canonical form for $\Omega$ - that
$((\Lambda/\Lambda_0)^\perp)^\perp = \Lambda/\Lambda_0 $ and hence
that $\tilde \Lambda^\perp/\Lambda_\epsilon$ is a nondegenerate
Heisenberg group.

Let us now specialize the above construction to model more closely
the situation in Maxwell-Chern-Simons theory. Accordingly, we take
$V= P \oplus Q$ where $Q$ is a vector space, $P=Q^*$ is the dual
space, and $\Lambda\subset Q$ is a subgroup. (In
Maxwell-Chern-Simons theory $P=\Omega^p(Y)$, $Q= \Omega^p(Y)$ and
$\Lambda= \Omega^p_{\IZ}(Y)$. )  Suppose, moreover, that
\eqn\speicalomeg{
\Omega = \Omega_c + \omega
}
where $\Omega_c$ is the canonical form on $P\oplus Q = T^*Q$ and $\omega$ is a translation
invariant symplectic form on $Q$, integral on $\Lambda$.
Denoting elements of $V$ by $(p,q)$ one
 easily computes that $\Lambda^\perp = \{ (p,q)\in V_0 \vert p + \omega(q,\cdot) \in
 \Lambda^*\}$,
where $\Lambda^* =\{ p\in P\vert p\cdot \lambda \in \IZ \qquad \forall \lambda\in \Lambda\}$.
(Note that if $\Lambda$ is not of full rank then $\Lambda^*$ is a vector space times a lattice.)
%
%From the nondegeneracy of $\Omega$
%it follows that $\tilde \Lambda^\perp/\Lambda_\epsilon$ is a nondegenerate
%Heisenberg group.
%

The Heisenberg group $\tilde \Lambda^\perp/\Lambda_\epsilon$ has a
unique irreducible representation (with the canonical representation
of scalars and a fixed choice of polarization). Recall from the
discussion following \ceone\ that we can construct this
representation by choosing a maximal Lagrangian subspace. We do this
as follows. Note that $p\mapsto (p,0)$ injects $\Lambda^*$ into
$\Lambda^\perp$. To see this note that $(p,0)\in V_0$. The reason is
that if $\lambda\in \Lambda_0$, then on the one hand $p\cdot
\lambda\in \IZ$, since $p\in \Lambda^*$, but since $\lambda\in
\Lambda_0$ is continuously connected to $0$ we have $p\cdot
\lambda=0$, and hence $(p,0)\in V_0$ by \vnought. In fact $\tilde
\Lambda^*$ injects into $\tilde \Lambda^\perp/\Lambda_\epsilon$ as a
maximal commutative subgroup.
 and moreover there is an exact sequence
\eqn\inengt{
0 \rightarrow \tilde \Lambda^* \rightarrow \tilde \Lambda^\perp/\Lambda_\epsilon \rightarrow Q/\Lambda \rightarrow 0
}
since   $  \Lambda^*\backslash \Lambda^\perp/\Lambda  \cong
Q/\Lambda$. \foot{One way to prove these statements is by using a
canonical form for $\Omega$ and reducing to the four-dimensional
case with $P= \IR^2, Q=\IR^2, \Lambda = \IZ^2$, and
$$
\Omega = \pmatrix{ 0 & 0 & 0 & 1 \cr 0 & 0 & 1 & 0 \cr 0 & -1 & 0 &
k \cr -1 & 0 & -k & 0 \cr}
$$
where the statements are straightforward to check. }

The Heisenberg representation of $\tilde
\Lambda^\perp/\Lambda_\epsilon$ is the representation induced by a
character $\rho$ of $\tilde \Lambda^*$. Geometrically, the induced
representation is nothing other than the sections of the line bundle
\eqn\aslbdle{
\bigl( \tilde \Lambda^\perp/\Lambda_\epsilon \bigr) \times_{\tilde \Lambda^*} \IC \to Q/\Lambda.
}
Denoting this line bundle as $\CL_\epsilon$ we find that the Hilbert space is
\eqn\ltwsosct{
\CH = L^2(Q/\Lambda; \CL_\epsilon) .
}

A few comments are in order concerning a closely related Hilbert space.
Using the constructions explained above it is easy to see that
the   data $(Q,\Lambda, \epsilon, \omega)$
also define a line bundle with connection $L_\epsilon \to Q_0/\Lambda$ whose curvature is $\omega$.
Here $Q_0$ is the zero-locus of the moment map associated to $\omega$.
Moreover, by  considerations analogous to those above, the
 automorphisms of the bundle with connection $L_\epsilon \to Q_0/\Lambda$ are given by
right-multiplication of the Heisenberg group $\tilde
\Lambda^{\perp,Q}/\Lambda_\epsilon$ where $\Lambda^{\perp,Q}= \{
q\in Q_0 \vert  \omega(q,\gamma) \in \IZ \qquad \forall \gamma \in
\Lambda\}$. This is a {\it finite} Heisenberg group. The Hilbert
space $L^2(Q_0/\Lambda; L_\epsilon)$ can be decomposed into
eigenspaces of the Laplacian $\nabla^2$ on $L_\epsilon$, and it
follows that these  eigenspaces are  representations of this finite
Heisenberg group.

Finally, we now return to Maxwell-Chern-Simons theory. We apply the above
discussion with $P = \Omega^p(Y)$, $Q= \Omega^p(Y)$,
and $\Lambda = \Omega^p_{\IZ}(Y)$.  From \symplef\  we see that
 $\omega(\alpha_1, \alpha_2) = 4\pi k \int_Y \alpha_1 \wedge \alpha_2$.
In the case of Maxwell-Chern-Simons theory the space $Q_0/\Lambda$
is the moduli space of topologically trivial flat fields (a
finite-dimensional torus). The Hilbert space \ltwsosct\ is a product
for the harmonic and oscillator modes. The wavefunction of the
harmonic modes lives in $L^2(Q_0/\Lambda; L_\epsilon)$ and the
finite Heisenberg group $\Gamma$ acting on eigenspaces of the
Laplacian  is an extension
\eqn\fininte{ 0 \rightarrow \IZ_k \rightarrow \Gamma \to H^p(Y,
 \IZ_k) \rightarrow 0 . }
This generalizes the familiar noncommuting Wilson line operators in
three-dimensional Chern-Simons theory and  makes contact with the
Heisenberg group discussed in \WittenWY\BelovHT\BelovZE.

To summarize, in the presence of a Chern-Simons term the grading by electric and magnetic
flux is significantly changed. The grading by magnetic flux is replaced by one (or at most
a finite) number of components from the tadpole constraint, while instead of grading by
electric flux the Hilbert space is a representation of a Heisenberg group of
noncommuting Wilson line operators.

\newsec{Comments on   noncompact spatial slices and charge groups}

In this section we make some preliminary remarks aimed at
generalizing our story in various directions.

\subsec{Noncompact $Y$}

A very natural generalization we might wish to consider is the case
where  the   spatial slice $Y$ is  noncompact. Let us focus in
particular on geodesically complete spaces $Y$ so that we discuss
asymptotic boundary conditions on the fields.  We have seen that
electro-magnetic duality is closely related to Poincar\'e duality of
differential cohomology groups.  Now, in the noncompact case there
is a perfect pairing \harveylawson
\eqn\noncompt{ \c H^{\ell}_{\rm cpt}(Y) \times \c H^{n-\ell}(Y)
\rightarrow \c H^{n}_{\rm cpt}(Y) \cong \IR/\IZ. }
(A differential character is compactly supported if there is a $K$
such that $\chi(\Sigma)=1$ for $\Sigma$ outside of $K$.) This seems
to introduce a fundamental asymmetry in the theory between $\ell$
and $n-\ell$, which appears to ruin electric-magnetic duality. It is
therefore natural to try to formulate an $L^2$ version of $\c
H^\ell$ which is Pontrjagin self-dual (under $\ell \to n-\ell$).
This raises many new issues and is beyond the scope of this paper,
but it should be pursued.

\subsec{Inclusion of Sources }

First let us include external sources. When working with sources we
cannot work just within the framework of differential cohomology. We
must introduce a notion of cochains and cocycles. In the formulation
of \HopkinsRD\freed\ we can consider   currents to be differential
cocycles. The electric current $\c j_e \in \c Z^{n-\ell+1}(M)$ and
the magnetic current $\c j_m \in \c Z^{\ell+1}(M)$. What is normally
called the current is - in this formulation - the ``fieldstrength''
\eqn\currfldstrg{J_m = F(\c j_m)\in \Omega^{\ell+1}(M), \qquad J_e =
F(\c j_e)\in \Omega^{n-\ell+1}(M).
}
Interaction with an external current is taken into account by
including a term in the action:
\eqn\interone{ S_{elec} = \langle [\c A], \c j_e \rangle = \int^{\c
H}_{Y\times \IR} \c A \star \c j_e }
\eqn\intertwo{ S_{mag} = \langle [\c A_D], \c j_m \rangle = \int^{\c
H}_{Y\times \IR}  \c A_D \star \c j_m }
Note that in   the topologically trivial sector we have $[A] \star
\c j_e =[ A \wedge J_e ]$ which gives the usual electric coupling.
We cannot add both $S_{elec}$ and $S_{mag}$ together in a local
Lagrangian.

In the Hamiltonian picture the action of a current on the
wavefunction is the operator
\eqn\pathordop{ P\exp \int^{\c H}_{Y\times I} \c A \star \c j_e }
where  $I$ is a time interval. Here $\c A$ is operator valued.  In
the limit that $\c j_e$ has support in an infinitesimal time
interval $I$ we have
\eqn\electricinteract{ \Biggl(P\exp \int^{\c H}_{Y\times I} \c A
\star \c j_e \Biggr) \psi(\c A) = e^{2\pi i \langle \int_I \c j_e,
\c A \rangle } \psi(\c A) }
\eqn\magneticinteract{ \Biggl( P\exp \int^{\c H}_{Y\times I} \c A_D
\star \c j_m \Biggr) \psi(\c A) =  \psi(\c A + \int_I \c j_m ) }
These satisfy an exchange algebra with phase
\eqn\exchange{ e^{2\pi i \langle \int_I \c j_e, \int_I \c j_m
\rangle } }
which is therefore related to an anomaly.

 The next step is to   include {\it backreaction} of the electric and magnetic
sources. This leads to the equations:
\eqn\sourceeom{ \eqalign{ dF & = J_m \cr d*F & = J_e \cr} }

The first remark is that, within the support of $J_m$ we can no
longer   consider the gauge field $\c A$ to be in $\c Z^\ell(M)$.
Similarly there is no $\c A_D $ in the support of $J_e$, so a full
account goes beyond the framework of this paper. Nevertheless, in
free field theories such as those we are considering we can always
use the principle of linearity. If we have a solution of the
equations \sourceeom\ then we can consider that solution as a
background. Then, other solutions obtained by physically realistic
deformations of this background differ by solutions of the vacuum
equations of finite energy. This brings us back to the quantization
of the system we considered above.  Roughly speaking, we are writing
$A = A_{classical} + A_{quantum}$ where $A_{classical}$ solves the
equations with sources.

Finally, we should consider dynamical charge sources. If we consider
the branes to be dynamical then they can be pair-produced from the
vacuum and reannihilate. There are finite-action processes where a
brane-antibrane pair is produced, one brane moves around a
nontrivial cycle, and then they reannihilate. We will now explain
how such a process will lead to mixing between different flux
sectors.

Suppose, for example that in $Y$ there is a cycle of the form
$\Sigma_{n-\ell-2} \times S^1$. Then we can imagine wrapping
magnetic branes on $\Sigma_{n-\ell-2}$ in a $B\bar B$ pair, move one
brane around the $S^1$, then let them reannihilate. More generally,
if we have any cycle $\CS_{n-\ell-1}\subset Y $ then by slicing it
with a Morse function we can interpret this as the worldhistory of a
similar process. Such a process produces a magnetic current:
\eqn\prodcurnt{
 \c \delta(\CS_{n-\ell-1}) \in \c H^\ell(Y)
}
 Now  consider a state $\Psi_i$ at the initial time in the cylinder
$Y\times I$. Suppose a process of the above type takes place in a
short time interval. Then the effect on the wavefunction is
\eqn\bbcreatone{ \Psi_i(\c A) \to  e^{-S} \Psi_i(\c A + \c
\delta(\CS_{n-\ell+1})) }
Here $e^{-S}$ is the tunneling amplitude, which to leading order
will be
$$S \sim  T \int_{\CS_{n-\ell-1}}\vol$$
 where $T$ is the tension
of the brane.

For this reason, the grading of the flux Hilbert space will not be
preserved by time evolution in the full theory with dynamical
branes, since the matrix elements \bbcreatone\ mix sectors $a$ and
$a + PD[\CS_{n-\ell+1}]$, in other words, {\it all} sectors mix!
\foot{Actually, there could be fermion zeromodes in the instanton
amplitude which kills mixing between some sectors. For example, one
might guess that the $\IZ_2$ grading of the self-dual Hilbert space
discussed in section 4 would be preserved. }

Similarly, if we consider a process involving electric
$(\ell-2)$-branes  from a cycle $\CS_{\ell-1}\subset Y$ we get
\eqn\bbcreatone{ \Psi_i(\c A) \to  e^{-S}e^{2\pi i
\int_{\CS_{\ell-1}} \c A }  \Psi_i(\c A  ) }
This will consequently modify the electric flux sector of the
wavefunction, $e\to e+ PD[\CS_{\ell-1}]$.

When the tunneling amplitudes are small, $e^{-S}\ll 1$, the Hilbert
space has an approximate grading in the sense that the support of an
eigenstate of the Hamiltonian can be taken to be predominantly in
one particular flux sector.

\subsec{Uncertainty principle for charges?}

It is natural to expect that an uncertainty principle for fluxes
leads to a similar uncertainty principle for charges.
Philosophically, at least in  the context of string theory, the
AdS/CFT correspondence and geometric transitions suggest that
classification of fluxes and branes should be treated
democratically. It is, however, difficult to make a precise
argument. One argument that suggests that an uncertainty principle
for fluxes implies a similar relation for charges  is the following.

Let us recall the relation between the charge group and the group of
fluxes as measured at infinity. (See, for example, \MooreGB\freed.)
We simply analyze  the basic equation $dF = J_m$ on a spacetime $M$.

Suppose that   $F, J_m$ are smoothly defined everywhere. First
$dJ_m=0$ but $J_m$ is trivialized by $F$ and hence $[J_m] \in
H^{\ell+1}(M)$ vanishes. This is similar to the the standard
vanishing of the total charge in a closed universe, but it applies
whether or not $Y$ is compact, given our assumption that $F$ is
everywhere defined.

Now, let
 $\CN_m$ be the support of $J_m$. Then the pair
 $(J_m, F)$ defines a cocycle in
$Z^{\ell+1}(M,M-\CN_m)$, however it is trivialized by $(F,0) \in
C^\ell(M,M-\CN_m)$. But $[(J_m,0)] = -[(0,F)] $ defines a
potentially nontrivial cohomology class. $[(J_m,0)]$ is in the
kernel of the map to $H^{\ell+1}(M)$. Hence, from  the long exact
sequence for the pair $(M, M-\CN_m)$,
\eqn\les{ \cdots \rightarrow H^{\ell}(M) \rightarrow
H^{\ell}(M-\CN_m) \rightarrow H^{\ell+1}(M,M-\CN_m) \rightarrow
H^{\ell+1}(M) \rightarrow \cdots }
 we find that the group of magnetic charges is
\eqn\magchrges{ \CQ_m = H^{\ell}(M-\CN_m)/\iota^* H^{\ell}(M) }%
The physical interpretation of this formula is that the group of
charges is measured by ``fluxes at infinity'' which cannot be
smoothly continued in to all of spacetime.  Similarly, the group of
electric charges is
\eqn\magchrges{ \CQ_e = H^{n-\ell}(M-\CN_e)/\iota^* H^{n-\ell}(M) }%
where $\CN_e$ contains the support of $J_e$.

Let us now apply this to the case of domain walls. That is, consider
the case $Y = \IR \times Z$, with $Z$ compact. Consider the case of
branes wrapping cycles within $Z$ located at a point $p\in \IR$.
Then we can take $\CN = (p-\epsilon, p+\epsilon) \times Z$. Note
that now the group of charges is the quotient by the diagonal, and
hence
\eqn\domainwall{ \eqalign{ \CQ_m& \cong H^\ell(Z;\IZ) \cr \CQ_e &
\cong H^{n-\ell}(Z;\IZ)\cr} }%
One would expect that one cannot simultaneously measure the electric
and magnetic charge if the corresponding fluxes cannot be
simultaneously measured.

The conclusion above  leads to some puzzles, and since the above
arguments are not very precise (and $Y$ is noncompact) we again
leave it as an open problem whether charges, in particular, whether
D-brane charges satisfy a Heisenberg uncertainty principle. Indeed,
one counterargument suggesting that there should be no uncertainty
principle is that D-branes with definite K-theory torsion charge
have already been described in the literature \WittenCD\BrunnerEG !
For example, \BrunnerEG\ study branes in   an interesting Calabi-Yau
manifold with torsion \FerraraYX. This manifold is the quotient
$\CZ= (S\times E)/\IZ_2$ where $S$ is a $K3$ surface with fixed
point free holomorphic involution double-covering an Enriques
surface, and $E$ is an elliptic curve. The  $\IZ_2$ acts as
$(\sigma, -1)$ where $\sigma$ is the Enriques involution. In
\BrunnerSK\ the torsion $K$-theory classes for $\CZ$ have been
carefully described, and a definite assignment of   specific BPS
states with definite $K$-theory classes is proposed. (This
assignment is, however, not entirely clear in the heterotic dual
theory, so the example definitely merits further investigation.)

\newsec{Conclusions}

\subsec{Implications}

One of the most striking conclusions we have found is that the
K-theory class of  RR flux  cannot be measured due to a quantum
mechanical uncertainty principle. When the manifold has torsion in
its K-theory there is no parameter, or effective $\hbar$ one can
introduce to ``turn off'' quantum effects, even in the long-distance
limit.

In this paper we have focussed on the theory of free gauge theories,
without discussing  sources in any detail. In section seven we were
only able to make some very tentative comments on the implications
of our results for D-brane charges.    Naively, one would think
that, while quantum effects might modify the standard picture of a
D-brane as a submanifold of spacetime equipped with vector bundle,
nevertheless,
 quantum effects  can be made arbitrarily small in the long-distance limit.
 This is typically
assumed when, for example, one identifies D-branes in Calabi-Yau
manifolds with  certain vector bundles (or sheaves) at large values
of the Kahler moduli. However if the effect we have discussed for
fluxes has a counterpart for D-brane charges, then it survives at
long distances. It would follow that the standard picture of
D-branes must be altered in the quantum-mechanical setting. We find
this possibility surprising and radical. On the other hand, if there
is no uncertainty principle for charges, but there is for fluxes,
then there is a fundamental asymmetry between charge and flux groups
which is also quite surprising and radical. It is highly desirable
to decide which of these two surprising conclusions is in fact
correct.

On the purely mathematical side, it might be interesting to explore
what implications there are for the formulation of topological
D-branes in terms of derived categories. This formulation is a
refinement of the $K$-theory classification of D-branes. Perhaps one
can formulate a noncommutative version of the derived category,
analogous to the Heisenberg group extensions of $K$-theory we have
studied above.

The Heisenberg groups studied in this paper are direct higher
dimensional generalizations of  vertex operator algebras which are
used in the construction of two-dimensional conformal field
theories. One severe limitation of the present paper is that the
``target space'' for these higher dimensional generalizations of
loop groups is abelian. It is natural to ask if there is a
higher-dimensional analog of the Frenkel-Kac-Segal construction of
representations of affine Lie algebras \FrenkelRN\SegalAP, or if
there are higher-dimensional analogs of higher level nonabelian
current algebras (i.e. WZW models). String theory predicts the
existence of nonabelian 5-brane theories, suggesting that such
generalizations indeed do exist.

\subsec{Applications }

One potential application of our considerations is to superstring
cosmology, and in particular to the   the Hartle-Hawking
wavefunction for   the RR fields of string theory. In general, in a
physical theory a  manifold $M$ with boundary $Y$ defines a state in
a Hilbert space $\CH(Y)$. In the context of supergravity this state
is the Hartle-Hawking wavefunction. We can understand the
wavefunction for the RR fields in the long-distance/weak-coupling
limit of the theory. In this limit the  wavefunction is essentially
a $\Theta$-function times some determinants. However,   writing
detailed formulae for these   two factors raises several subtle
topological issues.
 Clarifying the nature of the Hilbert space, as we have done, is a first step in
formulating this wavefunction precisely.

Another application of some interest is the generalization of
\GukovKN\ to spacetimes of the form $AdS_5 \times Z$ where $Z$ is a
Sasaki-Einstein manifold. These are expected to be holographically
dual to conformal field theories. Proposals have been made for the
holographically dual quiver gauge theories \FrancoSM. While many
Sasaki-Einstein manifolds of interest have torsion-free cohomology,
in general, Sasaki-Einstein manifolds will have torsion in their
cohomology and in their $K$-theory. A simple example is $S^5/\IZ_k$.
The states which are nonperturbative in the large-$N$ limit will
then be classified by nonabelian Heisenberg groups $QK(Z)$ of the
type we have studied. These Heisenberg groups should emerge as
groups of discrete symmetries of the quiver gauge theory. This is an
important test of the proposed holographic duals. Some recent work
elaborating this idea can be found in
\BurringtonUU\BurringtonAW\BurringtonPU.

\subsec{Relation to M-theory}

The noncommutativity of RR fluxes discussed above should also   be related to the noncommutativity of
Page charges discussed in \MooreJV.
Using the duality between $M$ theory on $Y \times S^1$ and IIA theory on $Y$ we find that
 for  the RR 6-form we expect a commutation relation such as
\eqn\sxfrm{
[\int_Y F_6 \omega^1_3, \int_Y F_6 \omega^2_3] = {i \over 2\pi} \int H \omega^1_3 \omega^2_3
}
where $\omega^1_3, \omega^2_3$ are two closed 3-forms on $Y$.
Similarly, in  IIB theory, using the duality between $M$ theory on  a $T^2$ fibration on $Y_8$ and IIB theory on
a circle fibration on $Y_8 $ we find that the self-dual 5-form must satisfy
\eqn\sdfrm{
[\int_{Y_8} F_5 \omega^1_3 , \int_{Y_8} F_5 \omega^2_3] = {i \over 2\pi } \int_{Y_8} F_2 \omega^1_3 \omega^2_3
}
where $F_2$ is the fieldstrength of the KK gauge field arising from reduction on $S^1$.
The relation of these commutation relations to those derived in type II string theory
will be discussed in  \belov.

\bigskip
\noindent{\bf Acknowledgements:} We would like to thank   J.
Distler, S. Gukov, M. Hopkins, and E. Witten for discussions and
correspondence.  GM would like to thank D. Belov for many
discussions, for collaboration on related material, and for comments
on the draft. The work of GM is supported in part by DOE grant
DE-FG02-96ER40949. The work of DF is supported in part by NSF grant
DMS-0305505. We would like to thank the Aspen Center for Physics and
the KITP for hospitality during part of this research. The KITP
workshop
 was supported in part by the National Science Foundation under
Grant No. PHY99-07949. GM would like to thank the Institute for
Advanced Study for hospitality and the Monell foundation for support
during completion of this paper.

\appendix{A}{Some mathematical background}

We include here some technical definitions which are not essential, but are included   for
completeness.

\subsec{A rough definition of differential K-theory}

One explicit model of  $\c K(M)$ is the following.   (We are
following the discussion of  \lott\HopkinsRD\freed.) We will define
``gauge potentials'' and an equivalence relation saying when these
are ``gauge equivalent.'' More mathematically, we indicate how one
can define a category whose isomorphism classes are elements of $\c
K(M)$. The category is a groupoid, i.e., all morphisms are
invertible. The objects in the category correspond to ``gauge
potentials.''
 Morphisms between objects express gauge equivalence.
An object is $(E,\nabla, c)$ where $E$ is a vector bundle, $\nabla$
is a connection on $E$ and $c \in \Omega(M)^{odd}$.  The
fieldstrength is $G= \ch(\nabla) + dc$. (Pre)-morphisms between $(E,
\nabla, c) $ and $(E', \nabla', c')$ are triples on $M \times I$
such that $\iota({\p \over \p t})G=0$.  In particular, this imposes
equivalence relations such as:  $c-c' = CS(\nabla, \nabla')$.

If $M$ is odd-dimensional and oriented for $\c K(M)$ (which entails
a choice of $Spin^c$ structure and  Riemannian metric on $M$) then
the pairing on objects can be written as:
\eqn\objectpairing{ \eqalign{ \langle (E_1, \nabla_1, c_1) , (E_2,
\nabla_2, c_2) \rangle & = \half\Biggl\{ \eta(\Dsl_{\nabla_1\otimes
1 + 1 \otimes \bar \nabla_2}) \cr & + \int c_1 {\Tr}
e^{-F(\nabla_2)} + \int \bar c_2 {\Tr} e^{F(\nabla_1)} + \int c_1
d\bar c_2\Biggr\}\mod \IZ }}
A proof that this is the same as the integration defined in
\HopkinsRD\ is being worked out in the thesis \klonoff.

An analogous formulation of $\c K^1(M)$ can be formed by integrating
classes in $\c K^0(M\times S^1)$, trivial when restricted to a point
in $S^1$,
 along $S^1$. This results in the following description. For
 fixed $M$ we   consider pairs $(g, c)$ where
$g: M \to U(N)$ for sufficiently large $N$ (depending on $M$), and
$c \in \Omega^{even}(M)$. The fieldstrength is again $G= \ch(g) +
dc$. To define $\ch(g)$ we consider the connection $\nabla_g := d +
s  g^{-1} dg$ where $s\sim s+1$ is a coordinate along $S^1$ and set
$\ch(g) = \int_{S^1} \ch(\nabla_g)$. This defines  $\ch(g)$ as a
pullback of a sum of suitably normalized Maurer-Cartan forms.
(Pre)-morphisms between $(g, c) $ and $(g', c')$ are triples on $M
\times I$ such that $\iota({\p \over \p t}) G =0$.  In particular,
this imposes equivalence relations such as:  $c-c' = \omega(g,g')$,
where $\omega(g,g')$ is a familiar object in physics from the
Polyakov-Wiegmann formula.

The above description is closely related to
 a physical interpretation in terms of brane-antibrane
annihilation in boundary string field theory \DHMunpub. (One replaces bundles by
$\IZ_2$-graded bundles for the brane/antibrane, and one replaces the
connection by a superconnection. The scalar part of the superconnection is
the tachyon field. ) Unfortunately, no such physical
description exists for twisted differential K-theory. The best available
definition is given in \HopkinsRD\ in terms of differential function spaces.
This formulation of   the theory stresses  maps into certain classifying spaces.

Briefly, one interprets a   $K^0$-theory class as a homotopy class of a map $\Psi: M \to \CF$,
where $\CF$ is the space of Fredholm operators. In the differential case we include
the data of the map itself, and moreover fix a representative $\omega^{\rm univ}$ of
the Chern character on $\CF$. Then the differential $K$-theory class is represented by a
triple $(\Psi, c,G)$ where $G \in \Omega^{\rm even}$ is the fieldstrength, $c\in \Omega^{\rm odd}$,
and $ G= \Psi^*(\omega^{\rm univ}) + dc$. Now, to introduce a twisting one wants to
replace the Fredholm operators by a {\it bundle} of Fredholm operators over $M$ (determined by
the twisting)  and  replace $\Psi$ by a section of that bundle. It is best to choose
a universal bundle $\CE$ of Fredholm operators over $K(\IZ,3)$. Now, choosing $\iota \in Z^3(K(\IZ,3),\IR)$
to represent the generator of $H^3$ we think of $\c B$ as a map $\psi_B: M \to K(\IZ,3)$,
together with $H\in \Omega^3(M)$ and $b\in C^2(M;\IR)$ with $H = \psi_B^*(\iota) + \delta b$.
A twisted differential $K$-theory class is defined by a lifting of this map into $\CE$.
A similar construction with $\CF$ replaced by skew-adjoint odd Fredholm operators on a $\IZ_2$-graded
Hilbert space leads to twisted odd differential $K$-theory.

It would be very nice to find a physical model for twisted differential $K$-theory
which is more obviously connected to the physics of D-branes.

\subsec{A heuristic description  of the twisted Chern character}

We give a simple heuristic definition of the
twisted Chern character $ch_B: K^{0 ,B}(M) \to H_{d_H}^{\rm even}(M)$.
 We represent a twisted
K-theory class by   local bundles $E_\alpha$ defined over an
open covering $\CU_\alpha$  of $M$.
\foot{In general we must use virtual bundles. If the
cohomology class $[H]$ is nontorision then, since the AHSS
differential is $Sq^3 + [H]$, the virtual dimension
of $E_\alpha$ must vanish if it is to represent a K-theory class.}
 On overlaps we have
 $E_\alpha \cong E_{\beta} \otimes L_{\alpha\beta}$.
The $E_\alpha$ are equipped with connections $\nabla_\alpha$,
and the twisting is thought of as a gerbe $B_\alpha$.
Note that $H= dB_\alpha$ is globally well-defined.
On overlaps we have $B_\alpha = B_\beta + F_{\alpha \beta} $
and $F_{\nabla_\alpha} = F_{\nabla_\beta} - F_{\alpha \beta} {\bf 1} $
where $F_{\alpha\beta}$ is fieldstrength of a  connection on the line
$L_{\alpha\beta}$.  With this data we can define
\eqn\twischernch{
ch(x)\vert_{U_\alpha} = e^{B_{\alpha}} {\Tr}_{E_\alpha} e^{F(\nabla_\alpha)}
}
This gives a globally-defined $d_H$-closed form. The odd case can be obtained,
formally, by considering $M\times S^1$, as usual.

\appendix{B}{Mathematical formulation of the tadpole condition}

The ``tadpole condition'' is the quantum mechanical implementation
of the familiar fact that the total charge in an abelian gauge
theory must vanish on a compact manifold $Y$. In this appendix we
take a point of view explained at length in \DiaconescuBM. See  also
\MooreJV\ for a brief account. We first sketch the physical idea and
then sketch how it can be made mathematically precise.

In the classical theory of \maxactthrd\ the equations of motion
are
\eqn\eqmot{
2\pi R^2 d *F + 4 \pi k F =0
}
and hence do not admit a solution unless the DeRham cohomology class
$[F]=0$. In the quantum theory, this is refined to a condition on
the characteristic class $a \in H^{p+1}(Y,\IZ)$. Quantum
mechanically, one wants to work with wavefunctions $\Psi(\c A)$,
where $\c A$ is some kind of ``gauge potential,'' which are gauge
invariant, i.e. satisfy the Gauss law:
\eqn\glone{
g \cdot \Psi(\c A) = \Psi(g \cdot \c A)
}
where $g$ is a ``gauge transformation.'' For the case $p>1$ this
terminology is  presently ambiguous but will be made precise below.
If $g$ is a ``global gauge transformation'' then $g \cdot \c A = \c
A$, i.e. $g$ is an automorphism of $\c A$. On the other hand, one
can {\it define} the total electric charge of a state $\Psi$ as the
character of the action of global gauge transformations:
\eqn\totchge{
g \cdot \Psi(\c A) = \rho_Q(g) \Psi(  \c A)
}
In the case $[\c A] \in \c H^{p+1}(Y)$ we have ${\rm Aut}(\c A) = H^{p-1}(Y,\IR/\IZ)$
and $\rho_Q(g) = \exp [ 2\pi i \langle g, Q\rangle ]$ for $Q\in H^{p+1}(Y, \IZ)$.
From \glone\ we learn that if $\Psi$ is to be nonzero then  $Q=0$, as expected.

Mathematically, we can make these notions precise as follows. As we
have mentioned in footnotes 8 and 11, $\c H^{\ell}(Y)$ may be
regarded as isomorphism classes of objects in a category. Unlike
nonabelian gauge theory there are different  but equivalent
physically useful categories we can use for the groupoid of gauge
potentials, but we will choose for definiteness the category of
differential cocycles $\c Z^{\ell}(Y)$ defined in \HopkinsRD. The
Chern-Simons term:
\eqn\csfunctor{
\c A \to \int_Y \c A \star  \c A
}
is proved in \HopkinsRD\ to be a {\it functor} from $\c Z^{p+1}(Y)
\to \c Z^2(pt)$ (recall $\dim Y =2 p$).  Moreover, the category $\c
Z^2(pt)$ is equivalent to the groupoid  of lines with invertible
linear transformations. In particular, automorphisms of $\c A$ map
to automorphisms of the line $L_{\c A}$. This map is a group
homomorphism $H^{p-1}(Y,\IR/\IZ) \to U(1)$. Applying the functor
\csfunctor\ to the universal connection gives an element of $\c
Z^2(\c Z^{p+1}(Y))$, that is, a line bundle with connection over the
groupoid $\c Z^{p+1}(Y)$.

In general, a line bundle over a groupoid $\Xi$ is a continuous
functor from $\Xi$ to the groupoid of lines with invertible linear
transformations. Let us spell out what this entails. If $\Xi_0$ is
the space of objects in $\Xi$ then we have, first of all, a line
bundle $L \to \Xi_0$. Next, if $\Xi_1$ is the set of morphisms then
there are two maps $p_0, p_1: \Xi_1 \to \Xi_0$ given by source and
target. We have an isomorphism $\xi: p_1^*L \to p_0^* L$,
equivalently, a nonzero element
\eqn\linegroupoid{
\xi^{\varphi}_{x_1, x_2} \in L_{x_1}^{-1} \otimes L_{x_2}
}
for each morphism $\varphi: x_1 \rightarrow x_2$. If $\varphi_2, \varphi_1$ are composable
morphisms then we require:
\eqn\compat{
\xi^{\varphi_1}_{x_1, x_2}\xi^{\varphi_2}_{x_2, x_3}= \xi^{\varphi_1\varphi_2}_{x_1, x_3}.
}
A section of a line bundle over a groupoid is a section $s$ of $L\to \Xi_0$ which is invariant,
i.e. $\xi(p_1s) = p_0 s$, or in other words $s(x_2) = s(x_1) \xi^{\varphi}_{x_1,x_2}$. Note,
in particular, that $s(x)$ can only be nonzero if ${\rm Aut}(x)$ acts trivially on $L_x$.

Now, a physical wavefunction must be a section of the line bundle over the groupoid
$\c Z^{p+1}(Y)$. It follows that nonzero wavefunctions only exist if the automorphisms
act trivially.

One way of producing physical wavefunctions is via the path integral over a manifold $X$ with
boundary $Y$. Denote the inclusion $\iota: Y \hookrightarrow X$.
 Fix an object $\c A_\p \in \c Z^{p+1}(Y)$, and consider a set of pairs
$(\c A, \theta)$ where $\c A \in \c Z^{p+1}(X)$ and $\theta: \iota^*\c A \to \c A_\p$ is an
isomorphism. Impose an equivalence relation $(\c A, \theta) \cong (\c A', \theta')$ if
there is a morphism $\Phi$ in $\c Z^{p+1}(X)$ with $\theta = \theta' \iota^* \Phi $.
Denote the resulting space of equivalence classes $\CB(\c A_\p)$.
The Chern-Simons action
\eqn\csaction{
{\rm CS}(\c A) = e^{2\pi i k \int_X \c A \star  \c A}
}
is an element of $L_{\c A_\p}$, and moreover the Chern-Simons action on a
manifold with boundary is in fact gauge invariant in the sense that
\eqn\gaugeinvt{
\xi^{\iota^* \Phi}_{\iota^* \c A, \iota^* \c A'} {\rm CS}(\c A) = {\rm CS}(\c A').
}
See \FreedVW\ for a detailed explanation of this.
By \compat\ the images under $\theta$ in $L_{A_\p}$ are equal.
It therefore makes (formal) sense to integrate the exponentiated action \maxactthrd\ over the
equivalence classes $\CB(\c A_\p)$ to produce an element $Z(\c A_\p) \in L_{\c A_\p}$.
This is indeed gauge invariant, i.e., a section of the line bundle over the
groupoid since, if $\varphi: \c A_\p \to \c A_\p'$ is a morphism then
$(\c A, \theta) \to (\c A , \varphi\theta)$ gives a map $\CB(\c A_\p) \to \CB(\c A_\p')$
mapping $Z(\c A_\p)$ appropriately, again by ``gauge invariance'' of the
Chern-Simons action.

Finally, one can give a general formula for the action of ${\rm Aut}(\c A_\p)$ on $L_{\c A_\p}$.
Consider $(\pi^* \c A_\p, \theta)$ on $Y\times [0,1]$ where $\pi^*$ is the pullback to the cylinder,
$\theta_0=1$ and $\theta_1 \in {\rm Aut}(\c A_\p)$. By the functorial definition
of a field theory this cobordism represents an action of the automorphism $\theta_1$ on
$L_{\c A_\p}$. This action must be a complex number, which, by gluing is given by
${\rm CS}(\pi^*\c A_\p, \theta)$ where we regard $(\pi^*\c A_\p, \theta)$ as a
``twisted connection'' on $Y \times S^1$. For further justification of this last step
see $\S{6.3}$ of \DiaconescuBM. Applying this construction to \csaction\ we find that
for $\theta\in H^{p-1}(Y,\IR/\IZ)$ the Chern-Simons functional of the twisted connection is
\eqn\homomorphism{
\exp[ 2\pi i k \langle \theta, a \rangle ]
}
and hence the tadpole condition is $ka =0$. As noted above, there is
a version of this argument when $k$ is half-integral. See \WittenVG
\HopkinsRD \BelovJD\ for a discussion of this.

\listrefs

\bye